\RequirePackage{fix-cm}
\documentclass[twocolumn]{svjour3}          
\smartqed  

\usepackage{graphicx}
\usepackage{amsmath}
\usepackage{empheq}
\usepackage{multirow}
\usepackage{dashrule} 
\usepackage{amssymb}
\usepackage{paralist}
\usepackage{fixmath}
\usepackage{hyperref}
\usepackage[round]{natbib}
\usepackage{xcolor}
\usepackage{tabularx}
\DeclareGraphicsExtensions{.eps}

\begin{document}

\title{A method to determine wall shear stress from mean profiles in turbulent boundary layers 
}

\author{Praveen Kumar   \and
        Krishnan Mahesh
}

\institute{Praveen Kumar \and Krishnan Mahesh \at
              Department of Aerospace Engineering and Mechanics, \\
              University of Minnesota, Minneapolis, Minnesota 55455, USA \\
              \email{kmahesh@umn.edu}
}

\date{Received: date / Accepted: date}

\maketitle

\begin{abstract}
The direct measurement of wall shear stress in turbulent boundary layers (TBL) is challenging, therefore requiring it to be indirectly determined from mean profile measurements. Most popular methods assume the mean streamwise velocity to satisfy either a logarithmic law in the inner layer or a composite velocity profile with many tuned constants for the entire TBL, both of which require reliable data from the inner layer. The presence of roughness and pressure gradient brings additional complications where most existing methods either fail or require significant modification. A novel method is proposed to determine the wall shear stress in zero pressure gradient TBL from measured mean profiles, without requiring near-wall data. The method is based on the stress model of Kumar and Mahesh [Phys. Rev. Fluids 6, 024603 (2021)], who developed accurate models for mean stress and wall-normal velocity in zero pressure gradient TBL.  The proposed method requires a single point measurement of mean streamwise velocity and mean shear stress in the outer layer, preferably between 20 to $50 \%$ of the TBL, and an estimate of boundary layer thickness and shape factor. The method can handle wall roughness without modification and is shown to predict friction velocities to within $3 \%$ over a range of Reynolds number for both smooth and rough wall zero pressure gradient TBL. In order to include the pressure gradients effects, the work of Kumar and Mahesh [Phys. Rev. Fluids 6, 024603 (2021)] is revisited to derive a novel model for both mean stress and wall-normal velocity in pressure gradient TBL, which is then used to formulate a method to obtain the wall shear stress from the profile data. Overall, the proposed method is shown to be robust and accurate for a variety of pressure gradient TBL. 

\end{abstract}

\section{Introduction}
\label{intro}

Flow over a solid surface or wall produces a thin region of high shear close to the wall due to no-slip conditions imposed  at surface by viscosity. This thin near-wall `boundary layer' has been extensively studied since the seminal work of Prandtl \citep{prandtl1904}. The high shear induces large tangential stresses at the wall, which leads to drag and energy expenditure in aerodynamic and hydrodynamic applications. Under most practical conditions, the boundary layer is turbulent, and is commonly described using first- and second-order flow field statistics \citep{pope}. 

In turbulent boundary layers, the skin-friction is represented by the friction velocity defined as $u_{\tau} = \sqrt{\tau_w/\rho}$ where $\tau_w$ is the mean wall-stress and $\rho$ is density of the fluid. $u_{\tau}$ is  a very important velocity scale in TBL due to its common use in most scaling laws. Unfortunately, the direct measurement of $u_{\tau}$ is often not feasible, requiring indirect methods to determine it. Although a number of indirect techniques are available for determining $u_{\tau}$, none are universally accepted. 

An indirect technique can be formulated if the behavior of the mean velocity profile is known in viscous units. Several so called `law of the wall' have been proposed for the inner layer (typically $y/\delta < 0.15$) that relate the inner scaled mean streamwise velocity ($U^+$) to wall normal coordinate ($y^+$), where `+' denotes normalization with $u_{\tau}$ and $\nu$. Table \ref{tab:lotw} list some of the popular formulae for the law of the wall. The readers are referred to \cite{zhang2021} for a more extensive list.

\begin{table*}
\caption{Popular formulae for the law of the wall.}
\label{tab:lotw}
\centering
\begin{tabular}{c@{\hskip 0cm}c@{\hskip 0cm}c@{\hskip 0cm}c@{\hskip 0cm}c}
\hline\hline
 Authors                & Formulae                               & Region where valid   \\ \hline
 \cite{Prandtl1910}    & $U^+ = y^+$                            & $0\leq y^+ \leq 11.5$   \\
 \cite{Taylor1916}     & $U^+ = 2.5\mathrm{ln} y^+$ + 5.5       & $y^+ \geq 11.5$ \\
 \cite{Karman1939}     & $U^+ = y^+$                            & $0\leq y^+ < 5$  \\
                        & $U^+ = 5\mathrm{ln} y^+$ - 3.05        & $5 \leq y^+ < 30$ \\
                        & $U^+ = 2.5\mathrm{ln}y^+$ + 5.5        & $ y^+ \geq 30$ \\
 \cite{rannie1956}     & $U^+ = 1.454\mathrm{tanh}(0.0688y^+)$  & $0\leq y^+ < 27.5$  \\
                        & $U^+ = 2.5\mathrm{ln} y^+ + 5.5 $               &  $ y^+  \geq 27.5$ \\
\cite{werner1993}      & $U^+ = y^+$                            & $0\leq y^+ < 11.81$ \\
                        & $U^+ = 8.3(y^+)^{1/7}$                 & $y^+ \geq 11.81$ \\
\cite{reichardt1951}   & $U^+ = 2.5 \mathrm{ln} (1 + 0.4y^+) + 7.8(1 - e^{-y^+/11} - (y^+/11)e^{-0.33y^+})$ & $y^+ \geq 0$ \\
\cite{spalding1961}    & $y^+ = U^+ + 0.1108 \bigg(e^{0.4\kappa U^+} - 1 -\kappa U^+ - \frac{(\kappa U^+)^2}{2} 
- \frac{(\kappa U^+)^3}{6} - \frac{(\kappa U^+)^4}{24} \bigg )$ & $y^+ \geq 0$ \\
\cite{monkewitz2007}   & $U^+_{inner} = U^+_{inner,23} + U^+_{inner,25}$, where & $y^+ \geq 0$ \\
                        & $U^+_{inner,23} = 0.68285472 \ \mathrm{ln}(y^{+2} + 4.7673096y^+ + 9545.9963)$ \\
& $+ 1.2408249 \ \mathrm{arctan}(0.010238083y^+ + 0.024404056)$ \\
& $+ 1.2384572 \ \mathrm{ln}(y^+ + 95.232690) - 11.930683 $ \\
                        & $U^+_{inner,25} = - 0.50435126 \ \mathrm{ln}(y^{+2} - 7.8796955y^+ + 78.389178)$\\
& $+ 4.7413546 \ \mathrm{arctan}(0.12612158y^+ - 0.49689982)$ \\
& $- 2.7768771 \ \mathrm{ln}(y^{+2} + 16.209175y^+ + 933.16587)$ \\
& $+ 0.37625729 \ \mathrm{arctan}(0.033952353y^+ + 0.27516982) $\\
& $+ 6.5624567 \ \mathrm{ln}(y^+ + 13.670520) + 6.1128254 $ \\
 \hline\hline
\end{tabular}
\end{table*}

\begin{table*}
\caption{Values of the log law constants in TBL.}
\label{tab:constants}
\centering
\begin{tabular}{c@{\hskip 1cm}c@{\hskip 1cm}c@{\hskip 1cm}c@{\hskip 1cm}c}
\hline\hline
 Authors                & Values                      & Region where valid   \\ \hline
 \cite{coles1956}      & $\kappa = 0.41$, $B=5$       & $50 < y^+ < 0.2Re_{\tau}$   \\
 \cite{osterlund1999}  & $\kappa = 0.38$, $B=4.1$     & $50 \leq y^+ \leq 0.15 Re_{\tau}$\\
 \cite{nagib2007}      & $\kappa = 0.384$, $B=4.173$  & $50 \leq y^+ \leq 0.15 Re_{\tau}$ \\
 \cite{marusic2013}    & $\kappa = 0.39$, $B=4.3$     & $3Re_{\tau}^{1/2}< y^+ < 0.15 Re_{\tau}$ \\
 \cite{monkewitz2007}  & $\kappa = 0.384$, $B=4.17$   & $50 \leq y^+ \leq 0.15 Re_{\tau}$ \\
 \cite{monkewitz2017}  & $\kappa = 0.384$, $B=4.22$   & $50 \leq y^+ \leq 0.15 Re_{\tau}$ \\
 \hline\hline
\end{tabular}
\end{table*}

Boundary layer researchers commonly use the Clauser chart method \citep{fernholz1996}, which assumes that the mean velocity satisfies the universal logarithmic law 
\begin{eqnarray} \label{eq:log}
U^+ = \frac{1}{\kappa}\mathrm{ln} y^+ + B
\end{eqnarray}
over the range $y^+ > 30$ and $y/\delta \leq 0.15$, where $\kappa \approx 0.4$ and $B \approx 5$. The exact values of the constants $\kappa$ and $B$, their universality, and the exact region of validity of Eq. \eqref{eq:log} have been widely debated in the literature \citep{TBLreview2011}. Table \ref{tab:constants} shows some of the popular values of the constants for zero pressure gradient TBL and region of validity. Using the Clauser chart method, $u_{\tau}$ is determined to be the value that best fits the measured mean streamwise velocity profile to Eq. \eqref{eq:log} in its range of validity. For example, \cite{degraaf} obtained $u_{\tau}$ using the Clauser chart method by fitting the mean velocity data to Eq. \eqref{eq:log} in the region $y^+ > 50$ and $y/\delta < 0.2$ with $\kappa = 0.41$ and $B=5$.

Several researchers have recognized and documented the limitations of the Clauser chart method. \cite{george1997} showed clear discrepancies between mean velocity profiles scaled using direct measurements of $u_{\tau}$ and approximations using the Clauser chart method. \cite{wei2005} explicitly illustrated how the Clauser chart method can mask subtle $Re$-number-dependent behavior, causing significant error in determining $u_{\tau}$. Several past works have attempted to improve the Clauser chart method by optimizing the constants of Eq. \ref{eq:log}. \cite{rodriguez2015} analyzed several such past efforts and proposed a better method by optimizing the various parameters appearing in the mean velocity profile of the entire TBL. However, as mentioned by the authors themselves, their method was sensitive to the assumed universal velocity profile, and the definition of the error which was used for the optimization. 

A more recent approach to determine $u_{\tau}$ \citep[e.g.][]{samie2018} is to fit the entire mean streamwise velocity profile to the composite profile of \cite{chauhan2009}:
\begin{equation}
U^+_{composite}   =   U^+_{inner} + \frac {2\Pi}{\kappa} W(\eta),
\end{equation}
where,
\begin{multline}
U^+_{inner}   =   \frac{1}{\kappa} \mathrm{ln}\bigg(\frac{ y^+ - a}{-a}\bigg) + \\
\frac{R^2}{a(4\alpha - a)} \bigg[(4\alpha + a) \mathrm{ln} \bigg (- \frac{a}{R} \frac{\sqrt{(y^+ - \alpha)^2 + \beta^2}}{y^+-a} \bigg ) \\
+ \frac{\alpha}{\beta} (4\alpha + 5a) \bigg (\mathrm{arctan} \bigg (\frac{y^+ -\alpha}{\beta} \bigg ) + \mathrm{arctan} \bigg (\frac{\alpha}{\beta} \bigg ) \bigg )\bigg ],
\end{multline}
with $\alpha = (-1/\kappa - a)/2$, $\beta = \sqrt{-2a\alpha - \alpha^2}$, and $R = \sqrt{\alpha^2 + \beta^2}$. The chosen values of $\kappa = 0.384$ and $a = -10.3061$. The wake function $W$ is given by
\begin{multline}
W (\eta)  = \\
\bigg [\bigg (1 - exp(-(1/4)(5a_2 + 6a_3 +7a_4)\eta^4 + a_2\eta^5 + a_3\eta^6 +a_4\eta^7) \bigg ) \\ \bigg /
\bigg(1 - exp(-(a_2 + 2a_3 + 3a_4)/4) \bigg ) \bigg ]
\bigg ( 1 - \frac{1} {2\Pi} \mathrm{ln} \eta \bigg ),
\end{multline}
where $a_2$ = 132.8410, $a_3=-166.2041$, $a_4=71.9114$, the wake parameter $\Pi = 0.45$ and $\eta = y/\delta$. \cite{chauhan2009} have also shown that the accuracy of the approach to be within $\pm 2 \%$ of those determined by direct measurements. Although the method appears attractive, it still requires accurate near-wall measurements to determine $u_{\tau}$ reliably. Moreover, the method can not be used for rough wall and pressure gradient TBL. 

Several recent studies \citep[e.g.][]{fik,mehdi2011,mehdi2014} have attempted to relate $u_{\tau}$ to measurable profile quantities using the governing equations of the mean flow. These methods often require additional measurements and assumptions to account for the unknown quantities. For example, \cite{mehdi2011} and \cite{mehdi2014} obtained $u_{\tau}$ using the measured mean streamwise velocity and Reynolds shear stress profiles. However, their method was sensitive to noisy or missing near-wall data. They overcome this limitation by assuming a shape for the total shear stress profile and by smoothing the measured data.

The presence of pressure gradient makes the indirect determination of $u_{\tau}$ from profile measurements more challenging. Clauser chart method requires significant modifications before it can be employed for obtaining $u_{\tau}$ for any pressure gradient TBL. Several non-dimensional parameters have been proposed in literature to quantify pressure gradient effects on TBL. These parameters only differ in length and velocity scales used to normalize the freestream pressure gradient ($dP_e/dx$). Some of the popular pressure gradient parameters are
\begin{eqnarray}
\beta & = & \frac{\delta^*}{\rho u_{\tau}^2} \frac{dP_e}{dx}, \\
K & = & -\frac{\nu}{\rho U_e^3}\frac{dP_e}{dx}, \\
p_x^+ & = & \frac{\nu}{\rho u_{\tau}^3}\frac{dP_e}{dx}.
\end{eqnarray}
Here, $\rho$ is the fluid density, $\delta^*$ is the displacement thickness, $\nu$ is the kinematic viscosity, $U$ is the streamwise velocity and the subscript `e' denotes the value at the edge of the boundary layer ($\delta$), which is defined as the wall-normal location where $U=0.99U_e$. In the present work, the Rotta--Clauser pressure gradient parameter ($\beta$) \citep{rotta, clauser} is used, which is common in literature for near-equilibrium TBL. It is important to understand that one requires knowledge of $u_{\tau}$ to obtain $\beta$, which is not available {\it apriori}. Therefore, experimental studies often use the acceleration parameter $K$ to quantify pressure gradient. It can be readily seen that they are related as
\begin{eqnarray}
\beta & = &  - K Re_{\delta^*} \frac{U_e^2}{u_{\tau}^2}, \label{eq:k}
\end{eqnarray}
where, $Re_{\delta^*}$ is the displacement thickness based $Re$.

\cite{volino2018} used integral analysis of the mean streamwise momentum equation to obtain a relation between mean velocity and Reynolds shear stress profiles, that was subsequently used to determine $u_{\tau}$ from measured profiles. The utility of the method was shown through application to experimental data including zero pressure gradient cases from smooth and rough walls, and smooth wall cases with favorable and adverse pressure gradients. Although their method is generally applicable, it has two main issues: (i) the formulation attempts to minimize dependence on streamwise gradients, but some dependence remains, making data from two or more streamwise locations necessary and the process of determining $u_{\tau}$ iterative, and (ii) since the method requires numerical integration of the profiles in the wall-normal direction, the accuracy deteriorates if near-wall data is omitted. In order to improve the accuracy in such cases, they either used the trapezoidal rule or an assumed velocity profile to account for the missing contribution. The requirement to have data at multiple streamwise locations to obtain streamwise gradients was mitigated by \cite{womack2019} by essentially using a skin-friction law and connecting the streamwise gradient to the wall-normal gradient using a universal $W$. However, this makes the method sensitive to the choice of $W$ and the constants in the law of the wall. 

It is clear that most of the issues in determining $u_{\tau}$ from profile data stem from the presumption of a universal mean velocity profile that is valid in the inner layer or throughout the TBL. Therefore, a method is proposed to determine $u_{\tau}$ from mean profile measurements without any assumption of a universal mean velocity profile, thereby avoiding all issues associated with uncertainties in the constants of the assumed profile. The method  utilizes profiles of the mean velocity and the mean shear stress without requiring any near-wall ($\eta < 0.2$) data, which is challenging to acquire at high $Re$ \citep{vallikivi2015,samie2018}. Presence of wall roughness and pressure gradients brings additional complications. The method is based on integral analysis of the streamwise momentum equation but does not require profiles at multiple streamwise locations, wall-normal integration of data or any iterative procedure to determine $u_{\tau}$ accurately. The method is also shown to be general enough to include wall roughness and is extended to include pressure gradient gradients.

The paper is organized as follows. The proposed method for zero pressure gradient (ZPG) TBL is described in Section \ref{sec:model}. Section \ref{sec:data} describes the databases used in the present work. The performance of the proposed method for smooth and rough wall ZPG TBL is discussed in Section \ref{sec:res}. In order to extend the method for pressure gradient TBL, a novel model each for mean stress and wall-normal velocity is first derived, which is subsequently used to propose a method to determine $u_{\tau}$ and assess its performance in a variety of pressure gradient TBL cases in Section \ref{sec:pg}. Section \ref{sec:conc} concludes the paper. 

\section{The proposed method for ZPG TBL} \label{sec:model}

The boundary layer approximations for the time-averaged Navier--Stokes equations in Cartesian coordinates yield
\begin{eqnarray}
\label{eq:bl1}
\frac{\partial U}{\partial x} + \frac{\partial V}{\partial y} & = & 0, \\
\label{eq:bl2}
U\frac{\partial U}{\partial x} + V\frac{\partial U}{\partial y} & = & -\frac{1}{\rho}\frac{dP}{dx} + \frac {\partial T} {\partial y},
\end{eqnarray}
where $U$ and $V$ are the streamwise ($x$) and the wall-normal ($y$) components of mean velocity vector, $\rho$ is the fluid density, $P$ is the mean pressure and $T$ is the mean total stress, i.e., the sum of the viscous stress ($\nu {\partial U}/{\partial y}$) and the Reynolds shear stress ($-\overline{u'v'}$). Using Eqs. \eqref{eq:bl1} and \eqref{eq:bl2}, and neglecting the terms containing pressure gradient and wall-normal velocity, it can be shown that
\begin{eqnarray}
\frac {\partial T} {\partial y} & = &
-U_e\frac{\partial V}{\partial y} + (U_e-U) \frac{{\partial V}}{\partial y}.
\label{eq:bl4}
\end{eqnarray}
($\delta$), which is defined as the wall-normal location where $U=0.99U_e$.   
Integrating Eq. \eqref{eq:bl4} from a generic $y$ to $y=\delta$ and normalizing in viscous units yield
\begin{eqnarray}
{T^+} & = & {U_e^+V_e^+} \bigg(1 - \frac{V}{V_e}\bigg) - \int_{y^+}^{\delta^+} (U_e^+-U^+) \frac{{\partial V^+}}{\partial y^+} dy^+
\label{eq:bl6}
\end{eqnarray}
where the second term requires modeling. \cite{kumar2021} used available TBL data to model the second term in the rhs of Eq. \eqref{eq:bl6} to obtain a modeled total stress,
\begin{eqnarray}
{T^+} & = & H (1 - V/V_e) + (H-1) (\eta - 1)
\label{eq:Tmodelin}
\end{eqnarray}
where, $H$ is the shape factor defined as the ratio of $\delta^*$ and momentum thickness ($\theta$). Note that the relation $U_e^+V_e^+=H$ \citep{wei2016, kumar2018} is used in Eq. \eqref{eq:Tmodelin}. The mean shear stress model was validated for a range of $Re$ using past simulations and experiments. Since, most past work do not report the $V$ profile, \cite{kumar2021} also provided a compact model for $V/V_e$ as function of $\eta$, i.e.,
\begin{eqnarray}
\frac{V}{V_e} = \mathrm{tanh} (a \eta + b \eta^3 ),
\label{eq:Vmodel}
\end{eqnarray}
where the model constants $a=0.5055$ and $b=1.156$. Now, Eq. \eqref{eq:Tmodelin}  can be rearranged to obtain 
\begin{eqnarray}
u_{\tau} & = & \sqrt{ \frac {T} {H(1 - V/V_e) + (H - 1) (\eta - 1)}}.
\label{eq:method}
\end{eqnarray}

\begin{figure}
\begin{center}
\includegraphics[height=55mm]{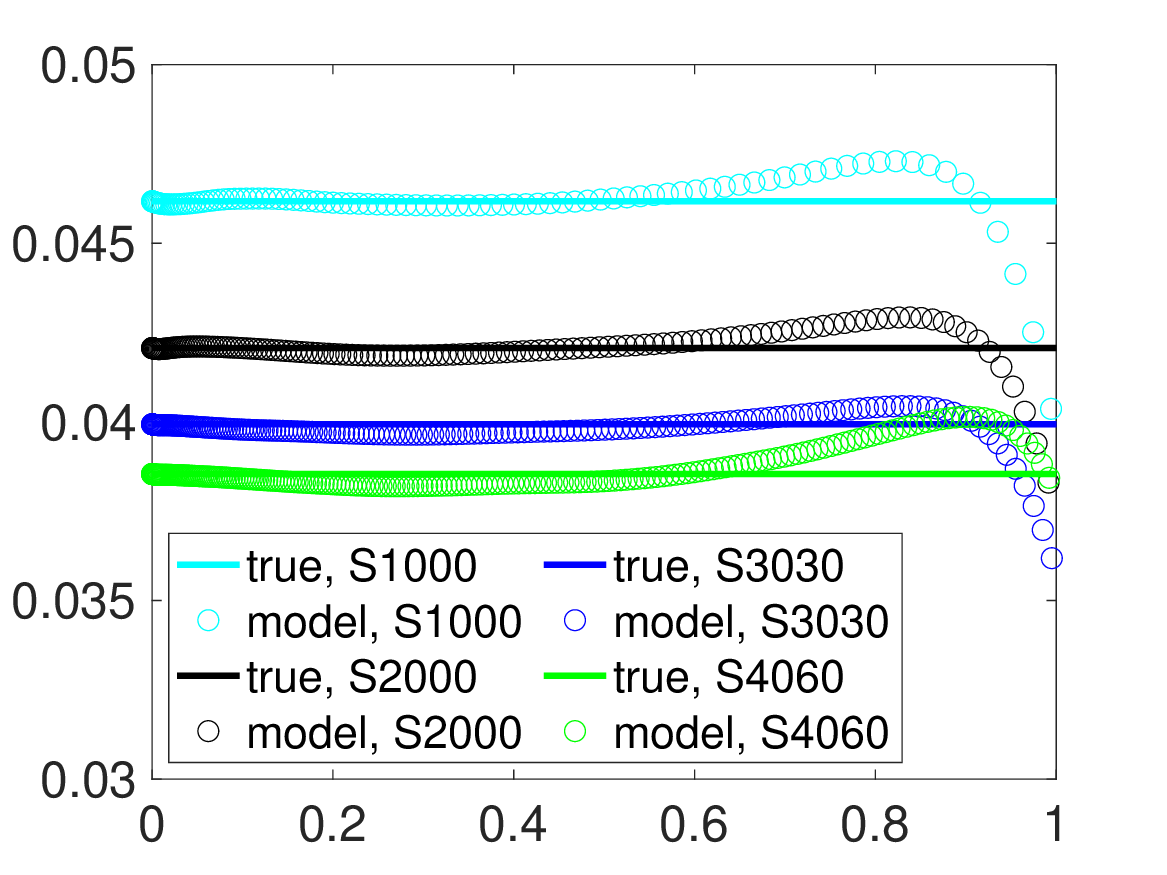}
\put(-220,90){$u_{\tau}/U_e$} \\
\includegraphics[height=55mm]{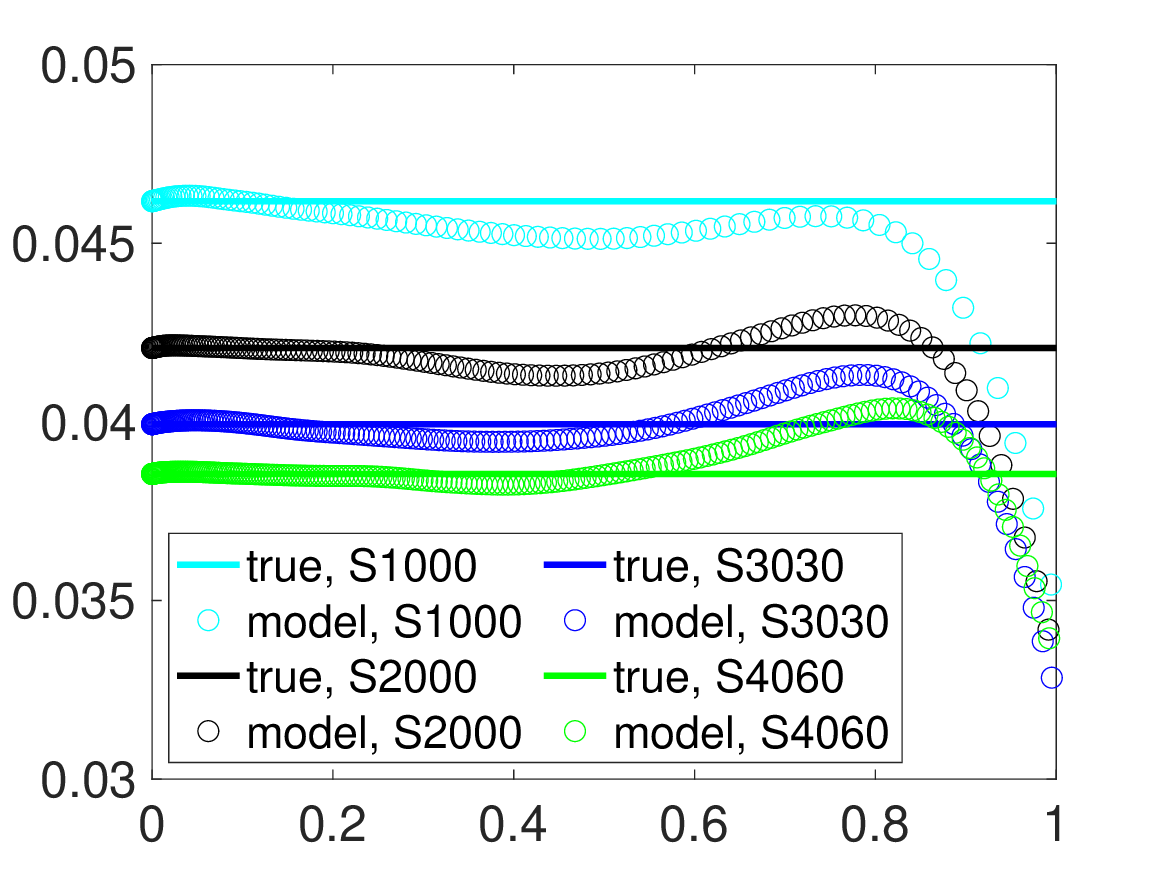}
\put(-110,-5){$\eta$}
\put(-220,90){$u_{\tau}/U_e$}
\end{center}
\caption{Profiles of rhs of Eq. \eqref{eq:method} (symbols) are compared to true $u_{\tau}$ (lines) for cases S1000, S2000, S3030 and S4060, using actual $V/V_e$ $(a)$ and modeled $V/V_e$ $(b)$ using Eq. \eqref{eq:Vmodel}.}
\label{fig:methodtest}
\end{figure}

Figure \ref{fig:methodtest} shows the rhs of Eq. \eqref{eq:method} for cases S1000, S2000, S3030 and S4060 (i.e. $Re_\theta =$ 1006 to 4061, (see Table \ref{tab:data}) using actual $V/V_e$ (Figure \ref{fig:methodtest} $(a)$) and modeled $V/V_e$ (Figure \ref{fig:methodtest} $(b)$) using Eq. \eqref{eq:Vmodel}. The true values of $u_{\tau}$ are also shown for comparison. Each symbol represents $u_{\tau}$ predicted using data point at only that location in the boundary layer. It is clear that Eq. \eqref{eq:method} predicts $u_{tau}$ accurately for any data point in the range $\eta < 0.6$. It is also evident that as $Re$ increases, the difference between the $u_{\tau}$ obtained using actual $V/V_e$ and modeled $V/V_e$ becomes increasingly smaller. Since $V$ is often not available, all subsequent results in this paper use modeled $V$ (Eq. \eqref{eq:Vmodel}).

The proposed method determines $u_{\tau}$ from Eq. \eqref{eq:method}. In principle, only one measurement location anywhere in the range $\eta<0.6$ is sufficient to obtain $u_{\tau}$ as observed in figure \ref{fig:methodtest}. However, if a coarse profile measurement is available, $u_{\tau}$ value is the horizontal line which best fits the rhs of Eq. \eqref{eq:method} data in the range $0.2 <\eta<0.5$. The region $\eta < 0.2$ is deliberately avoided since data in this region is difficult to acquire in experiments at high $Re$ \citep{vallikivi2015,samie2018}. 

\section{TBL databases} \label{sec:data}

Table \ref{tab:data} lists the relevant details of all the ZPG direct numerical simulation (DNS) and experimental databases used in this paper. For any TBL, momentum-based Reynolds number $Re_{\theta}$ and friction-based Reynolds number ($Re_{\tau}$) are defined as
$Re_{\theta} = U_{e}\theta/\nu$, and $Re_{\tau} = u_{\tau}\delta/\nu$. Note that the correlation $Re_{\tau} = 1.13 \times Re_{\theta}^{0.843}$ proposed by \cite{schlatter2010} is used to obtain $Re_{\theta}$ from the reported values of $Re_{\tau}$ in the experiments of \cite{morrill2015} and \cite{baidya}. All these smooth wall cases are named starting `S' or `E' to represent simulations or experiments respectively, followed by the rounded-off value of $Re_{\theta}$. The last three cases listed in the table are taken from \cite{flack2020} who performed experiments on rough wall TBL by systematically changing the surface skewness of a rough surface with same root-mean-square roughness height. These rough wall cases are named started `R' to represent rough, followed by `0', `-' or `+' for zero, negative and positive surface skewness respectively.  

Table \ref{tab:data1} lists the relevant details of all the APG TBL profiles used in this paper taken from the wall-resolved large eddy simulation (LES) databases of \cite{bobke2017}. \cite{bobke2017} had a large region of controlled $\beta$ making their data suitable for validation.

\begin{figure}
\begin{center}
\includegraphics[height=55mm]{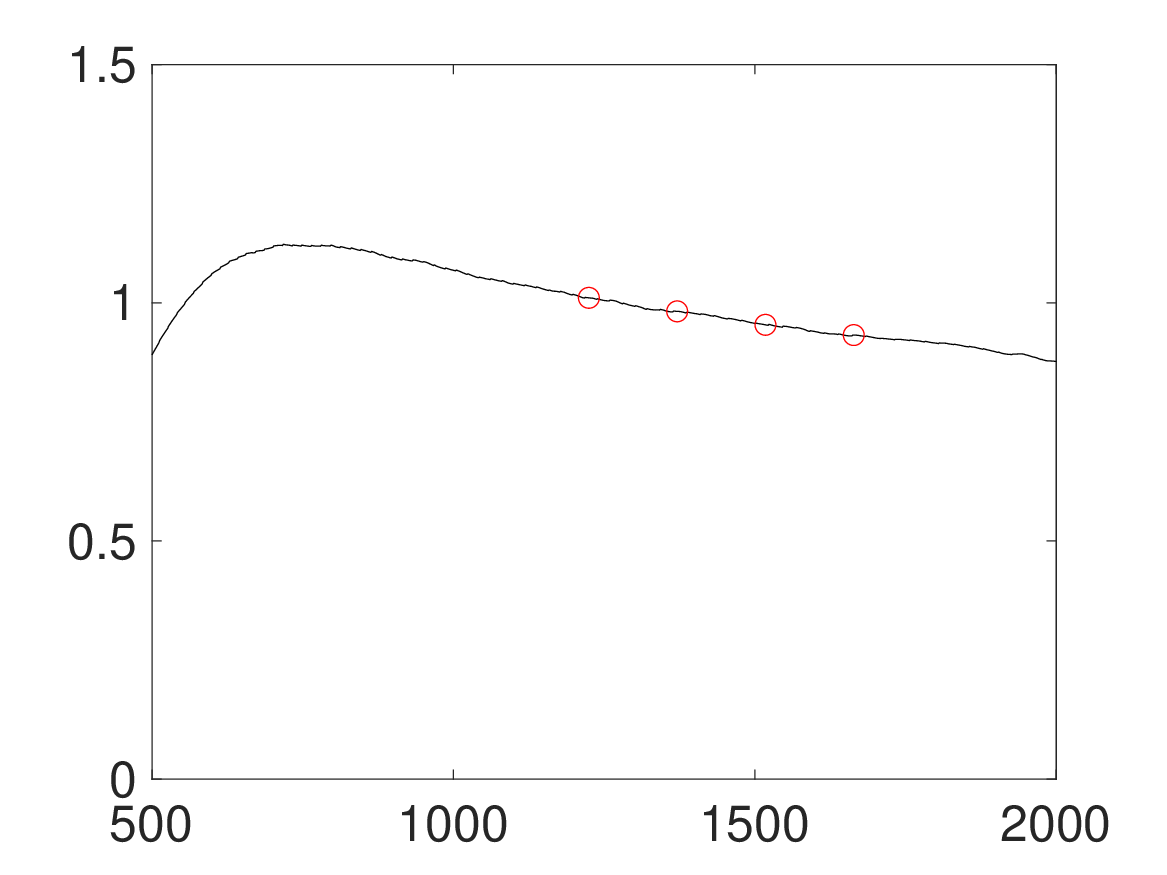}
\put(-220,140){$(a)$} 
\put(-220,90){$\beta$} \\
\includegraphics[height=55mm]{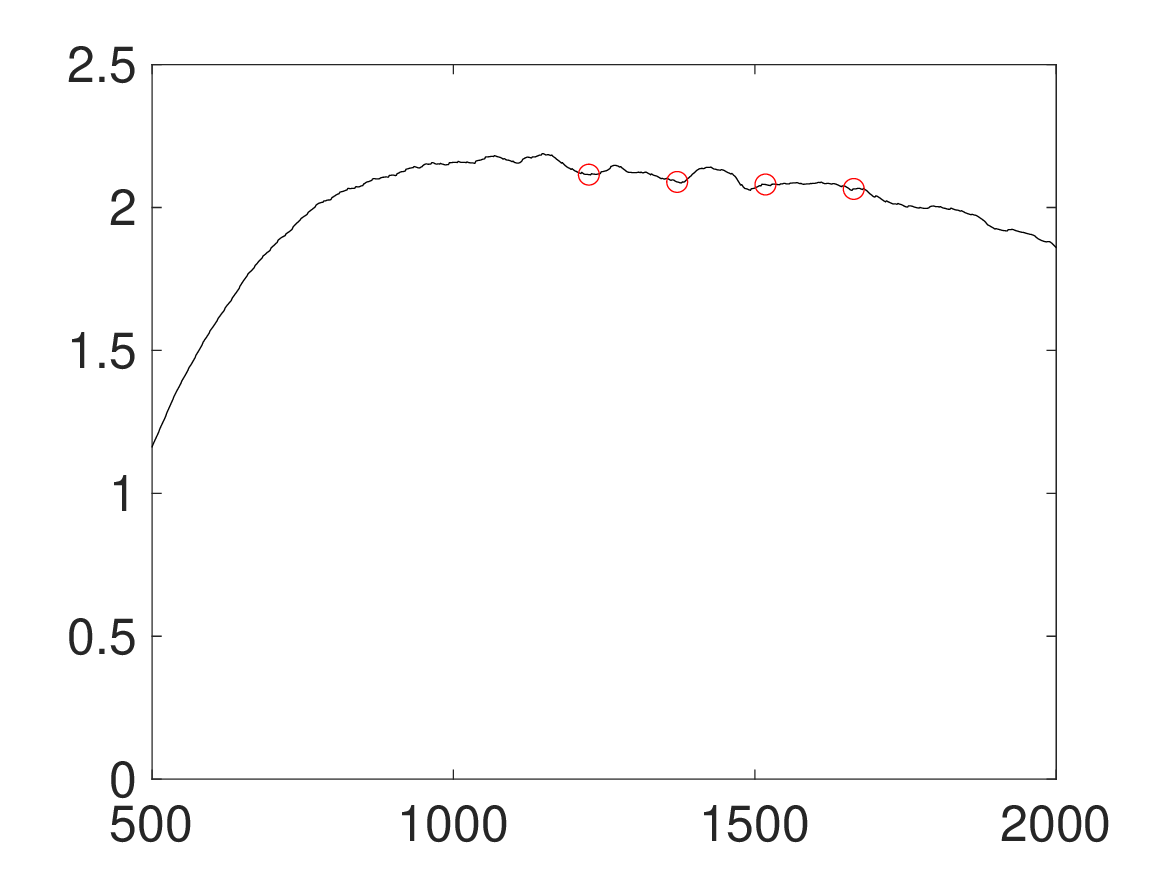}
\put(-220,90){$\beta$}
\put(-220,140){$(b)$} 
\put(-110,-5){$x$}
\end{center}
	\caption{The streamwise evolution of $\beta$ is  shown for $\beta \approx 1$ ($a$)  and $2$ ($b$) data. The streamwise stations are highlighted with {\color {red}$o$} where profiles listed in table \ref{tab:data1} are taken from.}
\label{fig:pgtbl1}
\end{figure}

\begin{figure}
\begin{center}
\includegraphics[height=55mm]{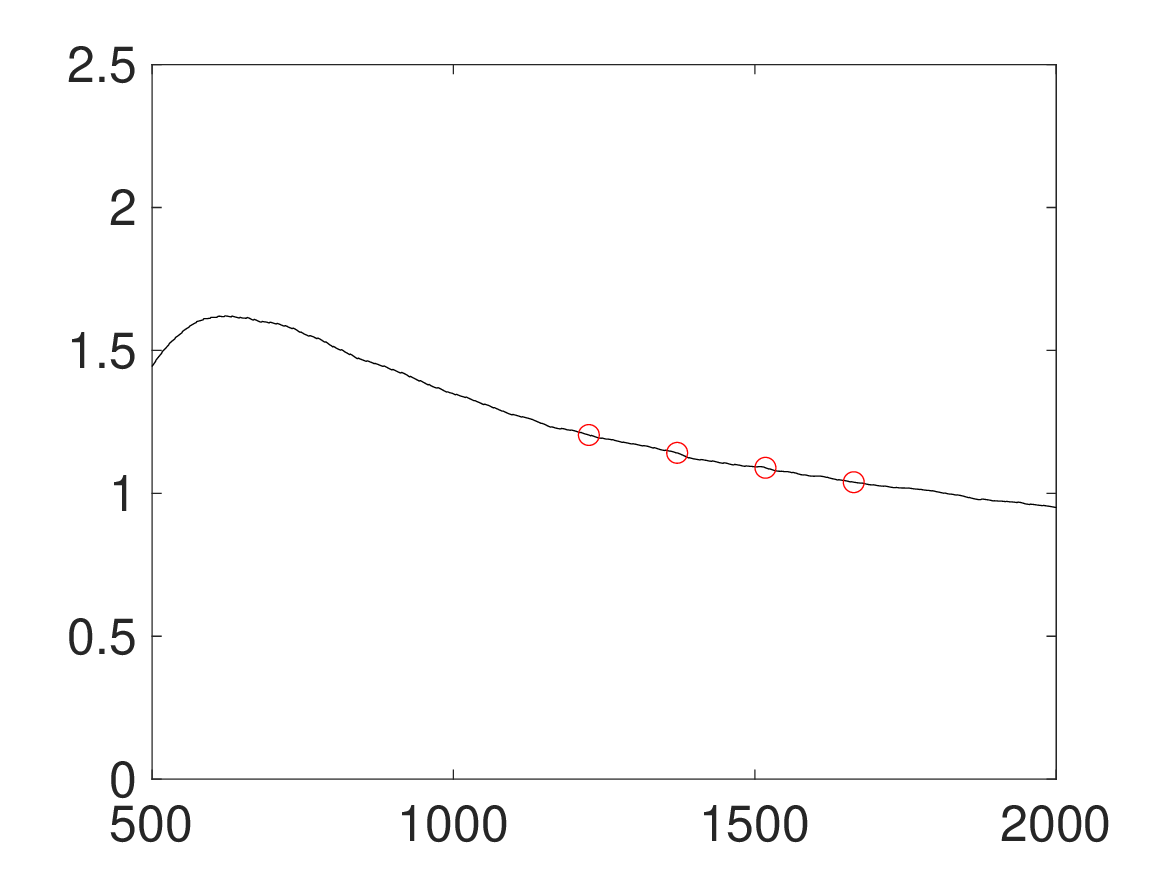}
\put(-220,140){$(a)$} 
\put(-220,90){$\beta$} \\
\includegraphics[height=55mm]{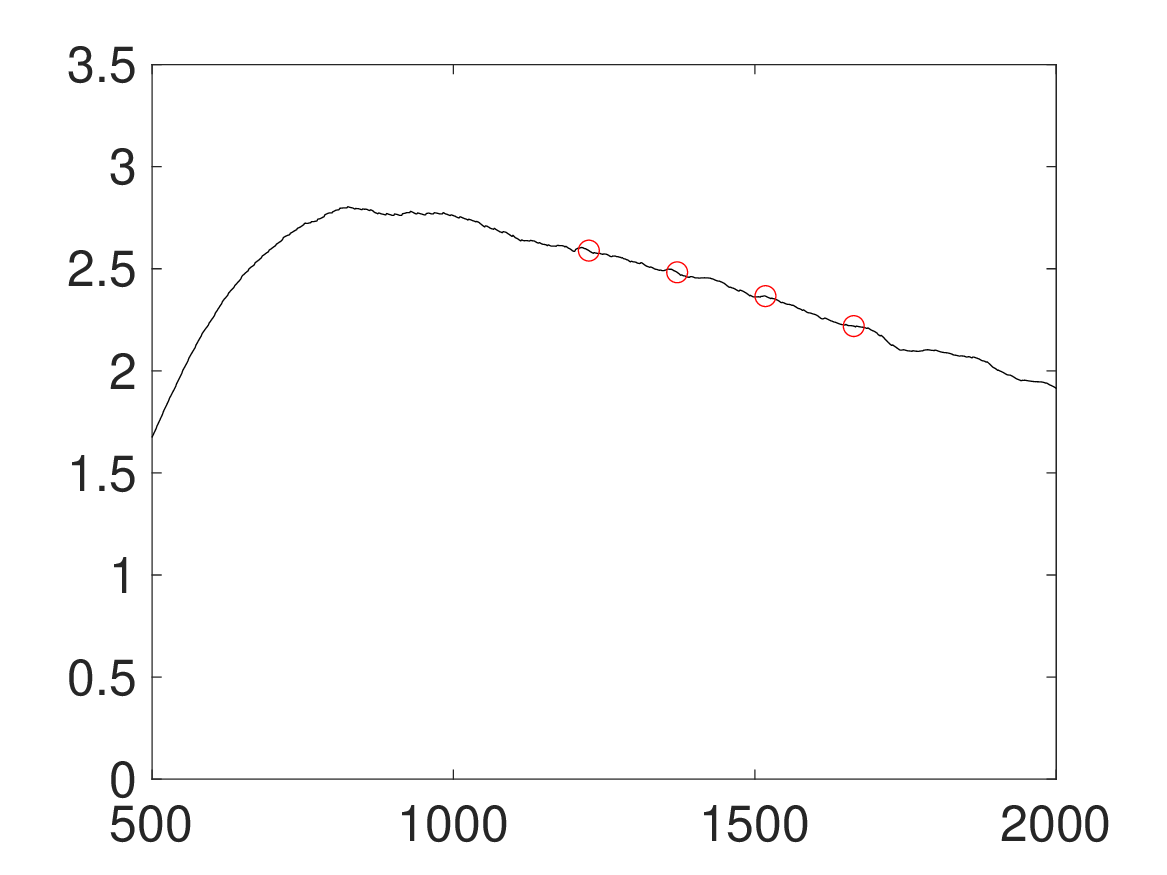}
\put(-220,140){$(b)$} 
\put(-220,90){$\beta$}
\put(-110,-5){$x$}
\put(-220,90){$\beta$}
\end{center}
	\caption{The streamwise evolution of $\beta$ is  shown for $m \approx -0.13$ ($a$) and $-0.16$ ($b$) data. The streamwise stations are highlighted with {\color {red}$o$} where profiles listed in table \ref{tab:data1} are taken from.}
\label{fig:pgtbl2}
\end{figure}

There are four different flat plate TBL databases listed in Table \ref{tab:data1}, with four profiles chosen from each case. The case name ``$\beta xy$'' denotes $y^{th}$ profile from the database where $\beta \approx x$. Similarly, the case name ``$mxxy$'' denotes $y^{th}$ profile from the database where the freestream follows a power-law $U_e \sim (x-x_0)^m$ with $xx$ indicating the percentage negative value of the exponent $m$, i.e., profiles with names starting with ``m13'' are taken from the database where $m = -0.13$ and so on. $Re_{\tau}$, $\beta$ and $H$ are listed for all the profiles. Figures \ref{fig:pgtbl1} and \ref{fig:pgtbl2} show the streamwise evolution of $\beta$ for the constant-$\beta$ and constant-{m} databases respectively, highlighting the stations where the profiles listed in Table \ref{tab:data1} are taken from. The readers are referred to the papers cited in the table for numerical and experimental details. Note that throughout the paper, `true $u_{\tau}$' is the value reported in the reference paper.

\begin{table*}
\caption{ZPG TBL datasets considered in this paper.}
\label{tab:data}
\centering\begin{tabular}{c@{\hskip 1cm}c@{\hskip 1cm}c@{\hskip 1cm}c@{\hskip 1cm}c}
\hline\hline
 Case  & $Re_{\theta}$ & $Re_{\tau}$ & $H$    & Database \\ \hline
 S1000 & 1006          & 359           & 1.4499   & DNS \citep{schlatter2010}\\
 S2000 & 2001          & 671           & 1.4135   & DNS \citep{schlatter2010}\\
 S3030 & 3032          & 974           & 1.3977   & DNS \citep{schlatter2010}\\
 S4060 & 4061          & 1272          & 1.3870   & DNS \citep{schlatter2010}\\
 S5000 & 5000          & 1571          & 1.3730   & DNS \citep{sillero2013one}\\
 S6000 & 6000          & 1848          & 1.3669   & DNS \citep{sillero2013one}\\
 S6500 & 6500          & 1989          & 1.3633   & DNS \citep{sillero2013one}\\
 \hline
 E6920  & 6919       & 1951          & 1.3741   & Expt \citep{morrill2015}\\
 E9830  & 9826       & 2622          & 1.3661   & Expt \citep{morrill2015}\\
 E11200  & 11200      & 2928          & 1.3366   & Expt \citep{morrill2015}\\
 E15120  & 15116      & 3770          & 1.3350   & Expt \citep{morrill2015}\\
 E15470  & 15469      & 3844          & 1.3476   & Expt \citep{morrill2015}\\
 E24140  & 24135      & 5593.         & 1.3264   & Expt \citep{morrill2015}\\
 E26650  & 26647      & 6080          & 1.2938   & Expt \citep{morrill2015}\\
 E36320  & 36322      & 7894          & 1.2774   & Expt \citep{morrill2015}\\
 E21630 & 21632      & 5100          & 1.3      & Expt. \citep{baidya}\\
 E51520 & 51524      & 10600         & 1.28     & Expt. \citep{baidya}\\
 \hline
 R0 & -           & 1918          & -        & Expt. \citep{flack2020} \\
 R- & -           & 1600          & -        & Expt. \citep{flack2020} \\
 R+ & -           & 2202          & -        & Expt. \citep{flack2020} \\
 \hline\hline
\end{tabular}
\end{table*}

\begin{table*}
\caption{APG TBL datasets considered in this paper.}
\label{tab:data1}
\centering\begin{tabular}{c@{\hskip 1cm}c@{\hskip 1cm}c@{\hskip 1cm}c@{\hskip 1cm}c}
\hline\hline
 Case  & $\beta$ & $Re_{\tau}$           & $H$      & Database \\ \hline
 \\ \hline
 $\beta$11 & 1.0104       & 515.5           & 1.6024    & LES \citep{bobke2017}\\
 $\beta$12 & 0.9820       & 558.3           & 1.5924     & LES \citep{bobke2017} \\
 $\beta$13 & 0.9537       & 599.1           & 1.5827     & LES \citep{bobke2017} \\
 $\beta$14 & 0.9322       & 637.9           & 1.5738     & LES \citep{bobke2017} \\
 \hline
 $\beta$21 & 2.1149       & 502.5           & 1.7196   & LES \citep{bobke2017} \\
 $\beta$22 & 2.0894       & 545.4           & 1.7114    & LES \citep{bobke2017} \\
 $\beta$23 & 2.0805       & 589.9           & 1.7024     & LES \citep{bobke2017} \\
 $\beta$24 & 2.0649       & 640.3           & 1.6921    & LES \citep{bobke2017}  \\
 \hline
 $m131$ & 1.2041       & 527.2           & 1.6379     & LES \citep{bobke2017} \\
 $m132$ & 1.1416       & 573.1           & 1.6202     & LES \citep{bobke2017} \\
 $m133$ & 1.0894       & 612.2           & 1.6061     & LES \citep{bobke2017} \\
 $m134$ & 1.0389       & 654.5           & 1.5925    & LES \citep{bobke2017} \\
 \hline
 $m161$ & 2.5889       & 500.7          & 1.7831     & LES \citep{bobke2017} \\
 $m162$ & 2.4825       & 547.7           & 1.7669    & LES \citep{bobke2017} \\
 $m163$ & 2.3660       & 596.6           & 1.7459    & LES \citep{bobke2017} \\
 $m164$ & 2.2192        & 644.8          & 1.7259     & LES \citep{bobke2017} \\
 \hline
 \hline
\end{tabular}
\end{table*}

\section{Method performance for ZPG TBL} \label{sec:res}
\subsection{Smooth wall TBL}
\begin{figure}
\begin{center}
\includegraphics[height=55mm]{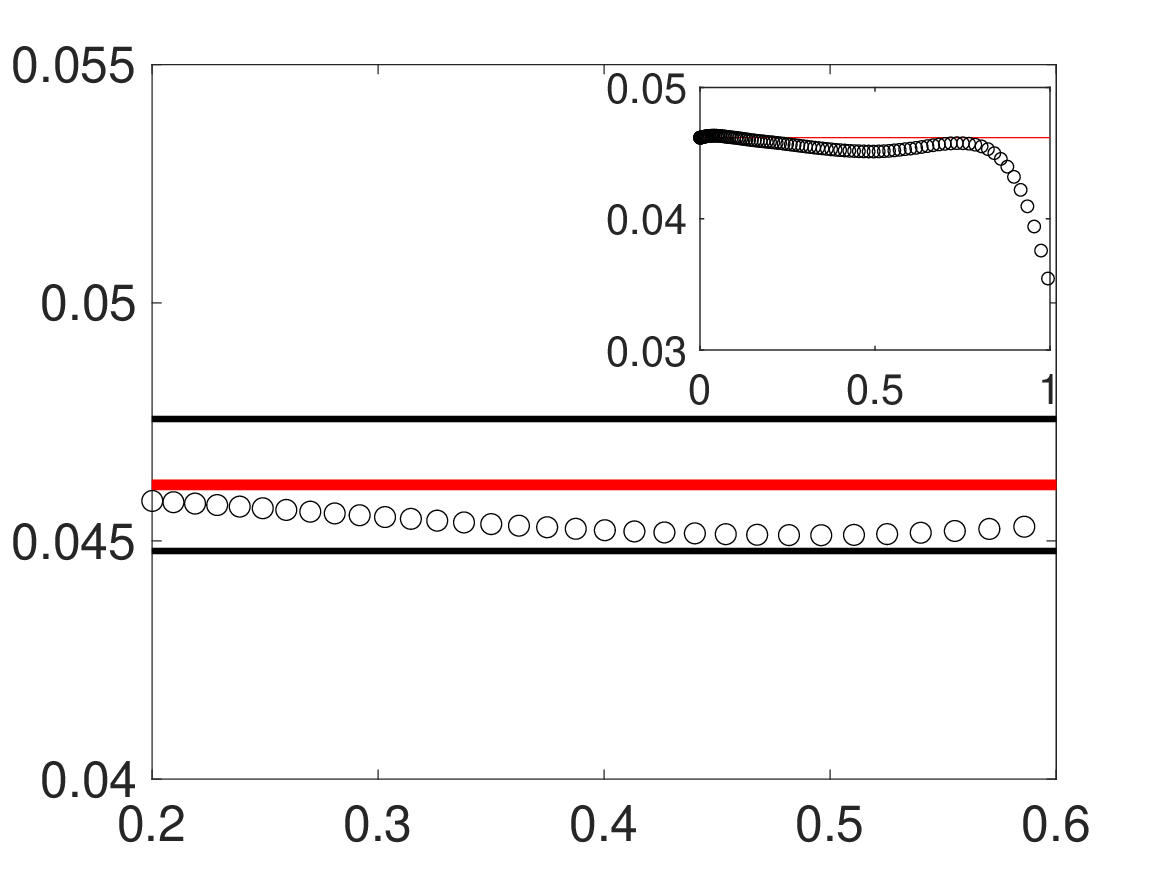}
\put(-220,140){$(a)$}
\put(-220,90){$u_{\tau}/U_e$} \\
\includegraphics[height=55mm]{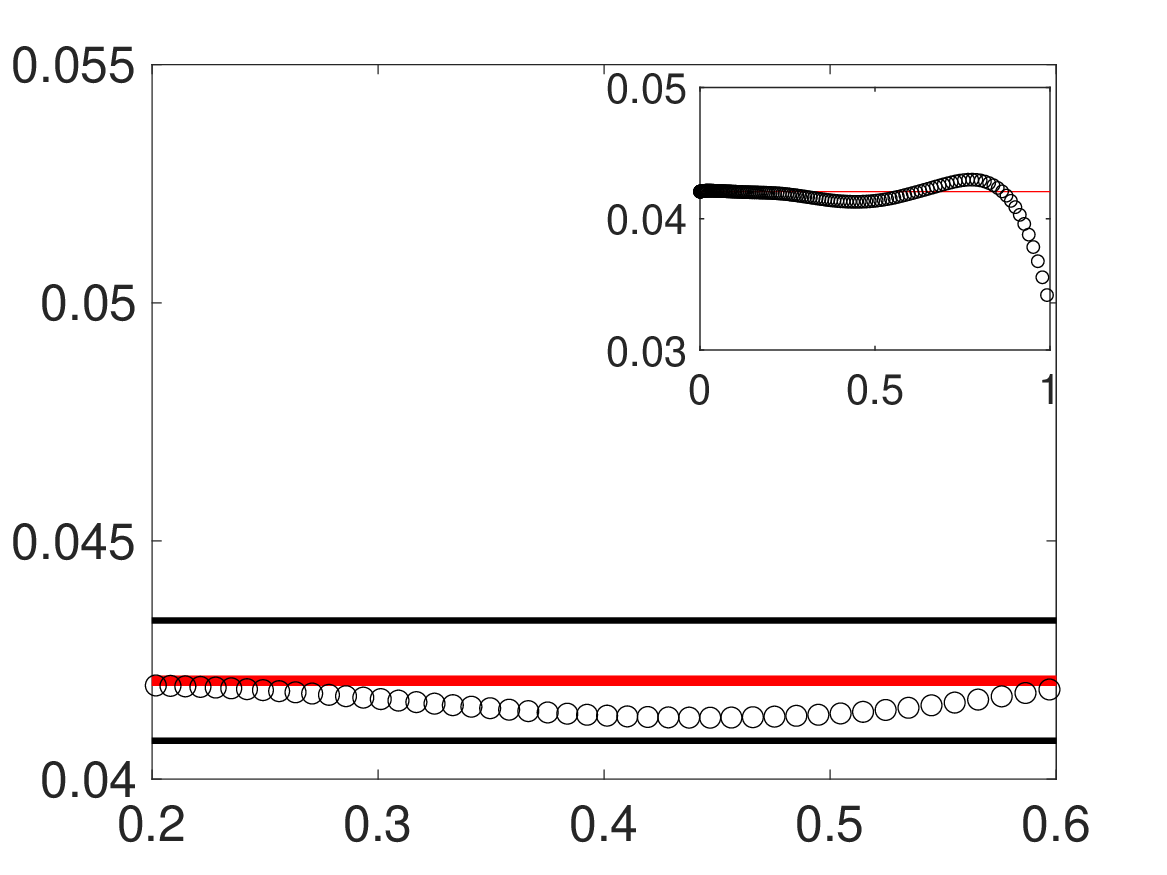} 
\put(-220,140){$(b)$}
\put(-220,90){$u_{\tau}/U_e$} \\
\includegraphics[height=55mm]{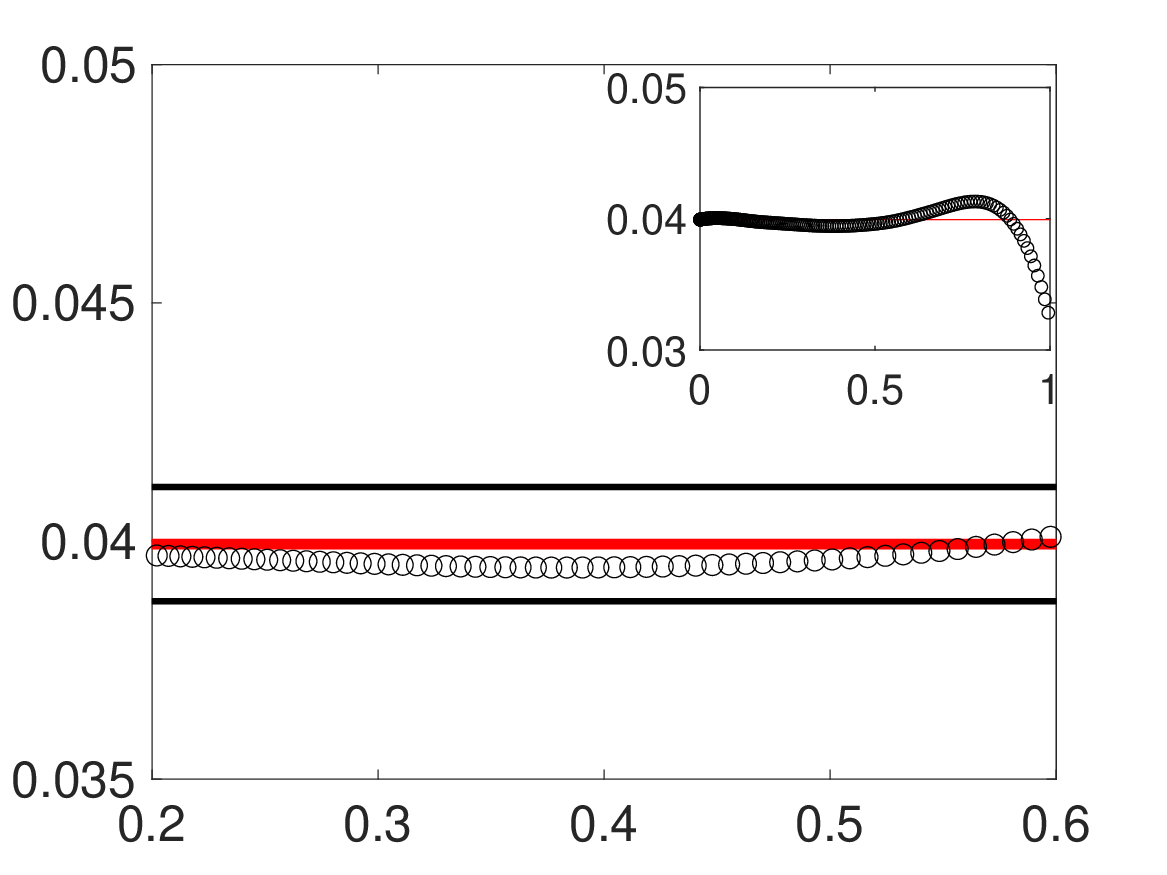}
\put(-220,140){$(c)$}
\put(-220,90){$u_{\tau}/U_e$} \\
\includegraphics[height=55mm]{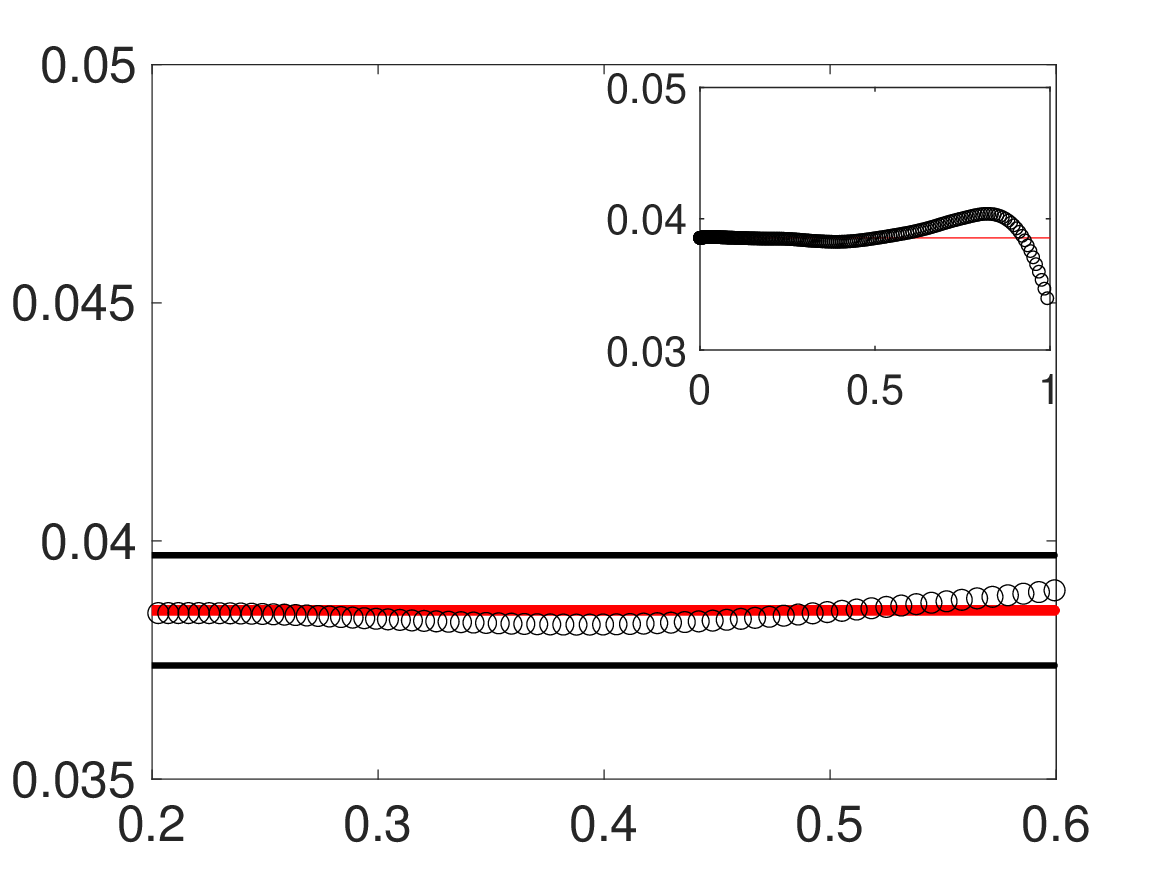}
\put(-220,90){$u_{\tau}/U_e$}
\put(-220,140){$(d)$}
\put(-110,-5){$\eta$}
\end{center}
\caption{Predicted $u_{\tau}$ is compared to true $u_{\tau}$ (red line) for cases S1000 $(a)$, S2000 $(b)$, S3030 $(c)$, and S4060 $(d)$. Note that the black lines show $\pm 3$ \% of of the red line.}
\label{fig:method1}
\end{figure}

\begin{figure}
\begin{center}
\includegraphics[height=55mm]{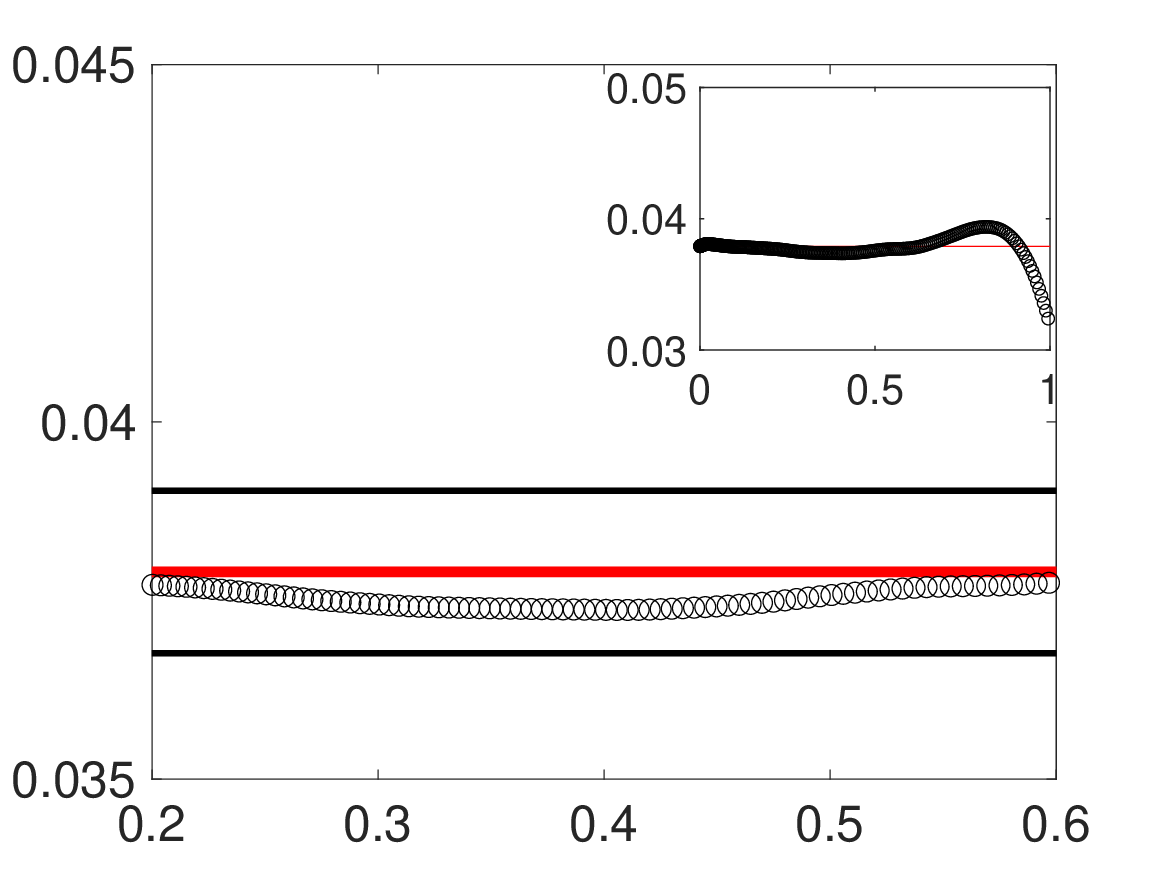}
\put(-220,140){$(a)$}
\put(-220,90){$u_{\tau}/U_e$} \\
\includegraphics[height=55mm]{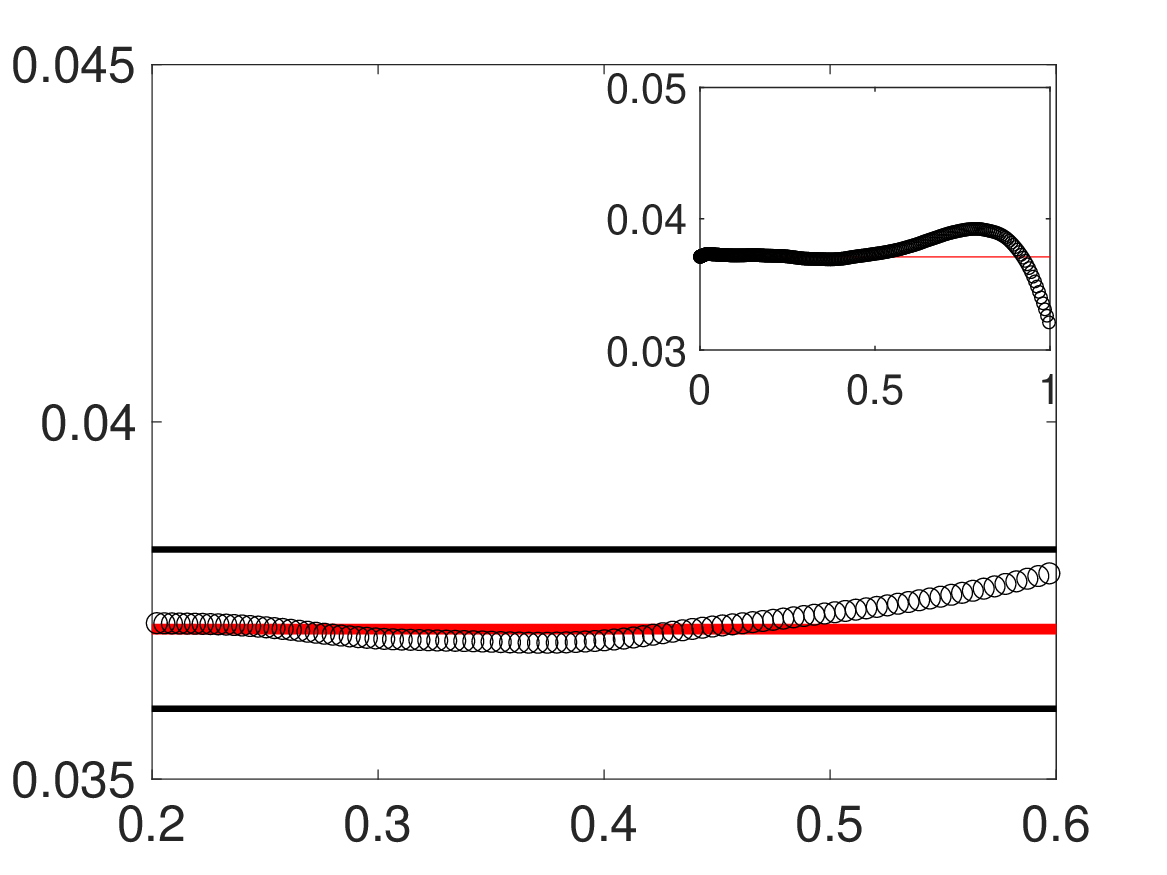}
\put(-220,140){$(b)$} 
\put(-220,90){$u_{\tau}/U_e$}\\
\includegraphics[height=55mm]{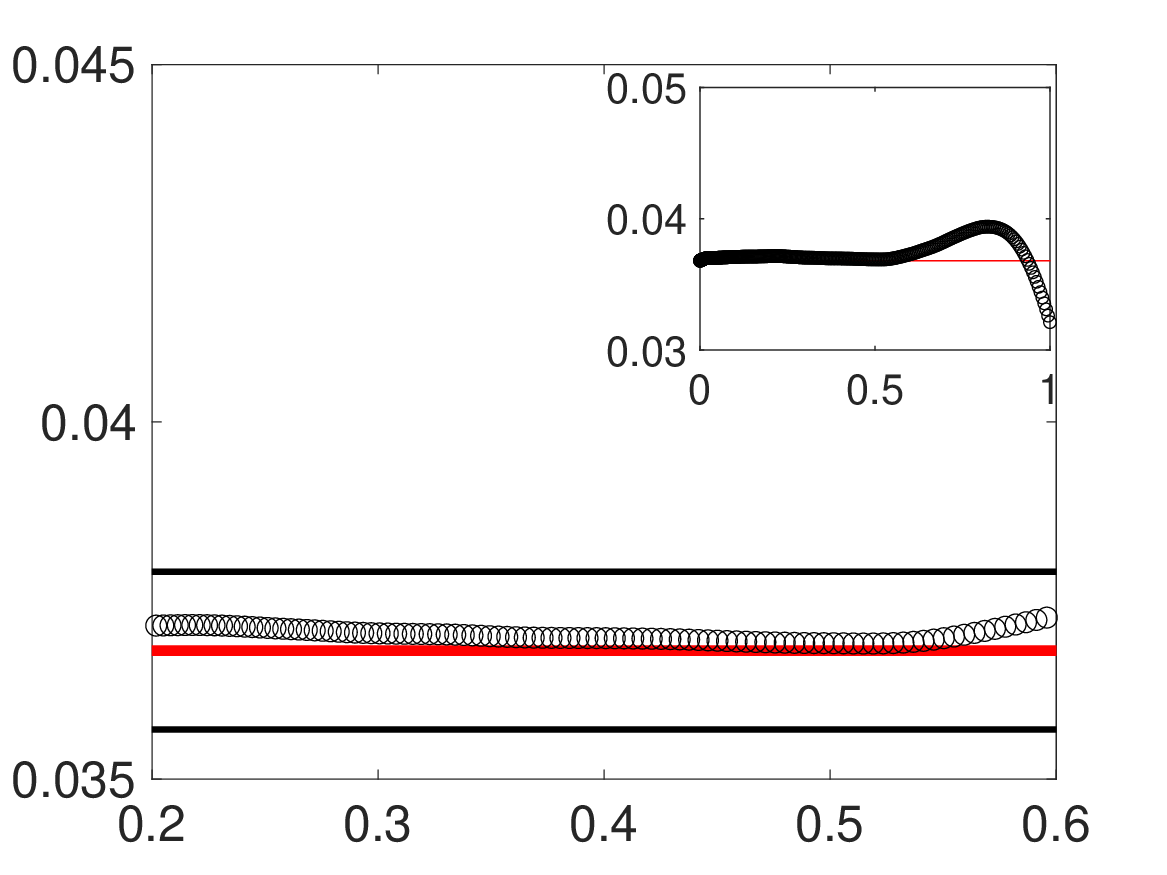}
\put(-110,-5){$\eta$}
\put(-220,90){$u_{\tau}/U_e$}
\put(-220,140){$(c)$}
\end{center}
\caption{Predicted $u_{\tau}$ is compared to true $u_{\tau}$ (line) for cases S5000 $(a)$, S6000 $(b)$, and S6500 $(c)$. Note that the black lines show $\pm 3$ \% of the red line.}
\label{fig:method2}
\end{figure}

Figure \ref{fig:method1} shows the predicted $u_{\tau}$ for TBL cases ranging from $Re_\theta =$ 1006 to 4061 using the DNS data of \cite{schlatter2010}. The black lines show the bounds of $\pm 3$ \% deviation from the true values. Figure \ref{fig:method2} shows similar results for TBL cases ranging from $Re_\theta = 5000$ to 6500 using the DNS data of \cite{sillero2013one}. In all these cases, the proposed method accurately predicts $u_{\tau}$ using the DNS data in the range $0.2 < \eta < 0.6$.

\begin{figure}
\begin{center}
\includegraphics[height=55mm]{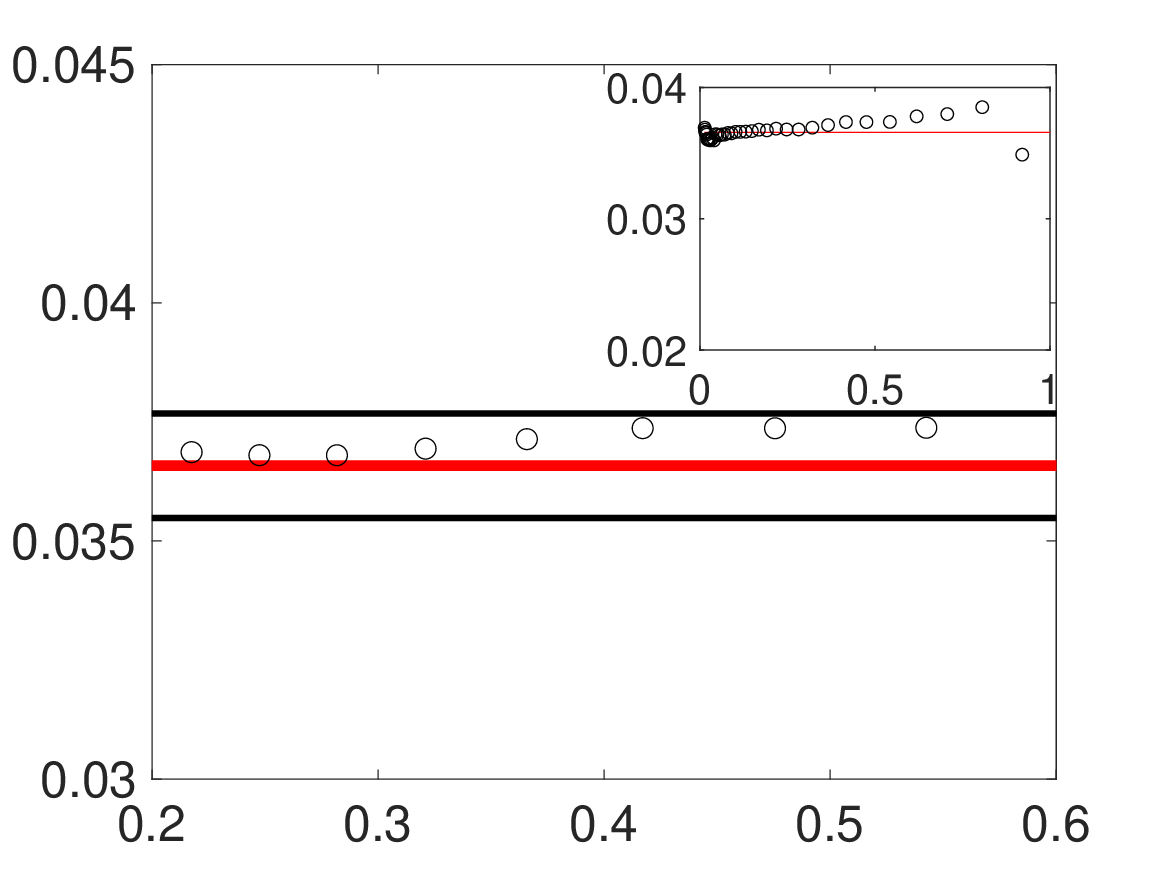}
\put(-220,140){$(a)$}
\put(-220,90){$u_{\tau}/U_e$} \\
\includegraphics[height=55mm]{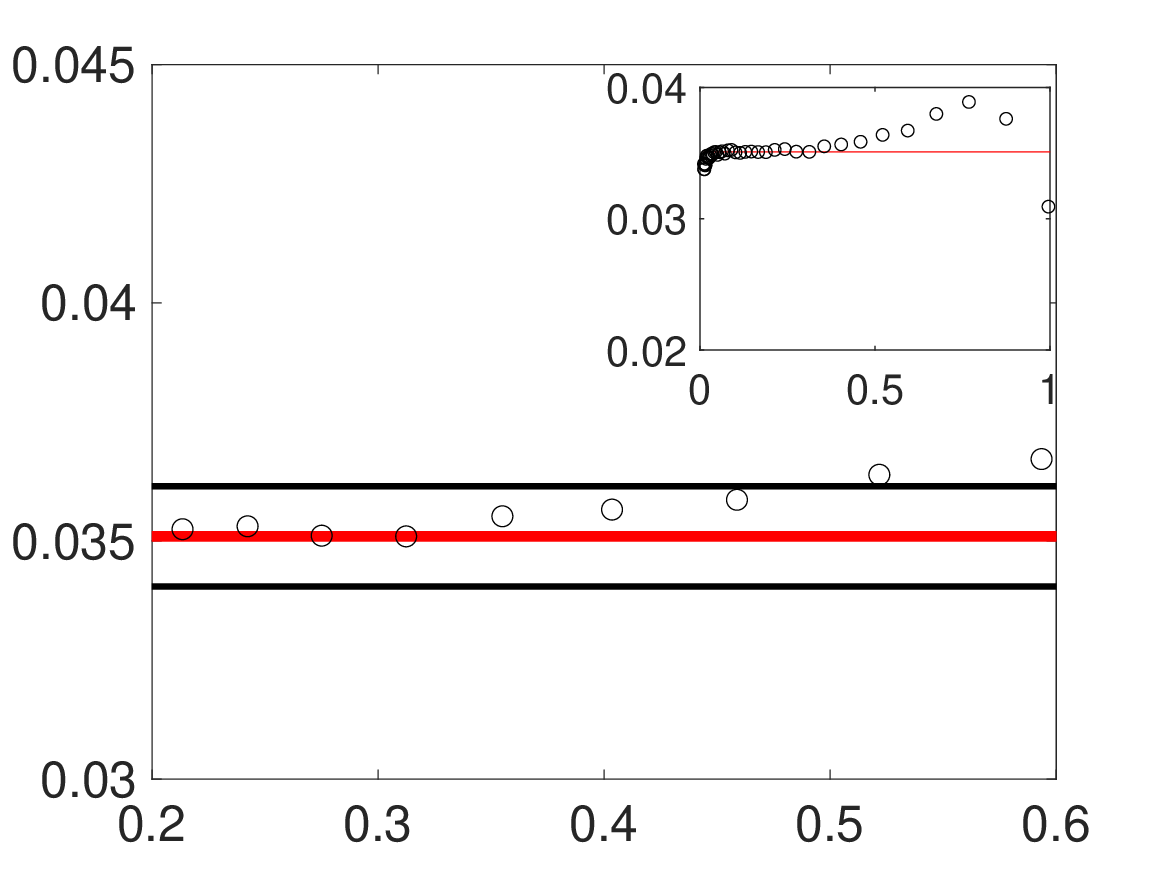} 
\put(-220,90){$u_{\tau}/U_e$} 
\put(-220,140){$(b)$} \\
\includegraphics[height=55mm]{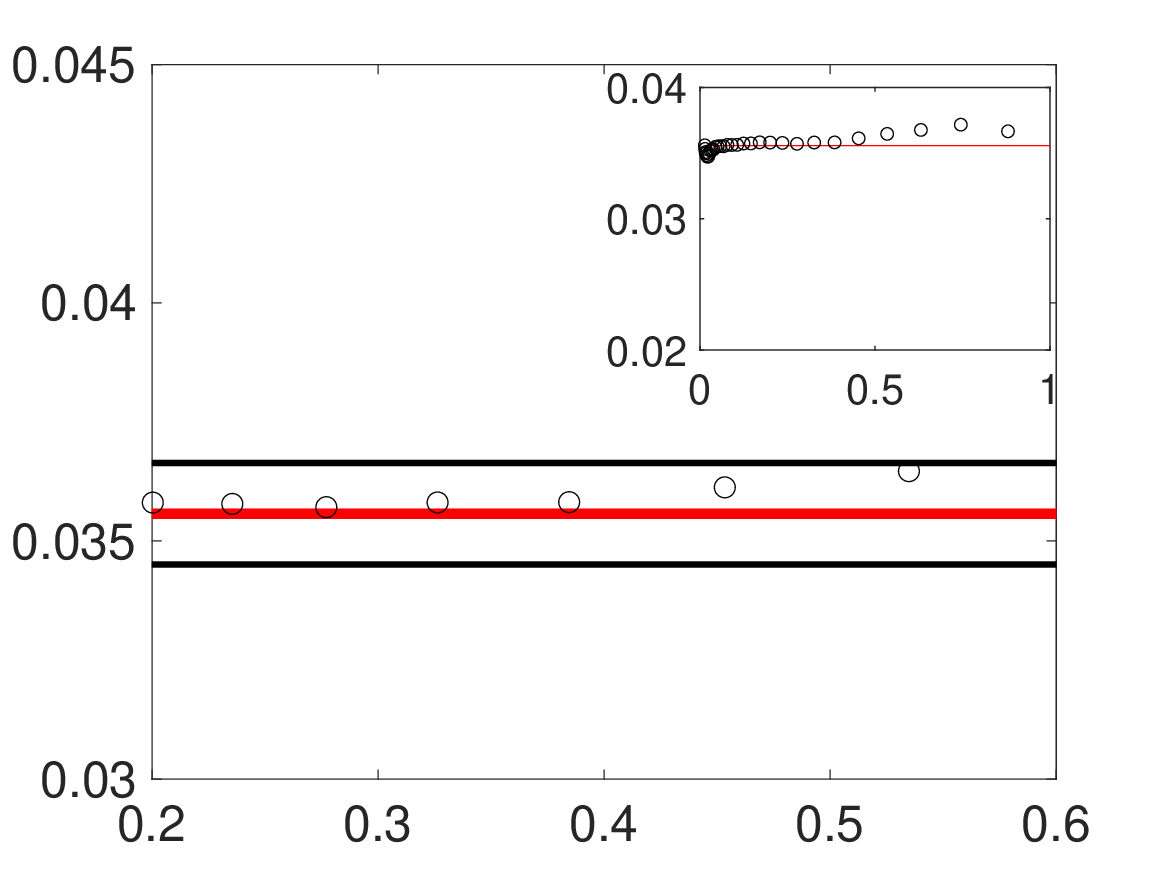}
\put(-220,140){$(c)$}
\put(-220,90){$u_{\tau}/U_e$} \\
\includegraphics[height=55mm]{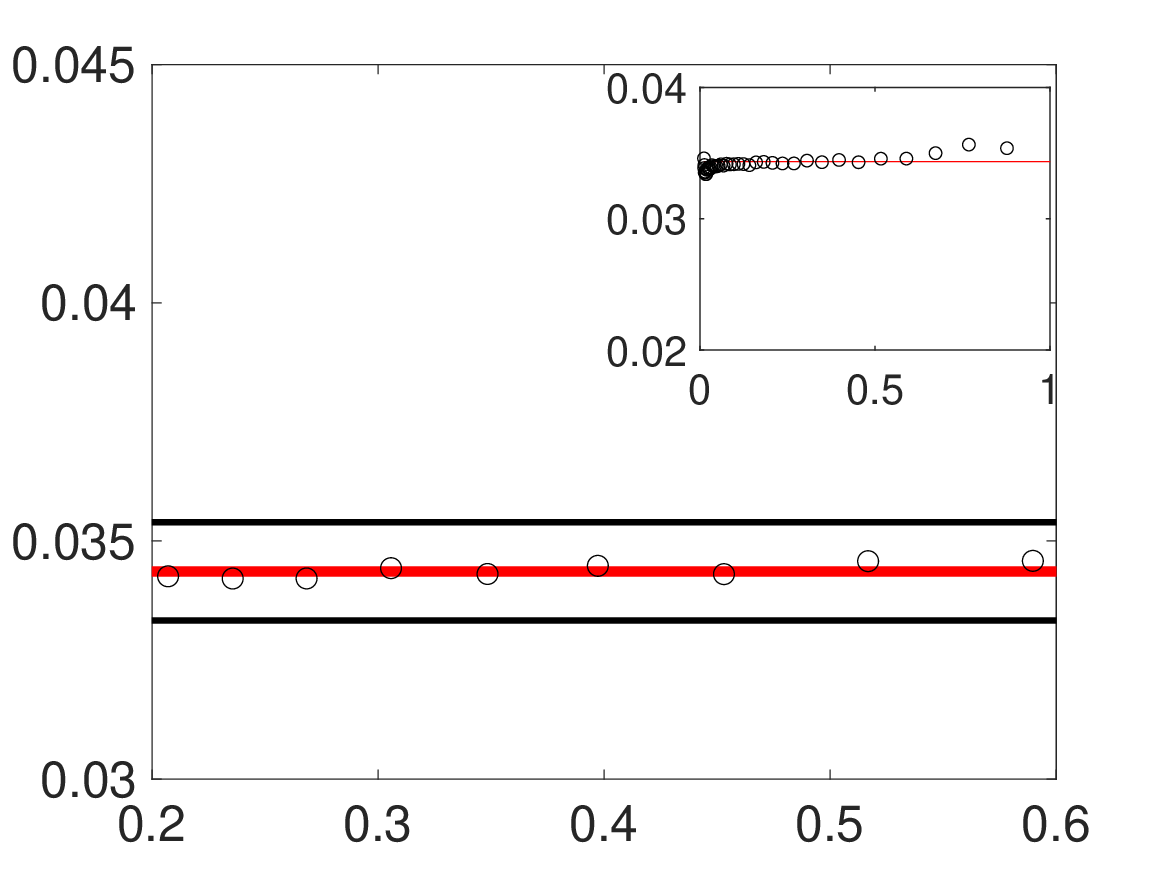}
\put(-220,140){$(d)$}
\put(-220,90){$u_{\tau}/U_e$}
\put(-110,-5){$\eta$}
\end{center}
\caption{Predicted $u_{\tau}$ is compared to $u_{\tau}$ obtained from the composite fit (red line) for cases E6920 $(a)$, E9830 $(b)$, E11200 $(c)$, and E15120 $(d)$. Note that the black lines show $\pm 3$ \% of the red line.}
\label{fig:MW1}
\end{figure}

\begin{figure}
\begin{center}
\includegraphics[height=55mm]{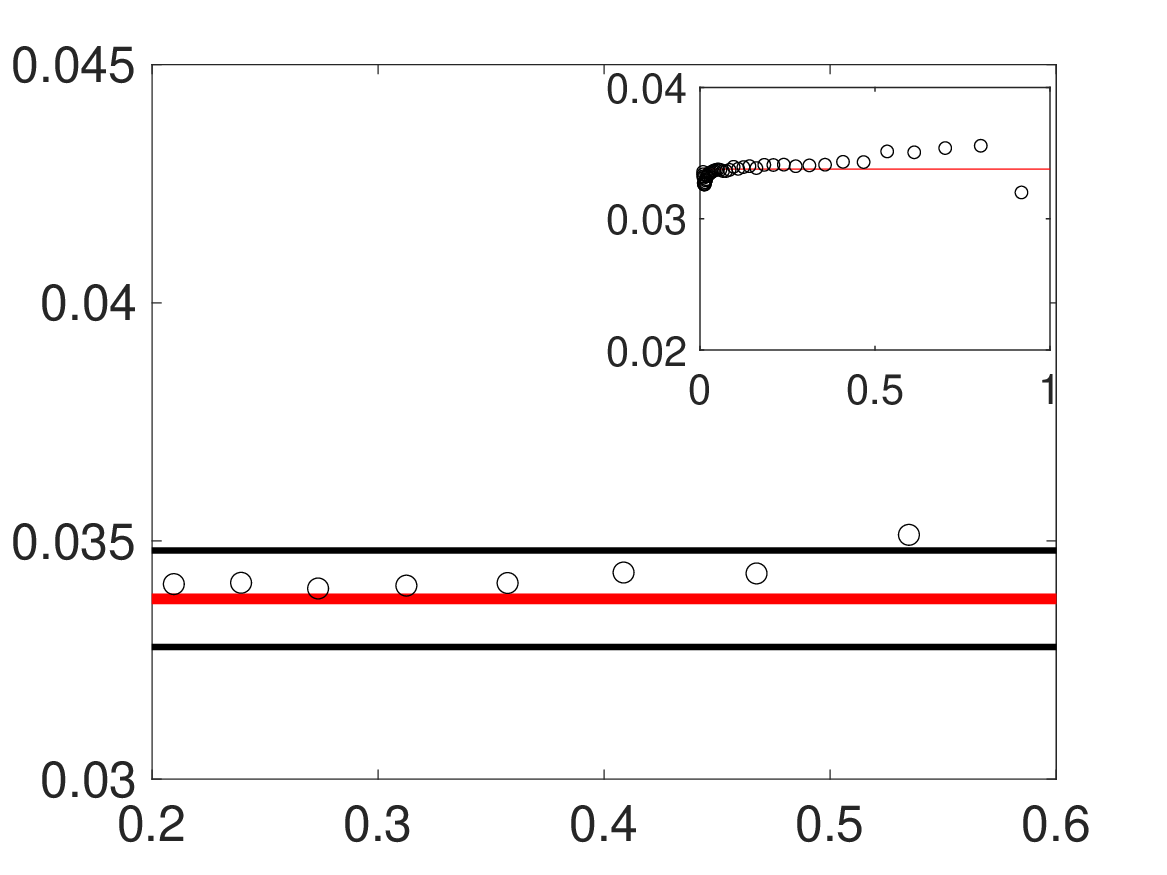}
\put(-220,140){$(a)$} 
\put(-220,90){$u_{\tau}/U_e$} \\
\includegraphics[height=55mm]{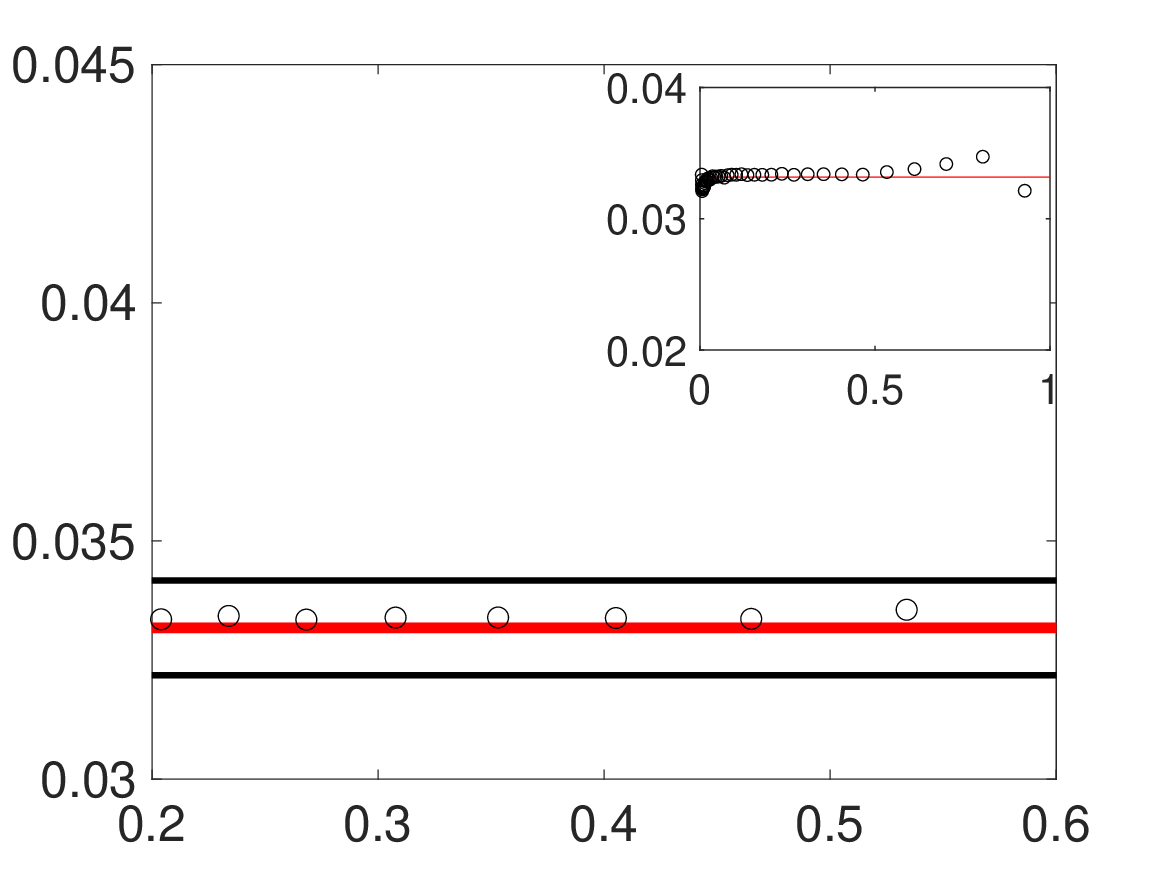} 
\put(-220,140){$(b)$} 
\put(-220,90){$u_{\tau}/U_e$} \\
\includegraphics[height=55mm]{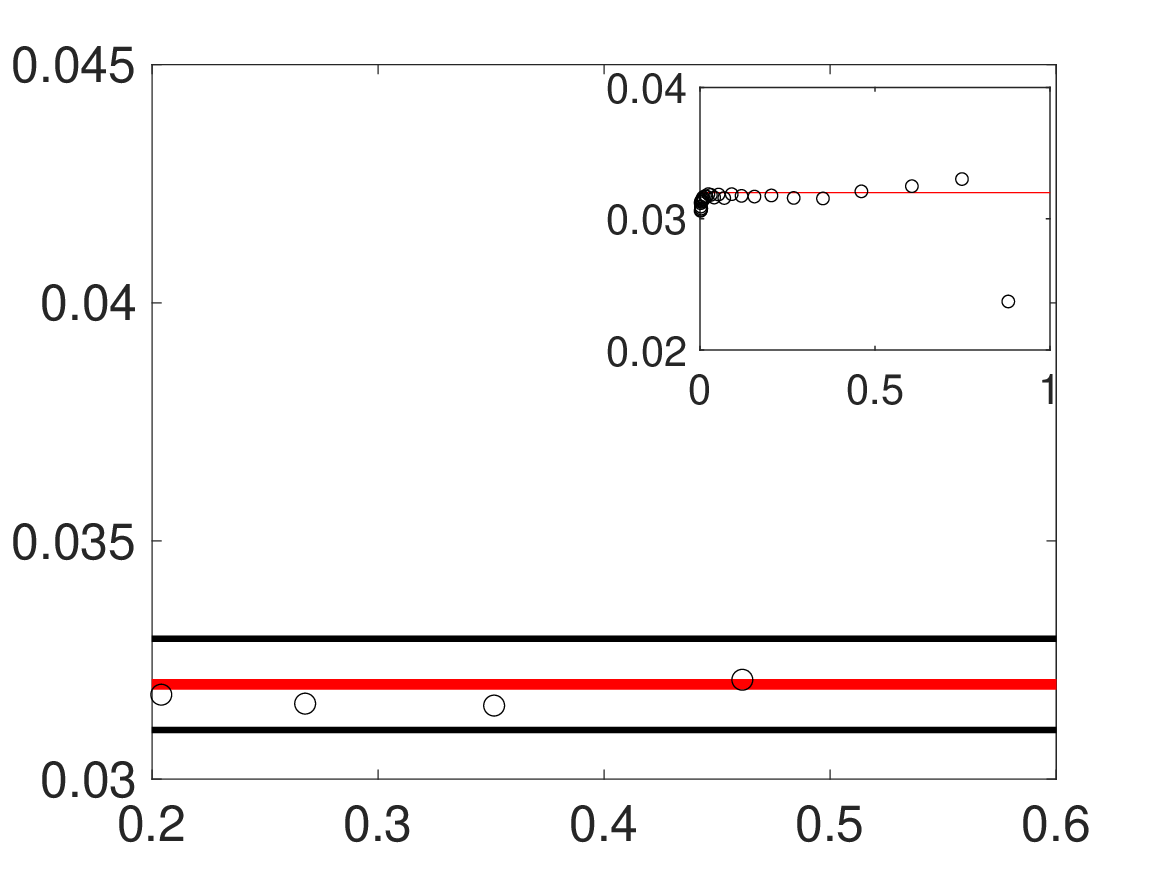}
\put(-220,140){$(c)$} 
\put(-220,90){$u_{\tau}/U_e$} \\
\includegraphics[height=55mm]{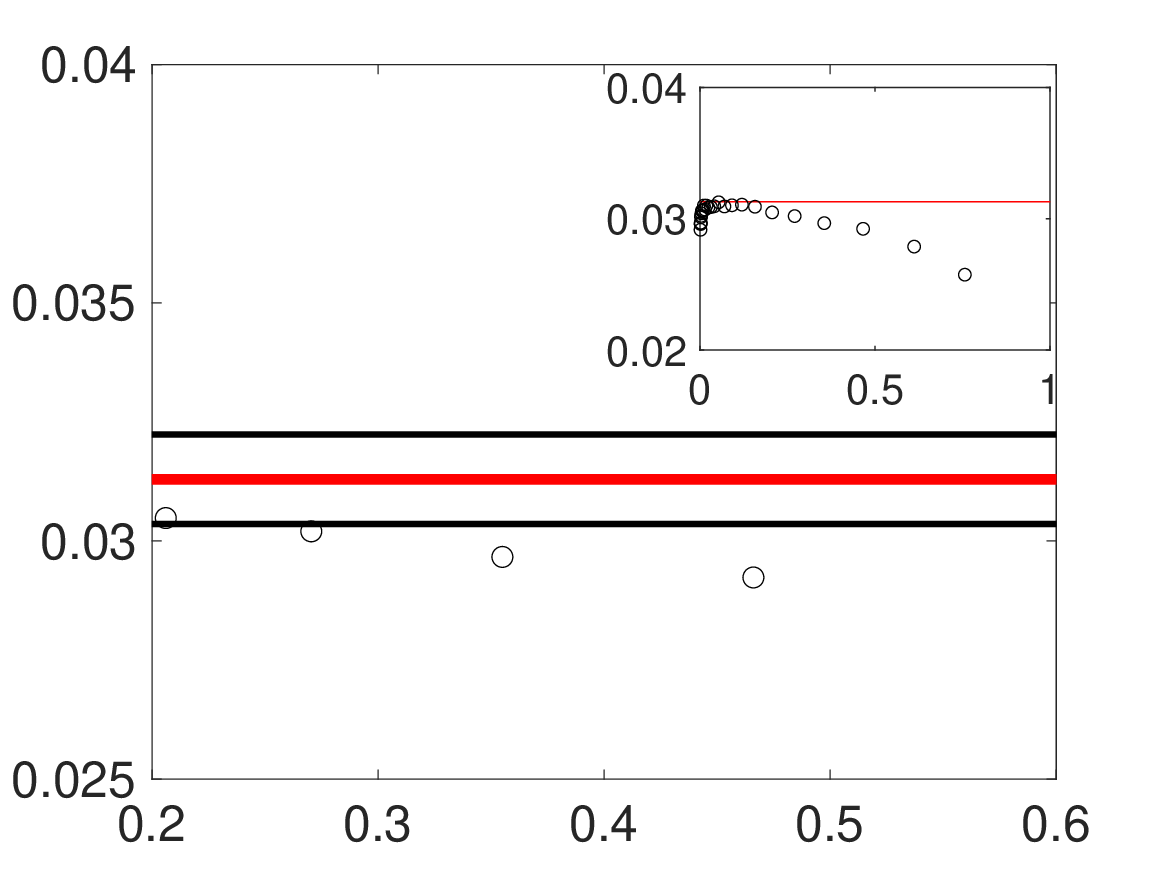}
\put(-220,140){$(d)$} 
\put(-220,90){$u_{\tau}/U_e$}
\put(-110,-5){$\eta$}
\end{center}
\caption{Predicted $u_{\tau}$ is compared to $u_{\tau}$ obtained from the composite fit (red line) for cases E15470 $(a)$, E24140 $(b)$, E26650 $(c)$, and E36320 $(d)$. Note that the black lines show $\pm 3$ \% of the red line.}
\label{fig:MW2}
\end{figure}

\begin{figure}
\begin{center}
\includegraphics[height=65mm]{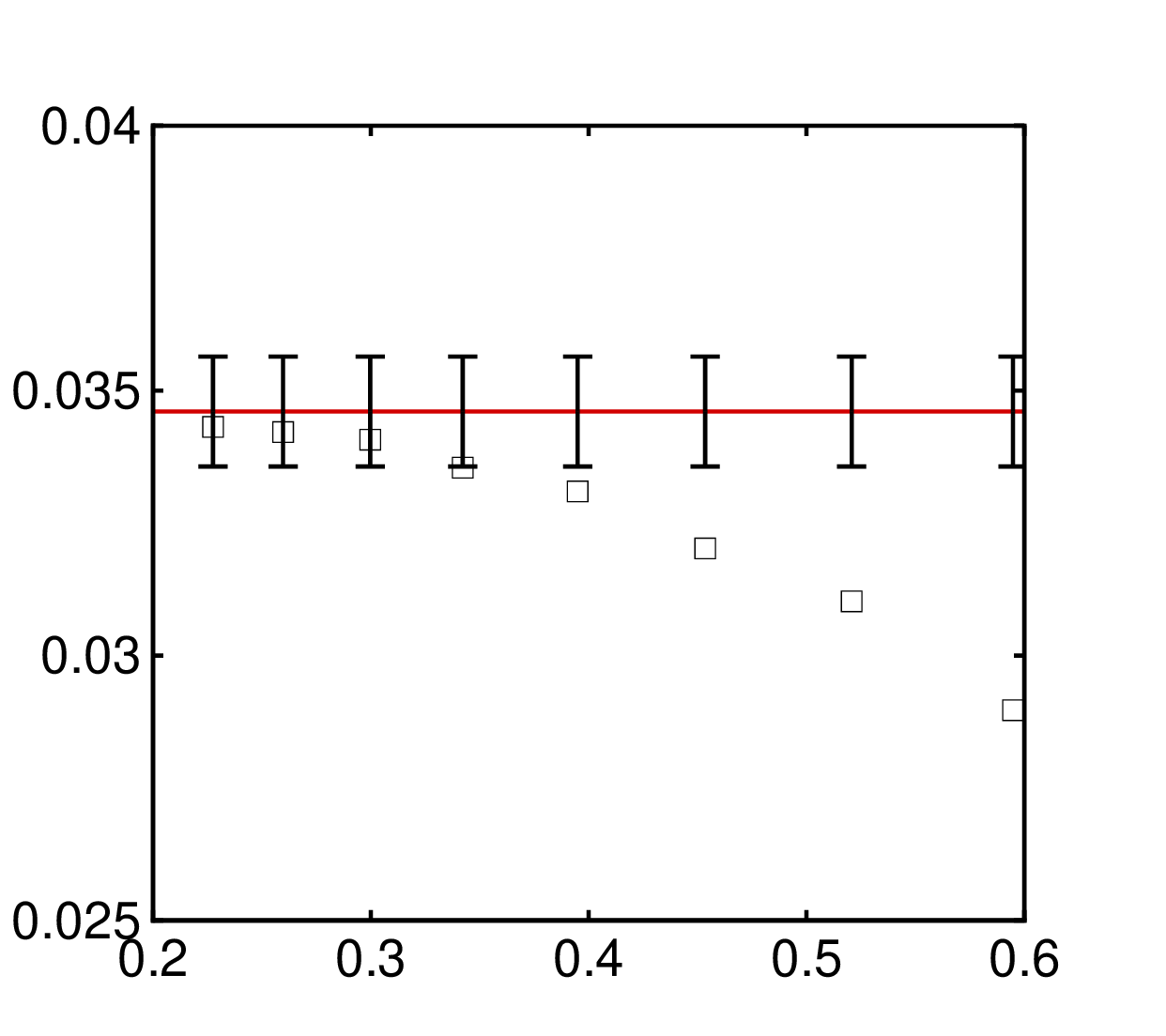}
\put(-220,160){$(a)$} 
\put(-230,100){$u_{\tau}/U_e$} \\
\includegraphics[height=65mm]{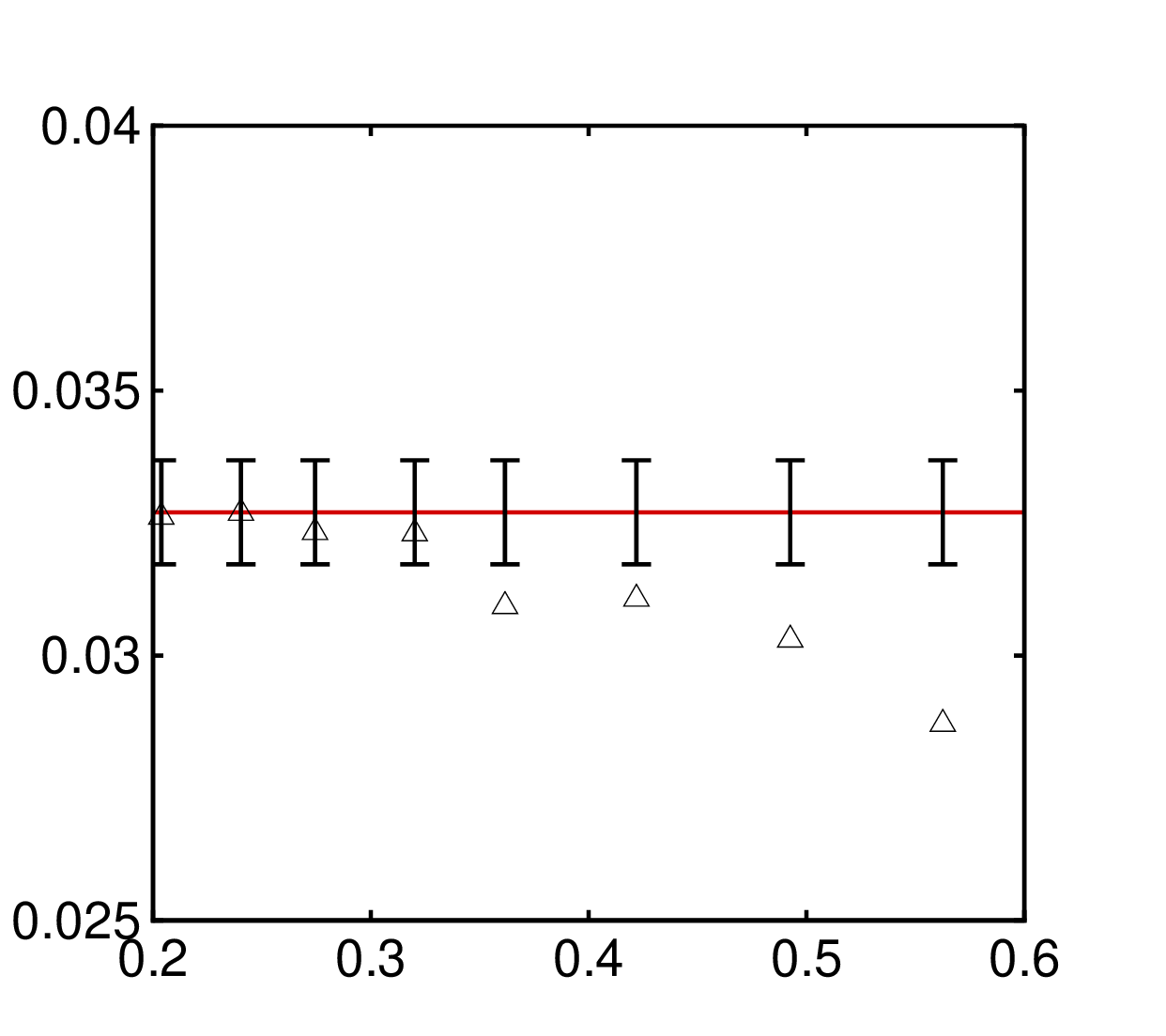}
\put(-110,-5){$\eta$}
\put(-220,160){$(b)$} 
\put(-230,100){$u_{\tau}/U_e$}
\end{center}
\caption{Predicted $u_{\tau}$ (symbol) is compared to true $u_{\tau}$ (line) for cases E21630 $(a)$ and E51520 $(b)$. Note that the error bar shows $\pm 3$ \% of the red line.}
\label{fig:Baidya}
\end{figure}

Next, the proposed method is used to predict $u_{\tau}$ using the experimental database of \cite{morrill2015} and \cite{baidya}. Note that both these experimental campaigns used the composite fit method to determine $u_{\tau}$. Figures \ref{fig:MW1} and \ref{fig:MW2} show the predicted $u_{\tau}$ compared for the experimental data of \cite{morrill2015}. Overall, the proposed method agrees to the data obtained from the composite fit approach to within $3 \%$ in the region $0.2 < \eta < 0.5$ for all the cases except E36320, where it is within $3 \%$ in the region $0.2 < \eta < 0.35$.

Figure \ref{fig:Baidya} shows the predicted $u_{\tau}$ compared to the experimental data of \cite{baidya}. Note that the contribution of viscous stress to the total stress is ignored for these data sets. Overall, the proposed method agrees with the composite fit approach to within $3 \%$ in the region $0.2 < \eta < 0.35$.

\subsection{Rough wall TBL}
\cite{flack2020} performed experiments on rough wall TBL where $u_{\tau}$ values were determined using the modified Clauser chart method of \cite{perry1990} as described in \cite{schultz2003} in detail. The method uses the velocity profile in the log-law region in an iterative procedure to obtain both $u_{\tau}$ and the wall offset i.e. the location in the roughness where $y=0$. The reported uncertainty in $u_{\tau}$ for the modified Clauser chart method was $\pm 4 \%$. 

\begin{figure}
\begin{center}
\includegraphics[height=55mm]{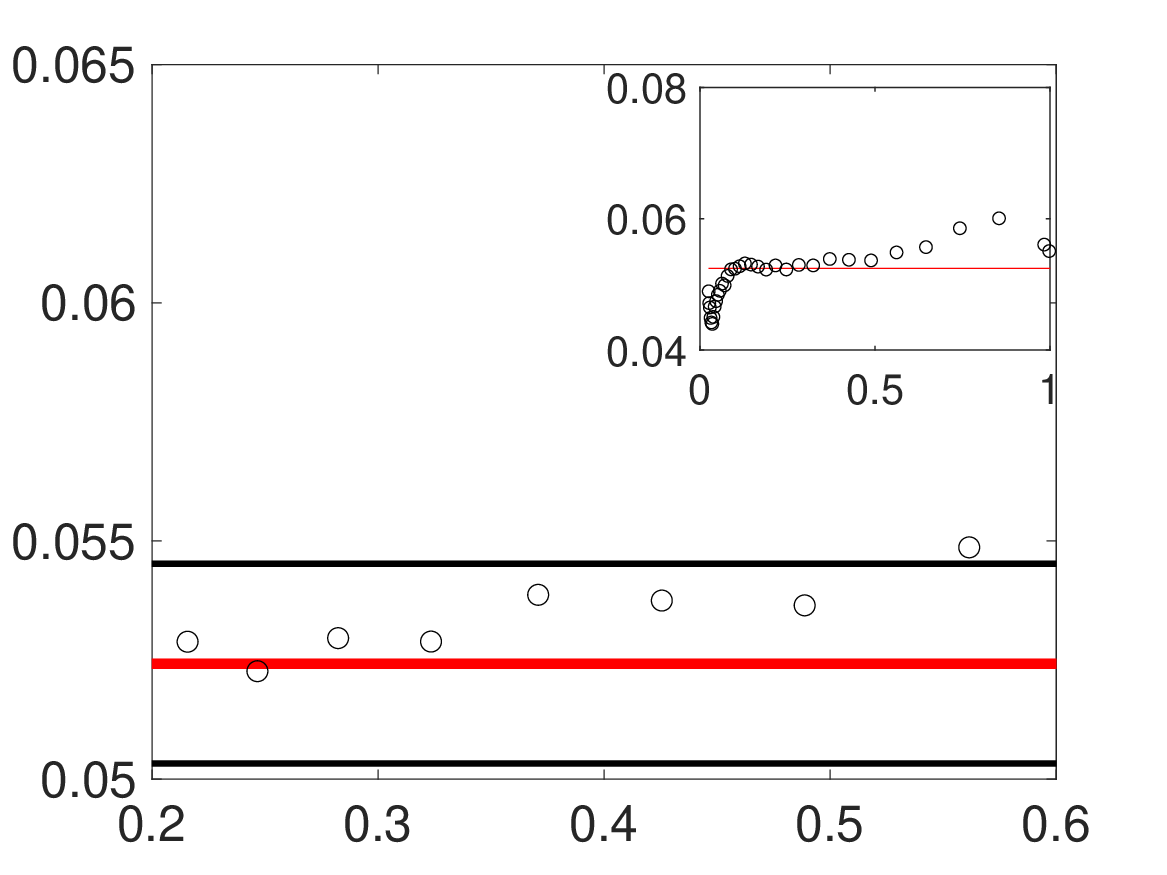}
\put(-220,90){$u_{\tau}/U_e$} \\
\includegraphics[height=55mm]{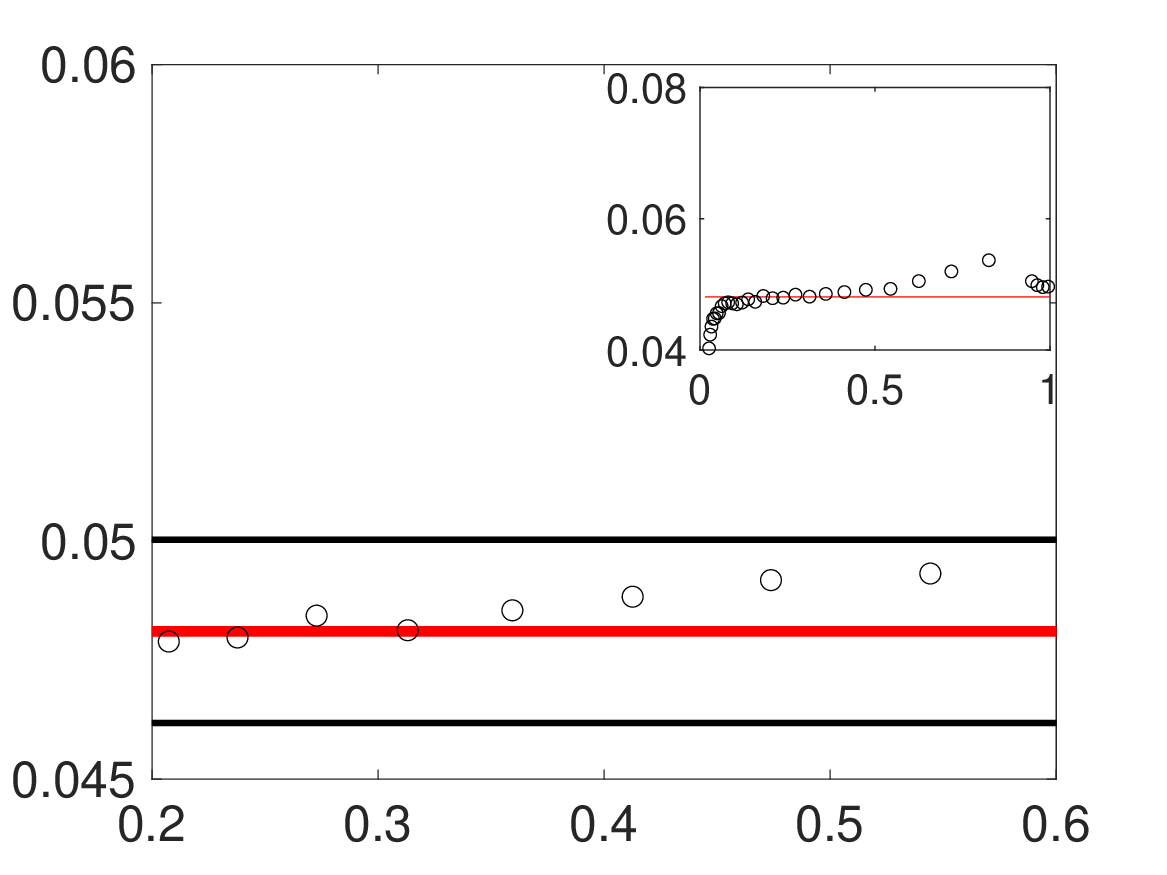} 
\put(-220,90){$u_{\tau}/U_e$} \\
\includegraphics[height=55mm]{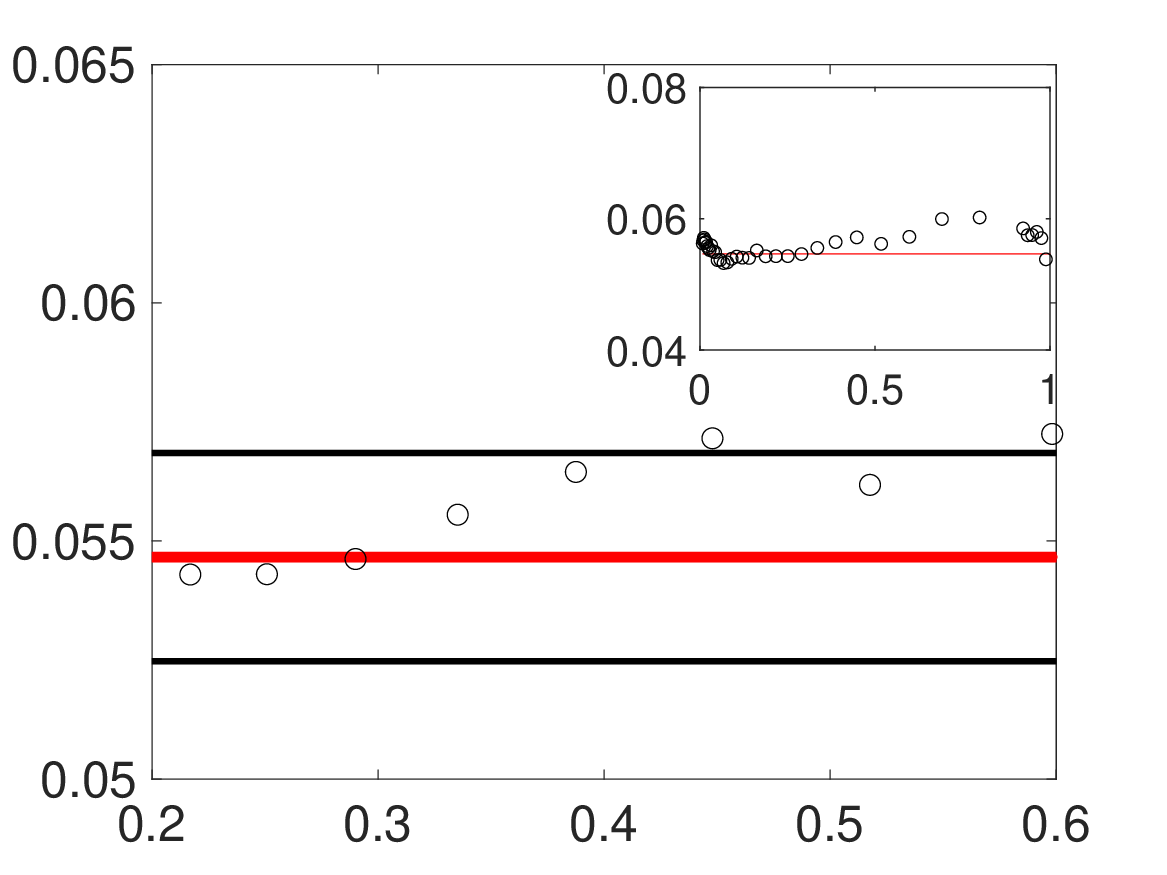}
\put(-220,90){$u_{\tau}/U_e$}
\put(-110,-5){$\eta$}
\end{center}
\caption{Predicted $u_{\tau}$ is compared to $u_{\tau}$ obtained from modified Clauser chart method (red line) for cases R0 $(a)$, R- $(b)$, and R+ $(c)$. Note that the black lines show $\pm 4$ \% of the red line.}
\label{fig:KF}
\end{figure}

Figure \ref{fig:KF} shows the performance of the proposed method for rough wall TBL data of \cite{flack2020}. Note that that the contribution of viscous stress in $T$ is neglected. As observed in Figure \ref{fig:KF}, the proposed method shows good performance in the range $0.2 < \eta < 0.5$ for all the rough wall TBL cases. Since the original paper does not report $H$, $H = 1.38$ is used for all these cases. This value is chosen as it is close to smooth wall case at similar $Re_{\tau}$ listed in Table \ref{tab:data}. It will be shown later in section \ref{sec:sensitivity} that the proposed method is relatively insensitive to $H$. It is remarkable that the method is able to handle wall roughness accurately without requiring any modification.  

\subsection{Sensitivity to TBL parameters} \label{sec:sensitivity}

\begin{figure}
\begin{center}
\includegraphics[height=55mm]{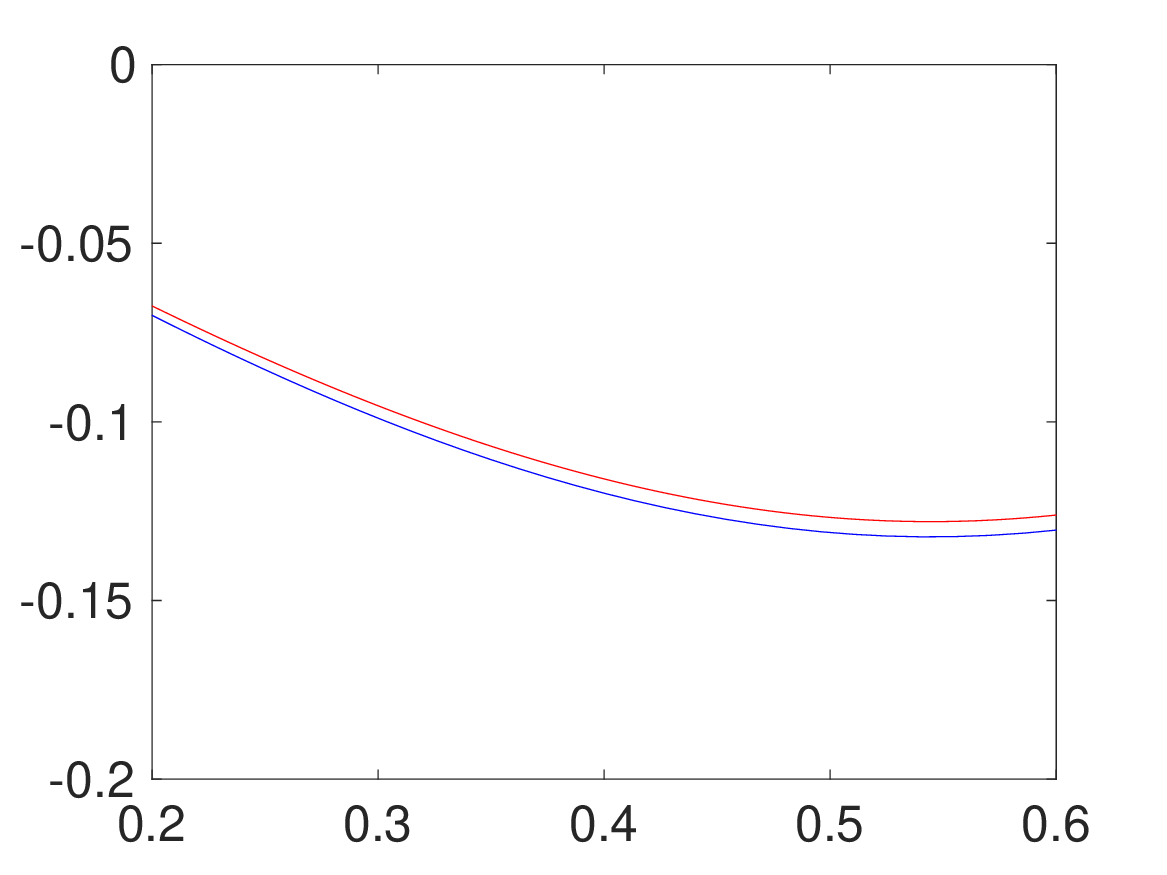}
\put(-110,-5){$\eta$}
\end{center}
\caption{Sensitivity of the proposed method to the error in $H$ is shown by plotting the term in brackets in Eq. \eqref{eq:error} for S1000 (blue) and S4000 (red).}
\label{fig:error}
\end{figure}

An important point to note is that the proposed method requires the knowledge of $H$ in addition to $\delta$. From the definition of $\delta^*$ and $\theta$, it is clear that the near-wall region contributes more to the former. Obtaining $H$ from the measured data in the absence of near-wall resolution can be challenging. It is tempting to assume the one-seventh power law for near-wall data to account for the contribution of missing data to the overall $H$. However, the goal here is to avoid using the inner layer altogether. Hence, the sensitivity of the obtained $u_{\tau}$ to $H$ is assessed using error analysis. It can be shown that the relative error in the friction velocity ($\epsilon_{u_{\tau}}$) is related to the relative error in the shape factor ($\epsilon_H$) as
\begin{equation}
\epsilon_{u_{\tau}} = \bigg (\frac {0.5H(f - \eta)}{H(1-f) + (H-1)(\eta-1)} \bigg )\epsilon_H. \label{eq:error}
\end{equation}
Figure \ref{fig:error} shows the term in parenthesis on the rhs of Eq. \eqref{eq:error} for S1000 and S4000 cases; note that it is of the order $10^{-1}$. The term in parenthesis shows a variation in the range $7-12 \%$, which implies that $10 \%$ error in $H$ would yield $0.7-1.2 \%$ error in the predicted $u_{\tau}$. Hence, it can be concluded that the predicted $u_{\tau}$ is relatively insensitive to the error in $H$. Instead of obtaining $H$ from measurements, one can use $H=1.36$ at high $Re$ (say $Re_{\theta} > 6500$) without compromising accuracy of the predicted $u_{\tau}$.

\begin{figure}
\begin{center}
\includegraphics[height=55mm]{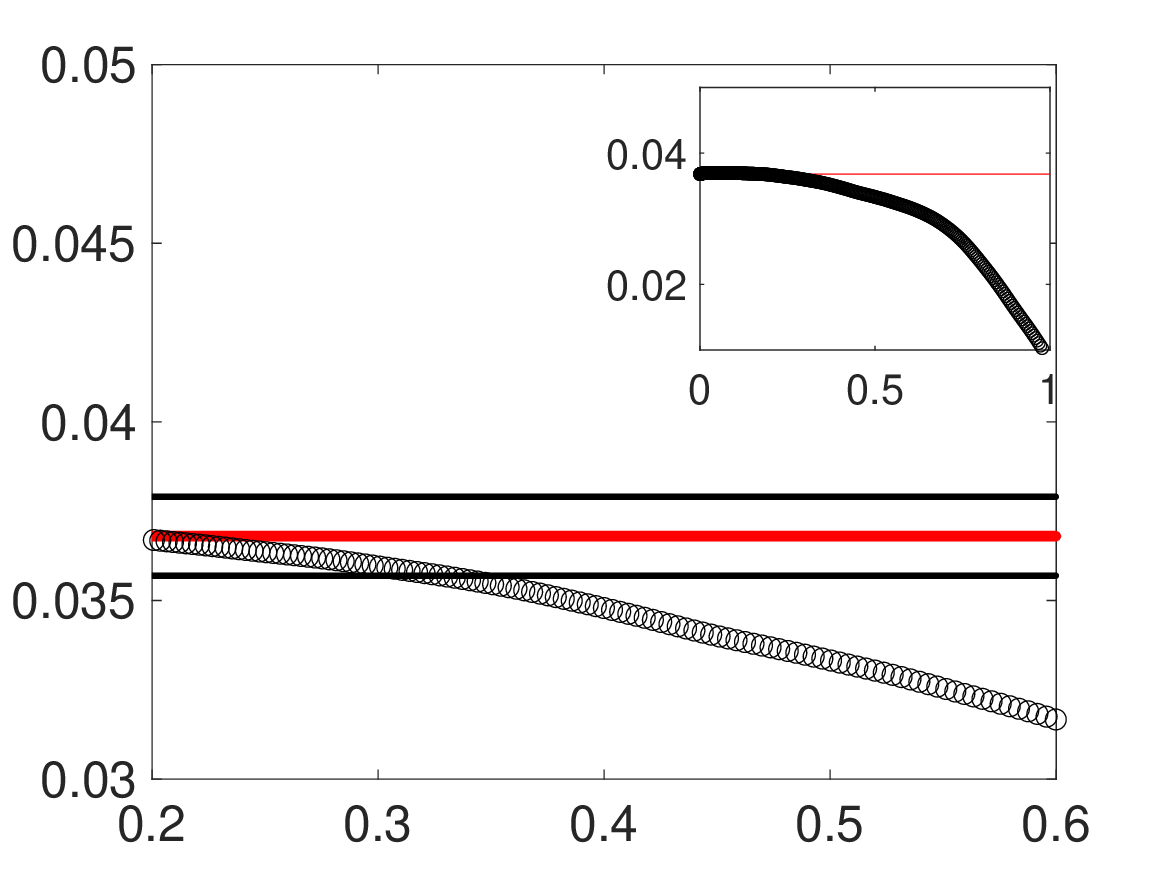}
\put(-110,-5){$\eta$} \\
\includegraphics[height=55mm]{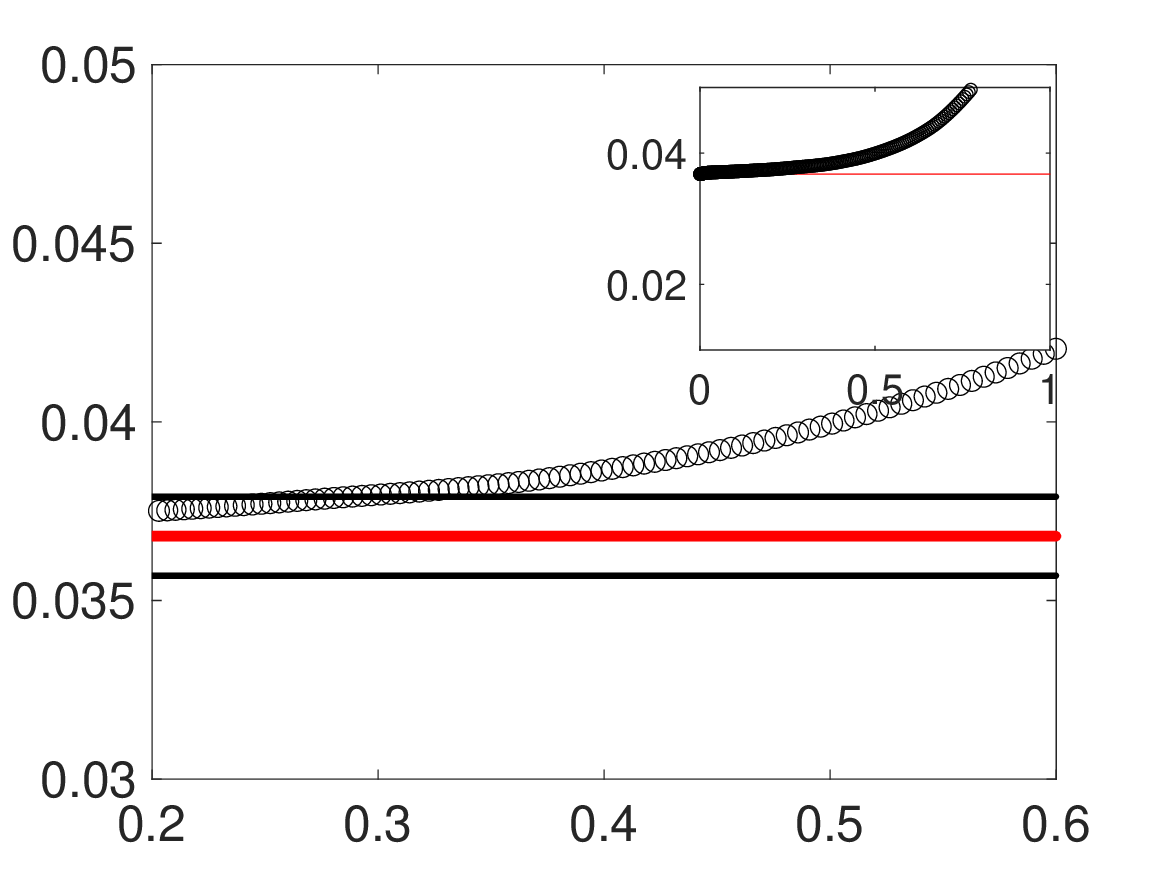}
\end{center}
\caption{Sensitivity of the proposed method to the error in $\delta$ is shown by using $\eta$ divided by 1.2 $(a)$ and 0.8 $(b)$ as input to the method and comparing the prediction to true values for S6500 case. Note that the black lines show $\pm 3$ \% of the red line.}
\label{fig:errordelta}
\end{figure}

For TBL, $\delta$ is not strictly defined. Most studies including the present work assume $\delta \approx \delta_{99}$, i.e. the wall-normal location where $U = 0.99U_e$. However, it has been reported in literature that the $\delta$ used in the composite profiles is typically larger than $\delta_{99}$ \citep{samie2018}. Also, iterative integral methods such as that of \cite{perry1990} yield $\delta$ that are up to $\sim 30 \%$ larger than $\delta_{99}$. Hence, the sensitivity of the obtained $u_{\tau}$ to $\delta$ needs to be assessed. Figure \ref{fig:errordelta} shows the predicted $u_{\tau}$ compared to the reference data for S6500 case if $\eta$ is divided by a factor of 1.2 (Figure \ref{fig:errordelta}$a$) or 0.8 (Figure \ref{fig:errordelta}$b$), which would correspond to $\pm 20 \% $ change in $\delta$. Overall, the method appears robust to errors in $\delta$ as long as the data points in the range $\eta < 0.4$ are used.  Hence, it can be concluded that the proposed method is robust to the uncertainty in $\delta$ and hence $\delta \approx \delta_{99}$ can be used without compromising accuracy of the predicted $u_{\tau}$.

\section{Extension to pressure gradient TBL} \label{sec:pg}
A major drawback of the method described in Section \ref{sec:model} is that it can not be used for pressure gradient TBL. In this section, the stress and wall-normal velocity model derivations are revisited to include the effects of pressure gradient on the mean shear stress in TBL.  

\subsection{Model derivation for $T$}

Using Eqs. \ref{eq:bl1} and \ref{eq:bl2}, it can be shown that
\begin{eqnarray}
\frac {\partial T} {\partial y}  & = &
- U_e \frac{dU_e}{dx} -U\frac{\partial V}{\partial y} + V\frac{\partial U }{\partial y},
\label{eq:bl3}
\end{eqnarray}
where the pressure gradient term in Eq. \ref{eq:bl2} is replaced by $U_e dU_e/dx$ in Eq. \ref{eq:bl3}. Note that this substitution assumes that $dP/dy=0$ i.e. $P = P_e$ throughout the boundary layer, which may not be true for general non-equilibrium flows. Integrating Eq. \ref{eq:bl3} from $0$ to a generic $y$ yields,
\begin{eqnarray}
{T} - {\tau_w} & = &  -U_e\frac{dU_e}{dx} y + \int_{0}^{y} \bigg ( -U\frac{\partial V}{\partial y} + V\frac{\partial U }{\partial y} \bigg) dy,
\label{eq:bl5}
\end{eqnarray}
where $\tau_w$ is the mean shear stress at the wall. After normalizing in viscous units,  Eq. \ref{eq:bl5} becomes
\begin{eqnarray}
{T^+} & = & 1 + \eta \beta \frac{\delta}{\delta^*} + I^+,
\label{eq:newT}
\end{eqnarray}
where,
\begin{eqnarray} 
{I^+} & = &
\int_{0}^{y^+} \bigg ( -U^+\frac{\partial V^+}{\partial y^+} + V^+\frac{\partial U^+ }{\partial y^+} \bigg) dy^+ \label{eq:I}
\end{eqnarray}
needs modeling to obtain a model for $T^+$ via Eq. \eqref{eq:newT}. It is also clear that $I$ is negative and its magnitude should increase away from the wall, since $T^+ > 1$ near wall but 0 at the boundary layer edge. 

\begin{figure}
\begin{center}
\includegraphics[height=55mm]{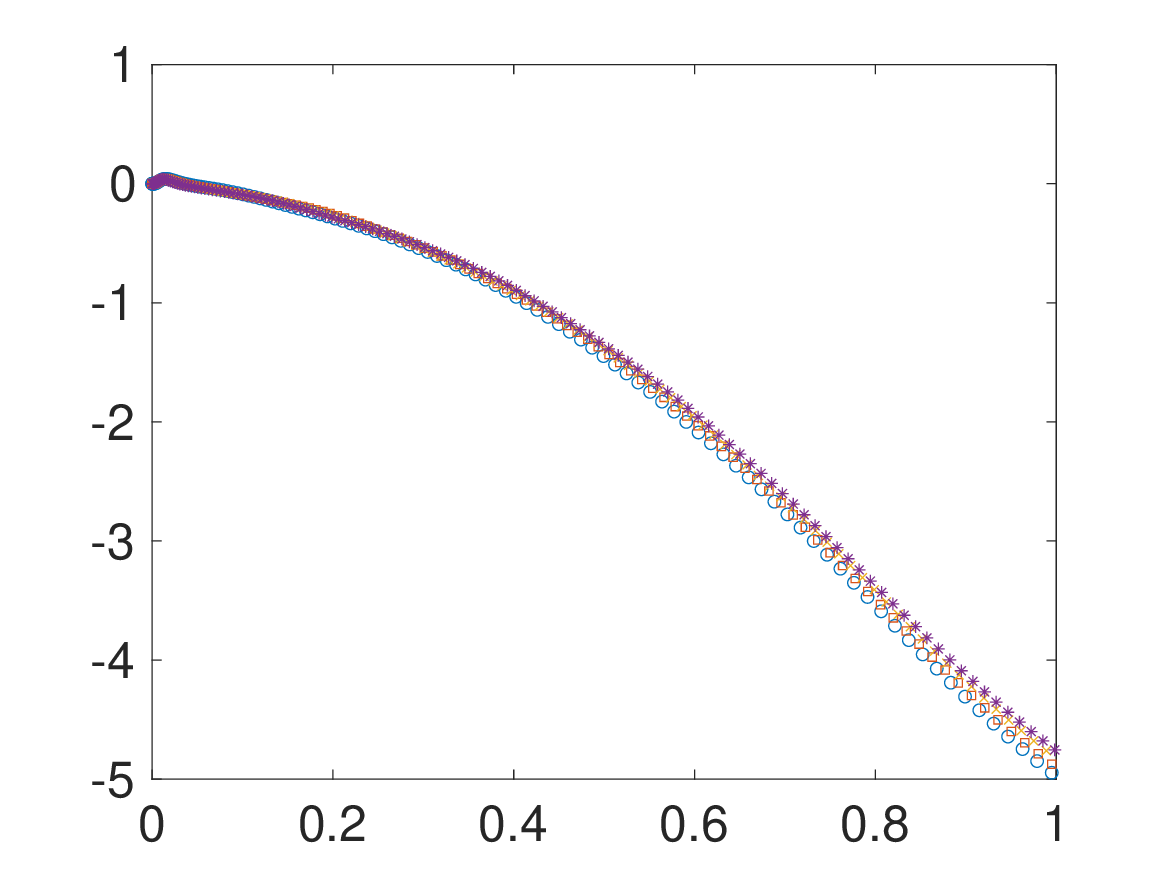}
\put(-220,90){$I$}
\put(-220,140){$(a)$} \\
\includegraphics[height=55mm]{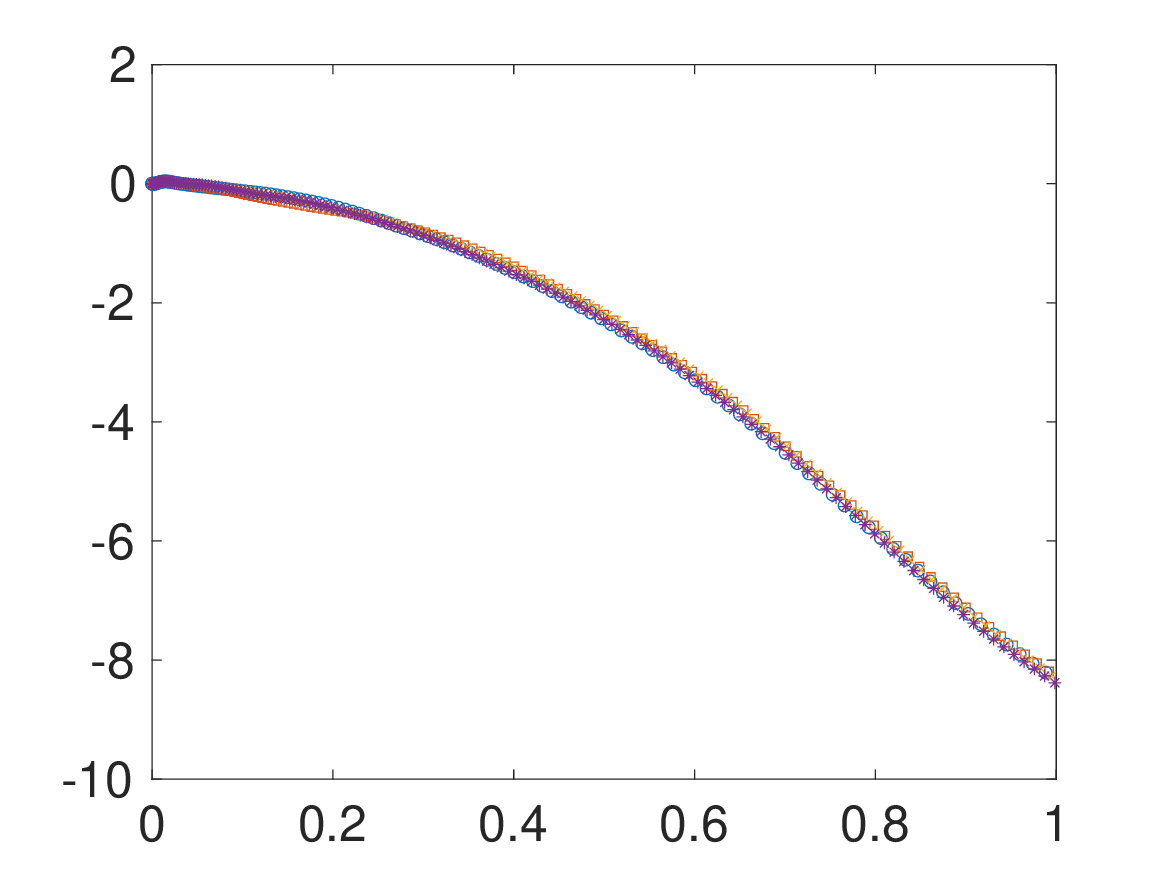}
\put(-110,-5){$\eta$}
\put(-220,90){$I$}
\put(-220,140){$(b)$}
\end{center}
\caption{$I$ profiles for $\beta \approx 1$ ($a$) and $\beta \approx 2$ ($b$) are shown.}
\label{fig:I}
\end{figure}

\begin{figure}
\begin{center}
\includegraphics[height=55mm]{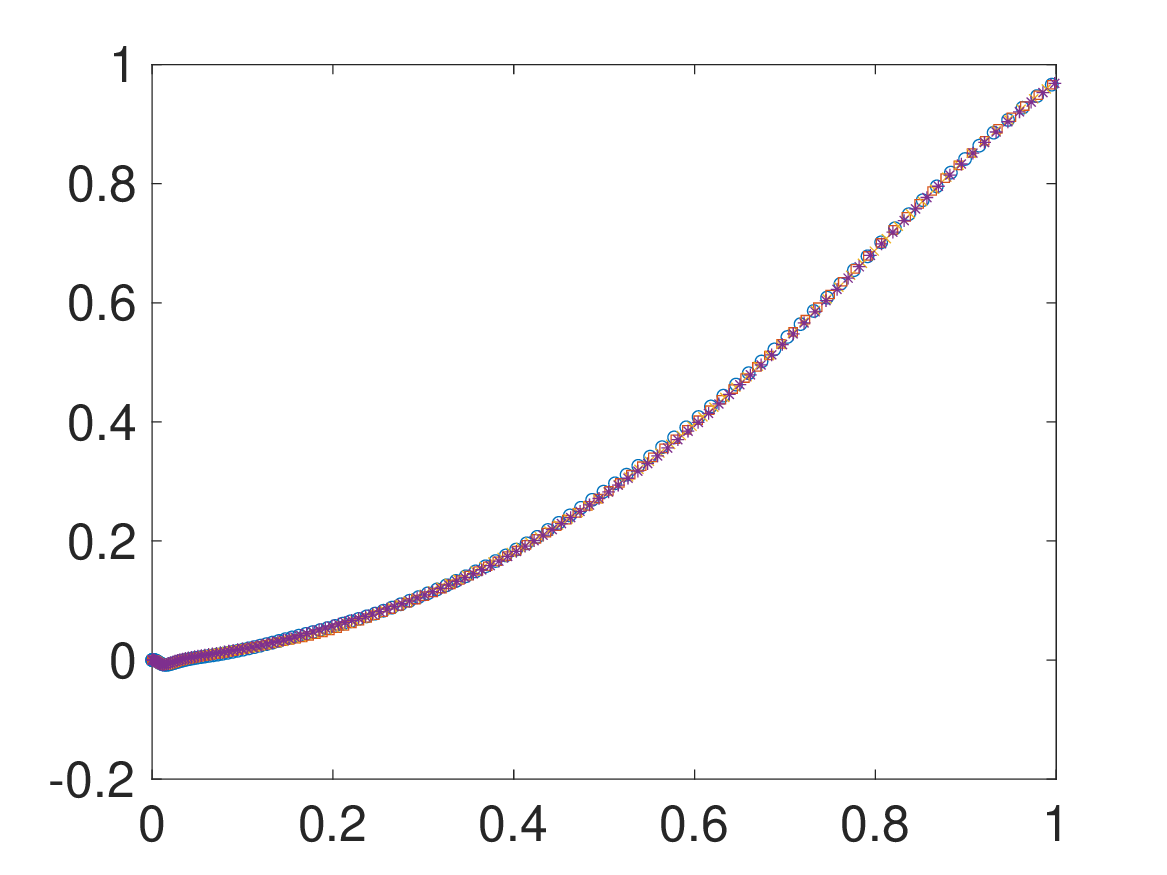}
\put(-220,140){$(a)$}
\put(-220,90){$\frac{I}{I(\delta)}$} \\
\includegraphics[height=55mm]{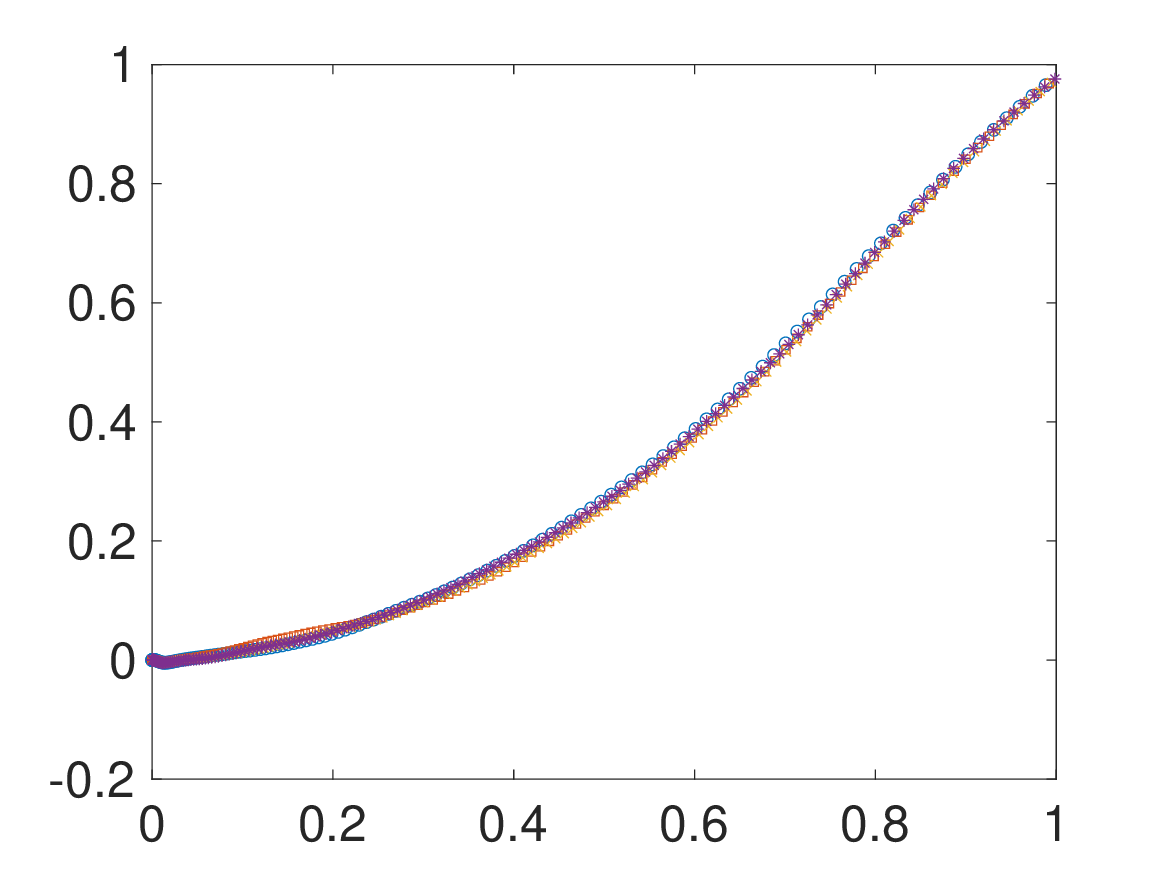}
\put(-220,140){$(b)$}
\put(-220,90){$\frac{I}{I(\delta)}$}
\put(-100,-5){$\eta$}
\end{center}
\caption{$I/I(\delta)$ profiles for $\beta \approx 1$ ($a$) and $\beta \approx 2$ ($b$) are shown.}
\label{fig:Iscaled}
\end{figure}

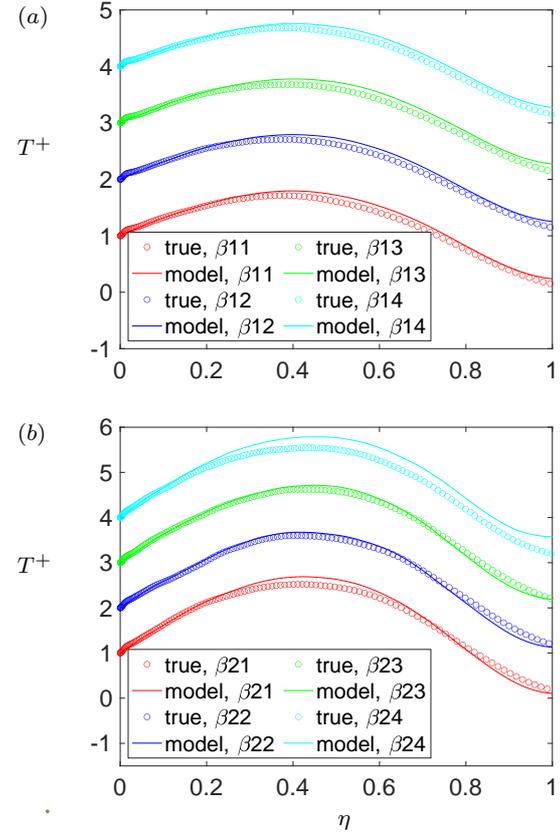
\begin{figure}
\begin{center}
\includegraphics[height=55mm]{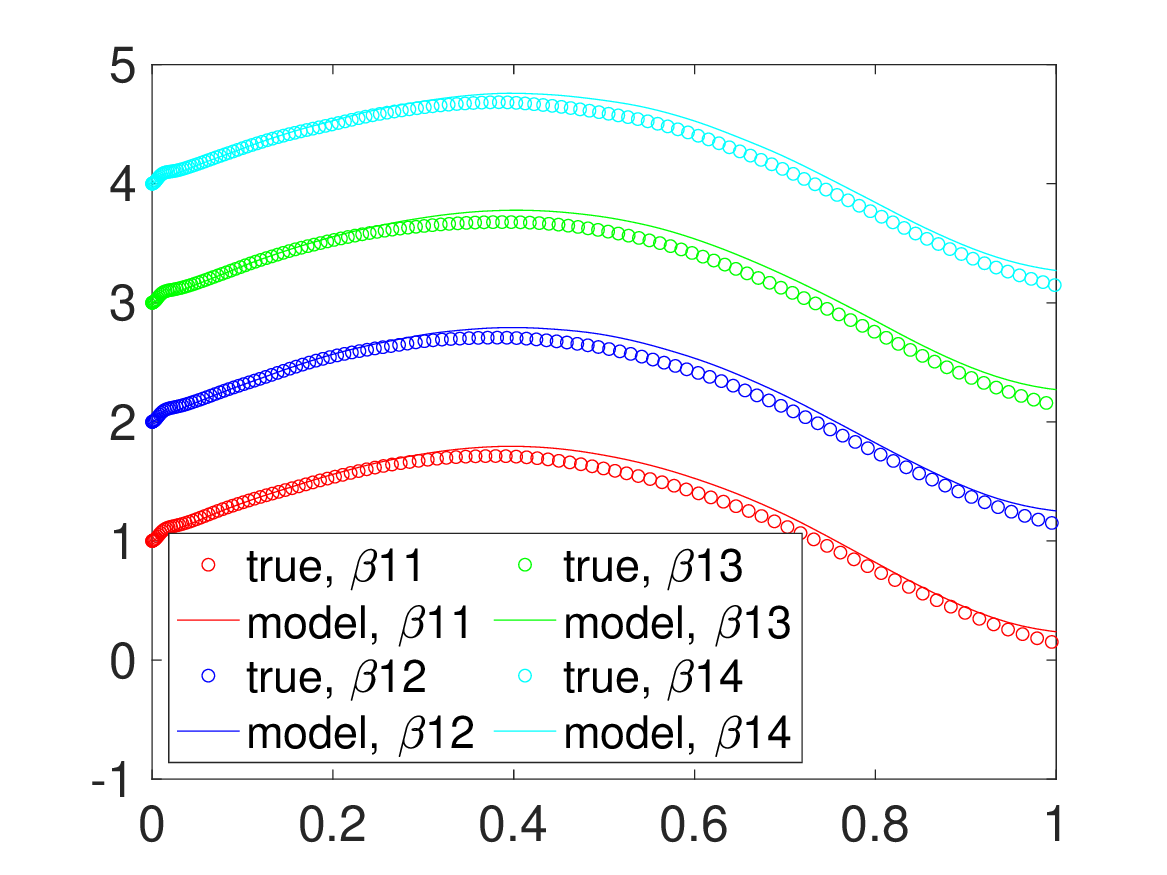}
\put(-220,140){$(a)$}
\put(-220,90){$T^+$} \\
\includegraphics[height=55mm]{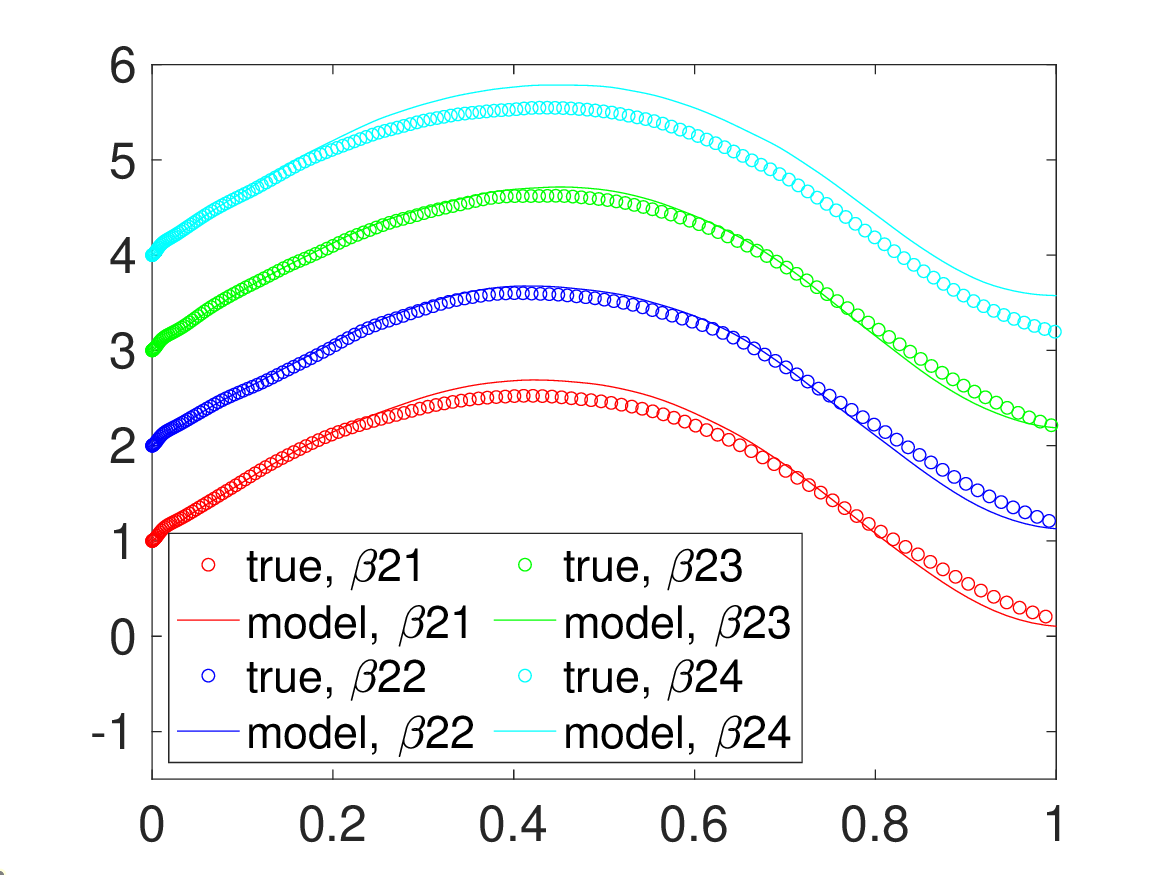}
\put(-220,140){$(b)$}
\put(-220,90){$T^+$}
\put(-100,-5){$\eta$}
\end{center}
\caption{Modeled $T^+$ profiles compared to the reference data for $\beta \approx 1$ ($a$) and $\beta \approx 2$ ($b$) cases are shown. Note that the profiles are shifted upwards by 1 for clarity.}
\label{fig:Tcomp}
\end{figure}

In order to model $I$, the first thing to note is that $I$ is a function of $y$ and using Eq. \eqref{eq:newT}, it can be shown that $I$ satisfies the boundary conditions
\begin{eqnarray}
I(0) & = & 0, \\ 
I(\delta) & = & - \bigg (1 + \beta\frac{\delta}{\delta^*} \bigg ).
\end{eqnarray}
The term $I$ is modeled using the existing TBL data from \cite{bobke2017} as described in Section \ref{sec:data}. Figure \ref{fig:I} shows $I$ profiles for $\beta11-\beta14$ and $\beta21-\beta24$ cases listed in table \ref{tab:data1}. As expected, $I$ is negative and its magnitude monotonically increases away from the wall. The profiles do not collapse. Figure \ref{fig:Iscaled} shows the same data after normalizing it with $I(\delta)$. The profiles show excellent collapse to a curve which monotonically increases from 0 to 1 in TBL. The excellent collapse of the data suggests that scaled $I$ is relatively insensitive to $Re_{\tau}$, which is encouraging for the purpose of model development. Moreover, Eq. \eqref{eq:I} suggests that $I^+ \sim U^+V^+$ implying $I/I(\delta) \sim UV/U_eV_e$. Therefore, a simple modeling choice for $I$ is
\begin{eqnarray}
I  & = &  I(\delta) \frac{UV}{U_eV_e},
\label{eq:Imodel} 
\end{eqnarray}
which combined with Eq. \eqref{eq:bl6} yields a simple model for $T^+$ as 
\begin{eqnarray}
{T^+}  & = &  1 + \eta \beta \frac{\delta}{\delta^*}  - \bigg (1 + \beta \frac{\delta}{\delta^*}\bigg )\frac{UV}{U_eV_e}.
\label{eq:Tfinal} 
\end{eqnarray}
It is worth mentioning that the modeling choice of Eq. \eqref{eq:Imodel} is not necessarily optimal. It is chosen just for simplicity and it will be shown later that this modeling choice is appropriate for a variety of pressure gradient TBL. The model prediction for $T^+$ (Eq. \eqref {eq:Tfinal}) is compared to the reference data in Figure  \ref{fig:Tcomp} for cases $\beta11-\beta14$ (Figure  \ref{fig:Tcomp}$a$) and $\beta 21-\beta24$ (Figure  \ref{fig:Tcomp}$b$) respectively. Overall, the model shows good performance despite its simplicity. 

Note that Eq. \eqref{eq:Tfinal} gives a model for $T^+$ in terms of $\beta$, which itself contains $u_{\tau}$ as mentioned earlier. Hence, it is useful to obtain $T^+$ in terms of $K$ using Eq. \eqref{eq:k} to obtain
\begin{eqnarray}
u_{\tau} & = & \sqrt{ \frac {T + K Re_{\delta} U_e^2 \bigg (\eta - \frac{UV}{U_eV_e}  \bigg)} {1  - \frac{UV}{U_eV_e}}}
\label{eq:pgmethod}
\end{eqnarray}
where, $Re_{\delta}$ is the boundary layer thickness based $Re$.  Eq. \eqref{eq:pgmethod} can be used to obtain $u_{\tau}$ for a pressure gradient TBL if all the terms on the rhs are known. Similar to the ZPG TBL, a model for $V$ is needed if unavailable. The $V$ model of \citet{kumar2021} (Eq. \eqref{eq:Vmodel}) was derived for zero pressure gradient TBL and hence, needs extension to include pressure gradient  effects. 

\subsection{Model derivation for $V$}
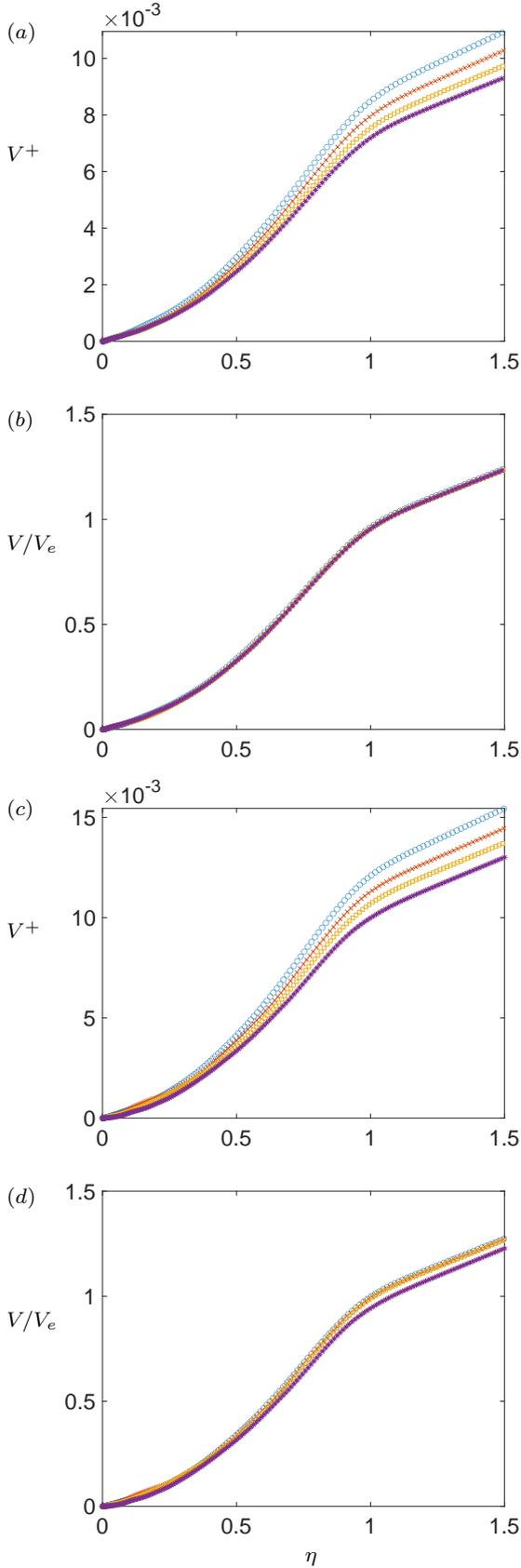
\begin{figure}
\begin{center}
\includegraphics[height=55mm]{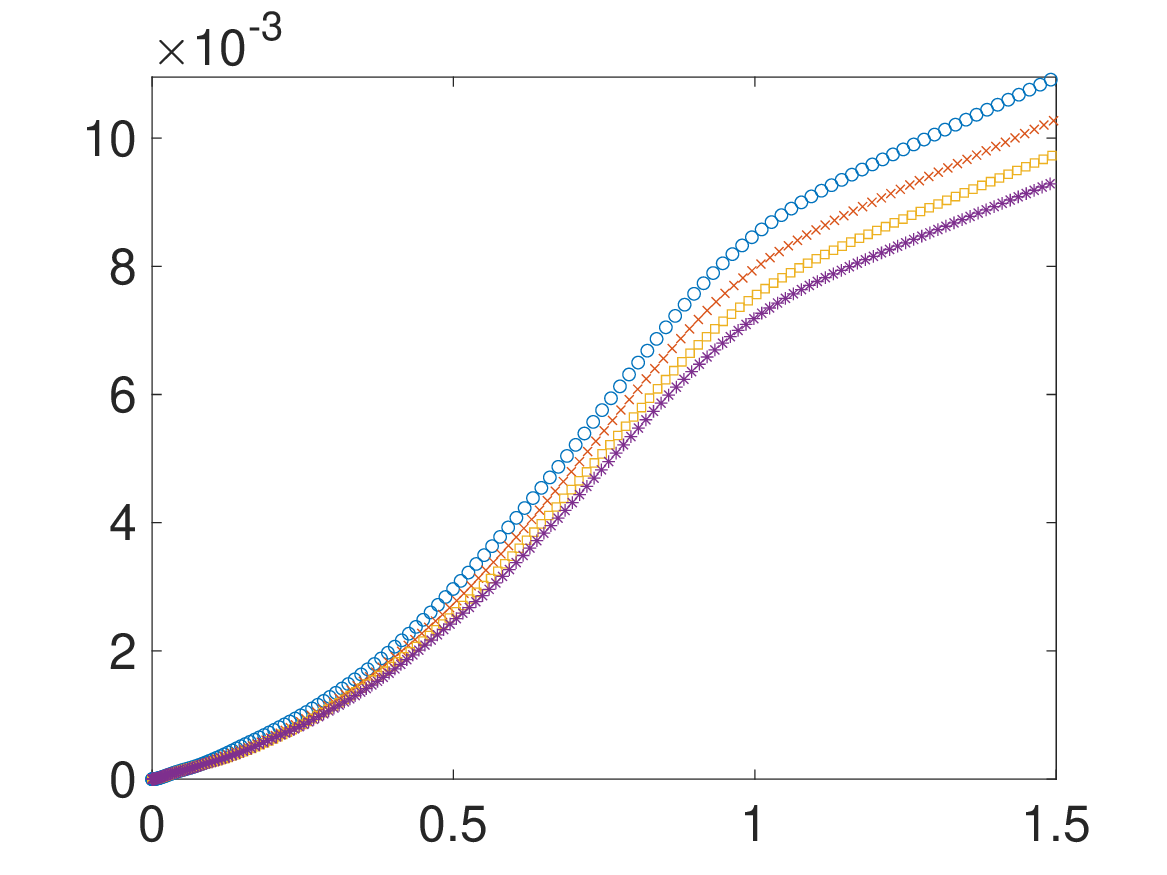}
\put(-220,90){$V^+$}
\put(-220,140){$(a)$} \\
\includegraphics[height=55mm]{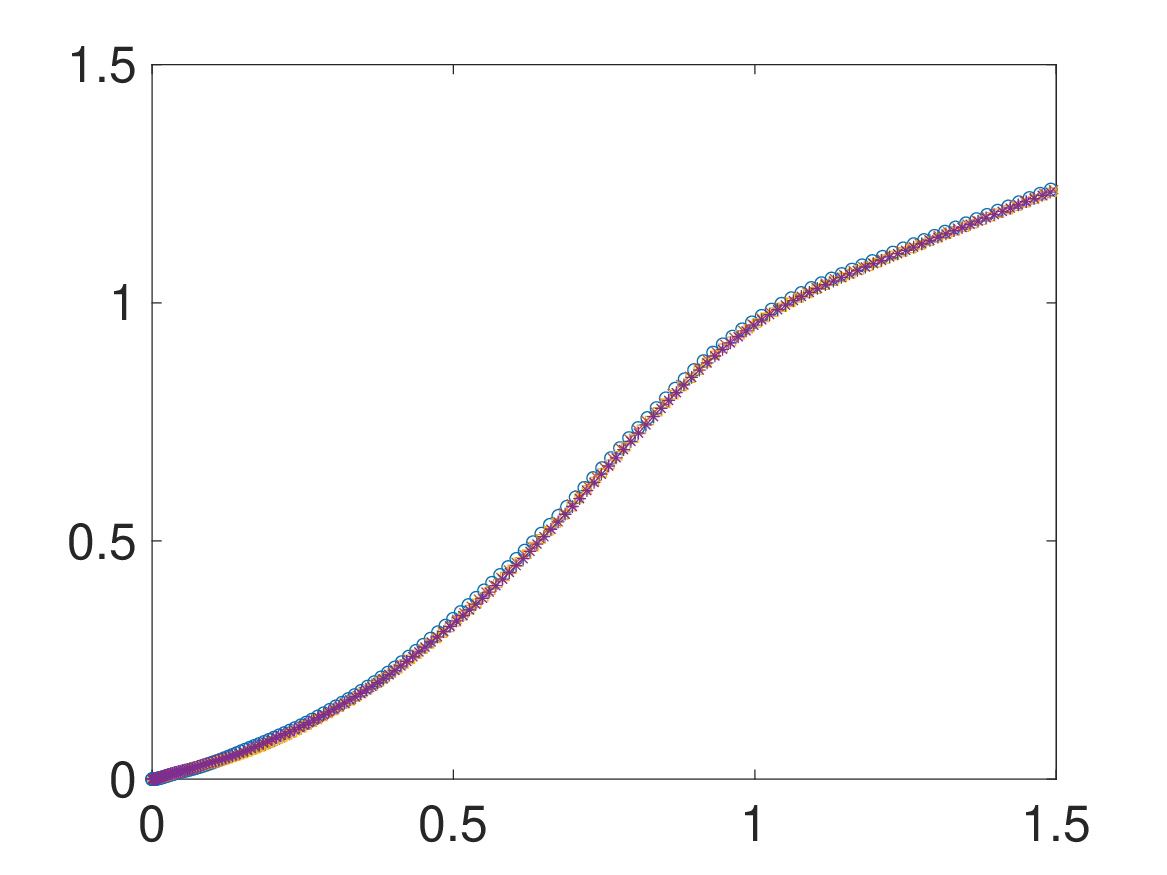}
\put(-220,90){$V/V_e$}
\put(-220,140){$(b)$} \\
\includegraphics[height=55mm]{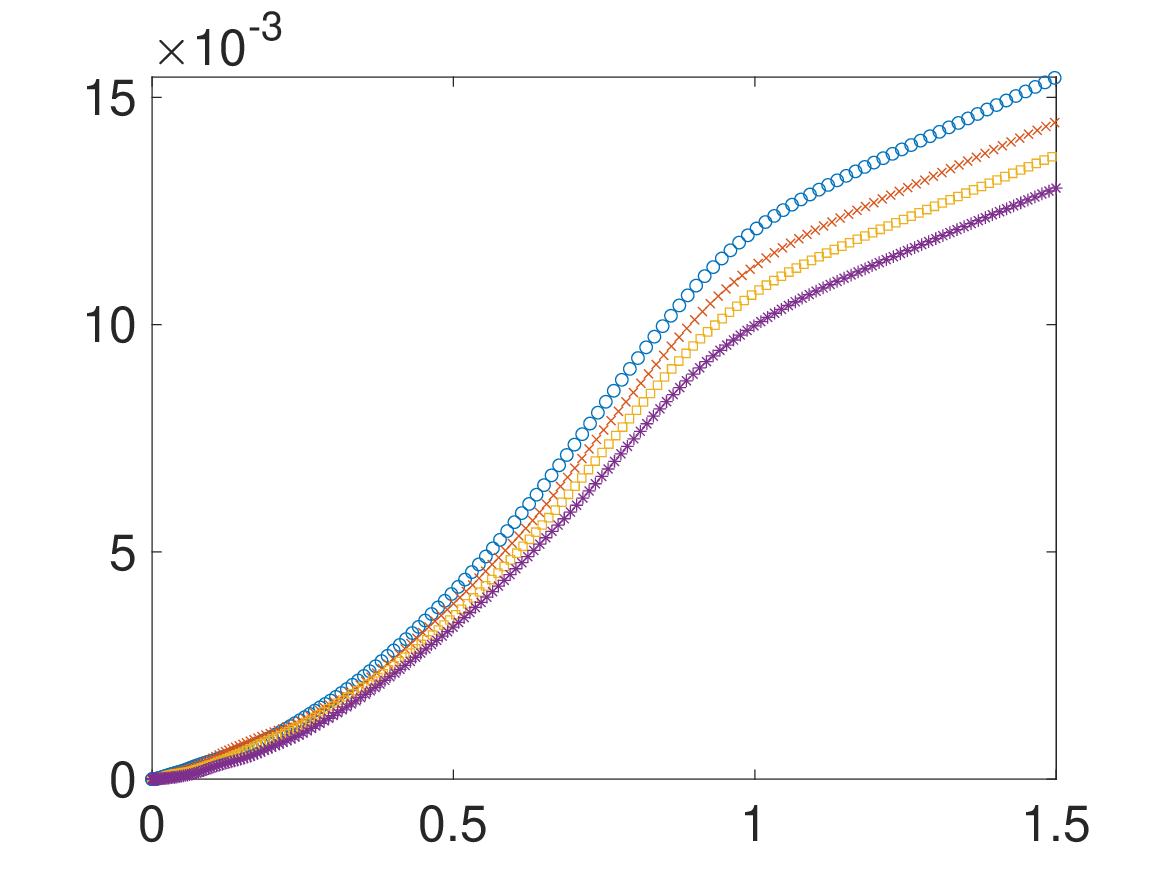}
\put(-220,90){$V^+$}
\put(-220,140){$(c)$} \\
\includegraphics[height=55mm]{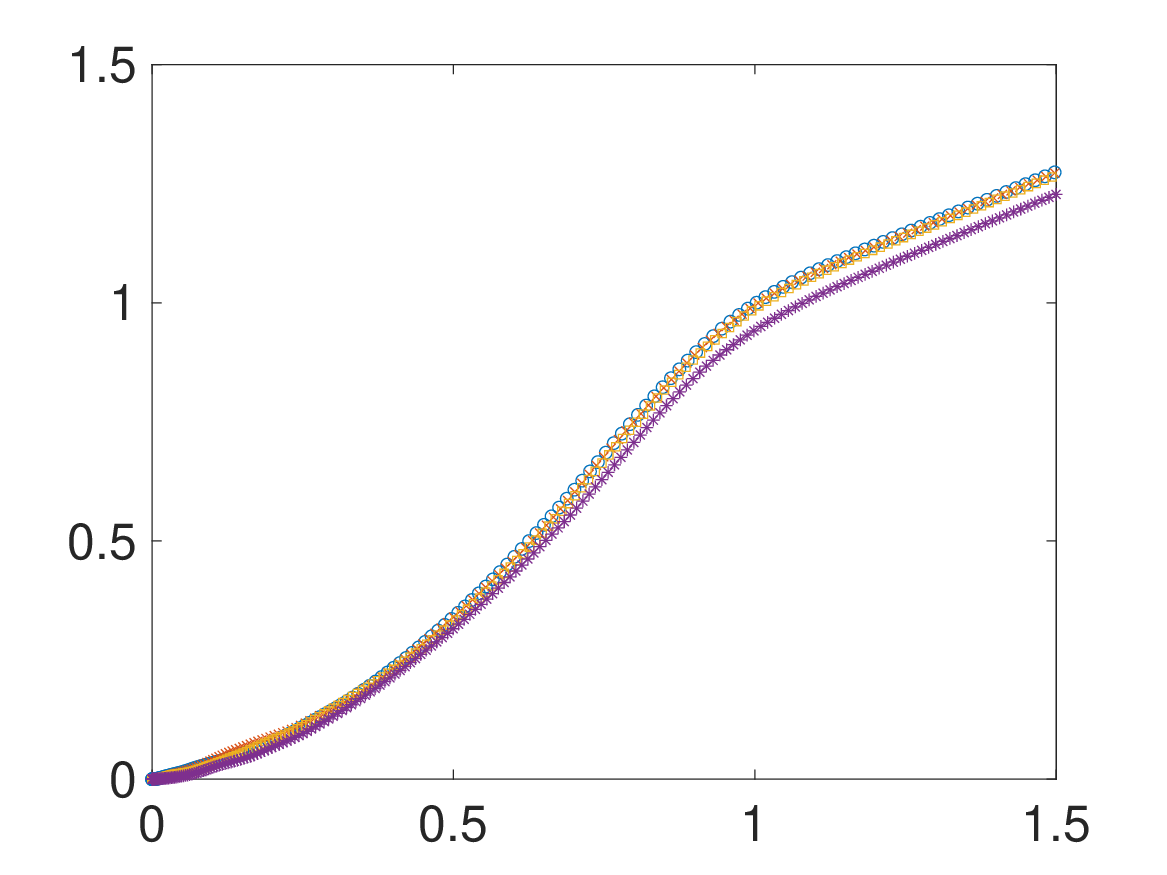}
\put(-100,-5){$\eta$}
\put(-220,90){$V/V_e$}
\put(-220,140){$(d)$}
\end{center}
\caption{$V^+$ ($a,c$) and $V/V_e$ ($b,d$) are shown for $\beta \approx 1 (a,b)$ and $ 2 (c,d)$.}
\label{fig:V}
\end{figure}

A key change in TBL behavior under pressure gradient is that $V$ is no longer constant outside TBL. In fact, using the continuity equation (Eq. \eqref{eq:bl1}) it can be shown that
\begin{eqnarray} 
\frac{\partial }{\partial \eta} \bigg (\frac{V}{V_e} \bigg ) & = &  \frac{\beta \delta}{U_e^+V_e^+ \delta^*} \label{eq:slope}
\end{eqnarray}
for all $\eta \geq 1$. Also, the edge velocities are related to the boundary layers integral parameters as
\begin{eqnarray} 
U_e^+ V_e^+ & = &  H + \beta \bigg (1 + H + \frac{\delta}{\delta^*} \bigg ), \label{eq:Ve}
\end{eqnarray}
which can be used to obtain $V_e$ \citep{wei2017, kumar2018}. Figure \ref{fig:V} shows $V$ for four profiles taken each from $\beta \approx 1$ and $2$ cases respectively. It is clear that the edge wall-normal velocity obtained using Eq. \eqref{eq:Ve} is able to collapse the profiles to a single curve. Therefore, a model for $V/V_e$ is sought as that can be expected to work for any $Re$. 

The model for $V/V_e$ has two basic requirements: (i) it should reduce to the model of \citet{kumar2021} for $\beta = 0$, and (ii) it should have the correct slope (Eq. \eqref{eq:slope}) for $\eta \geq 1$. A convenient choice for such model is 
\begin{eqnarray} 
\frac{V}{V_e} & = &  \frac{V}{V_e}\bigg |_0 \bigg ( 1 + \frac{\beta}{U_e^+V_e^+}   \frac{\delta}{\delta^*} (\eta -1)\bigg ) \nonumber \\
& = &  \frac{V}{V_e}\bigg |_0 \bigg ( 1 - K Re_{\delta} \frac{U_e}{V_e}(\eta - 1)\bigg ), \label{eq:Vmodelnew}
\end{eqnarray}
where $V/V_e|_0$ is the model of \citet{kumar2021} (Eq. \eqref{eq:Vmodel}) and Eq. \eqref{eq:k} is used to write $\beta$ in terms of $K$.

\begin{figure}
\begin{center}
\includegraphics[height=55mm]{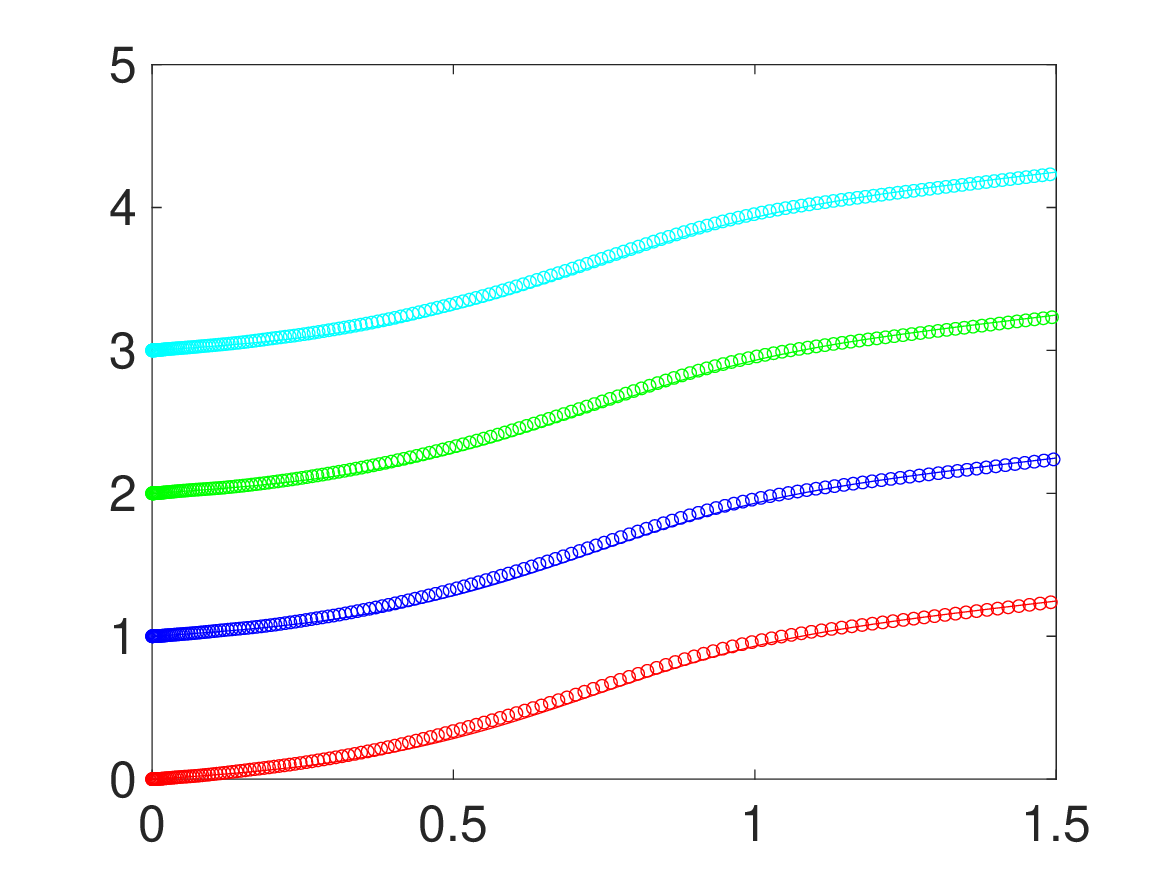}
\put(-220,90){$V/V_e$}
\put(-220,140){$(a)$} \\
\includegraphics[height=55mm]{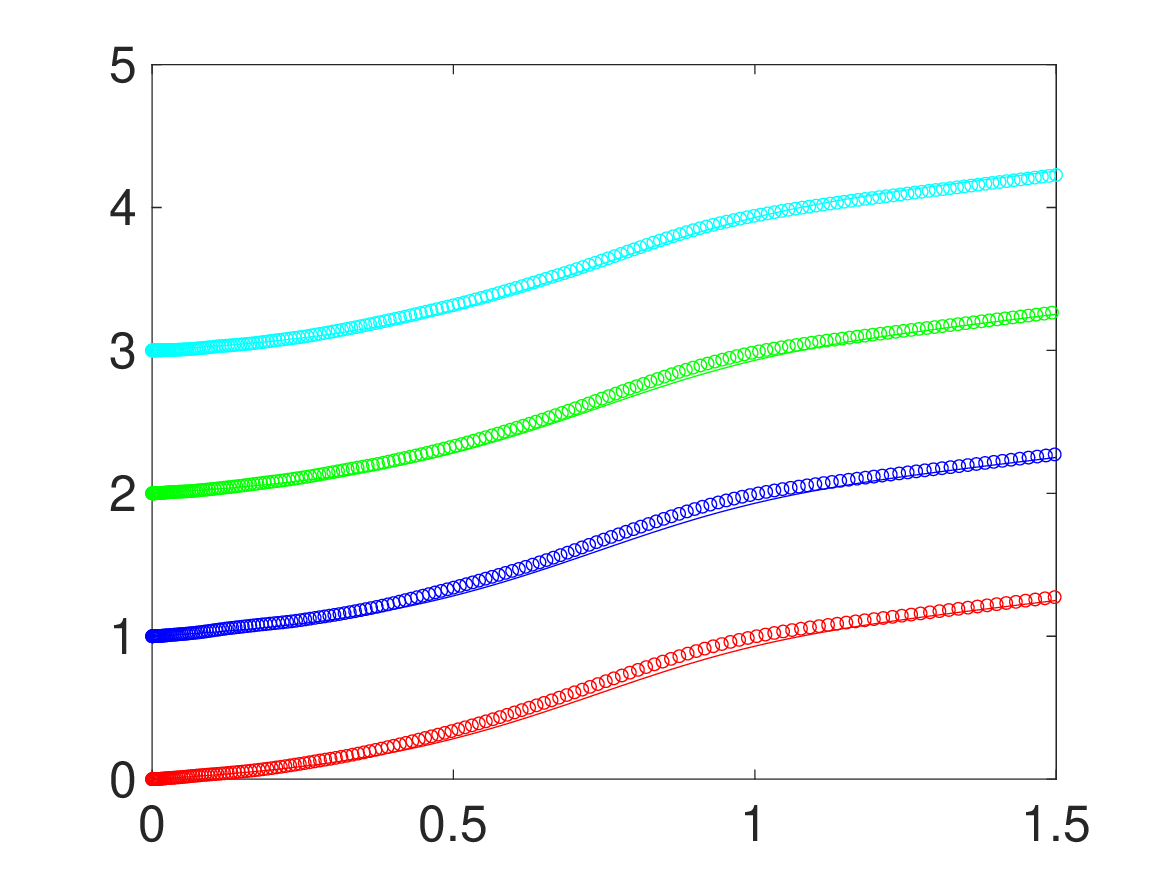}
\put(-220,140){$(b)$} 
\put(-220,90){$V/V_e$}
\put(-100,-5){$\eta$}
\end{center}
\caption{$V/V_e$ model performance is shown for $\beta \approx 1 (a)$ and $2 (b)$ respectively listed in table \ref{tab:data1}. Note that the profiles are
	shifted upwards by 1 from $\beta11-\beta14$ ($a$) and $\beta21-\beta24$ ($b$) respectively for clarity.}
\label{fig:Vmodel1}
\end{figure}

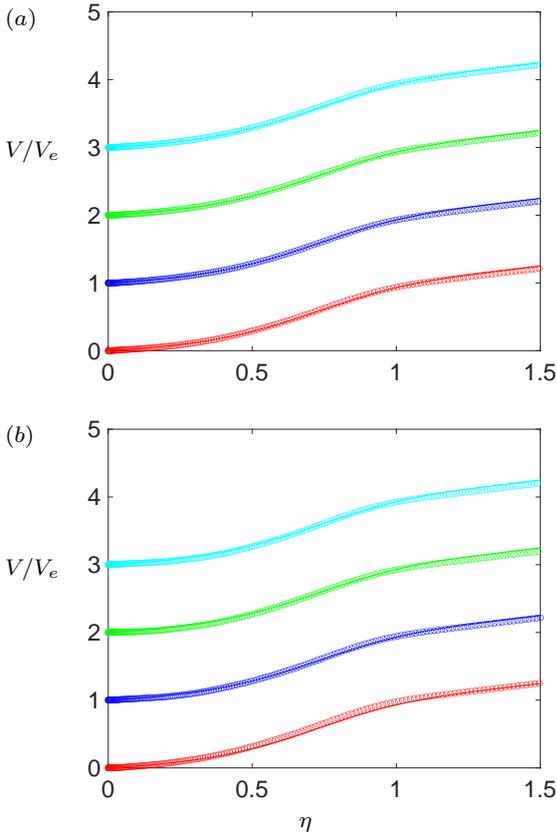
\begin{figure}
\begin{center}
\includegraphics[height=55mm]{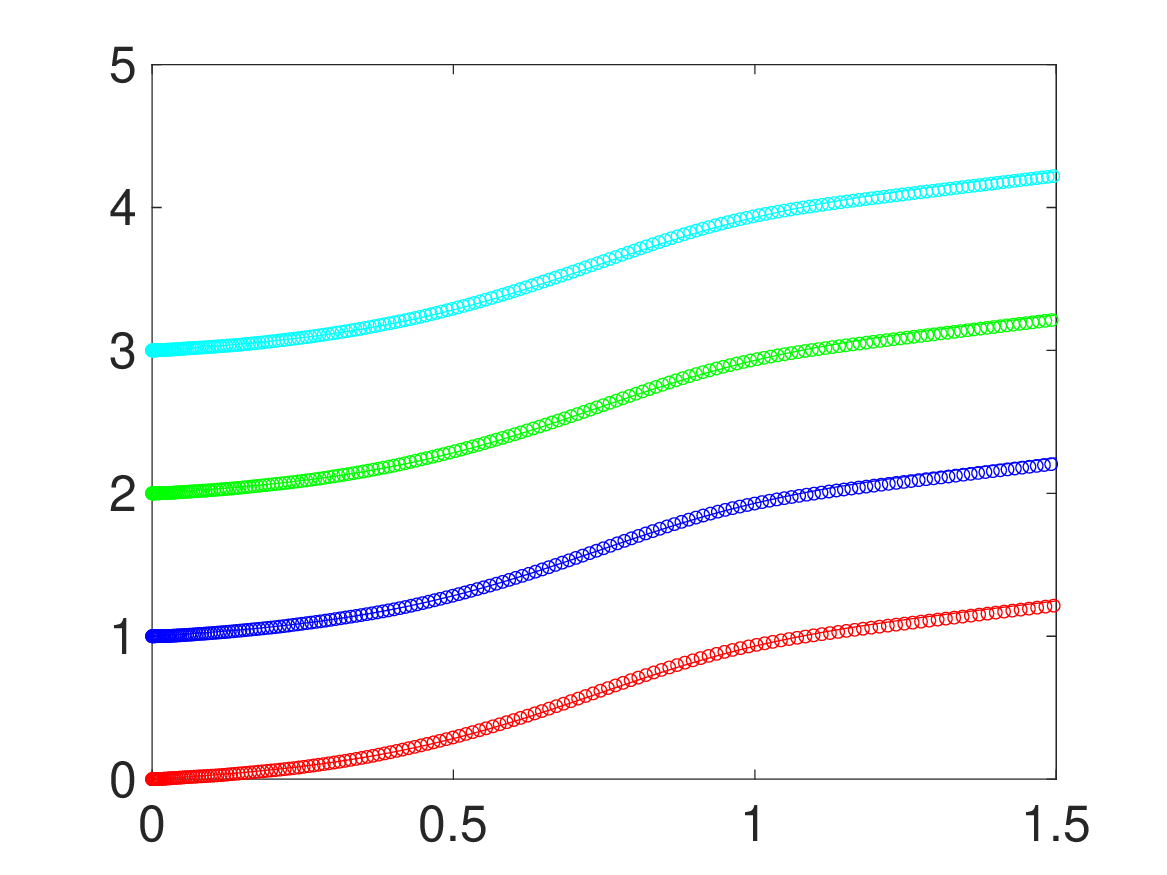}
\put(-220,90){$V/V_e$}
\put(-220,140){$(a)$} \\
\includegraphics[height=55mm]{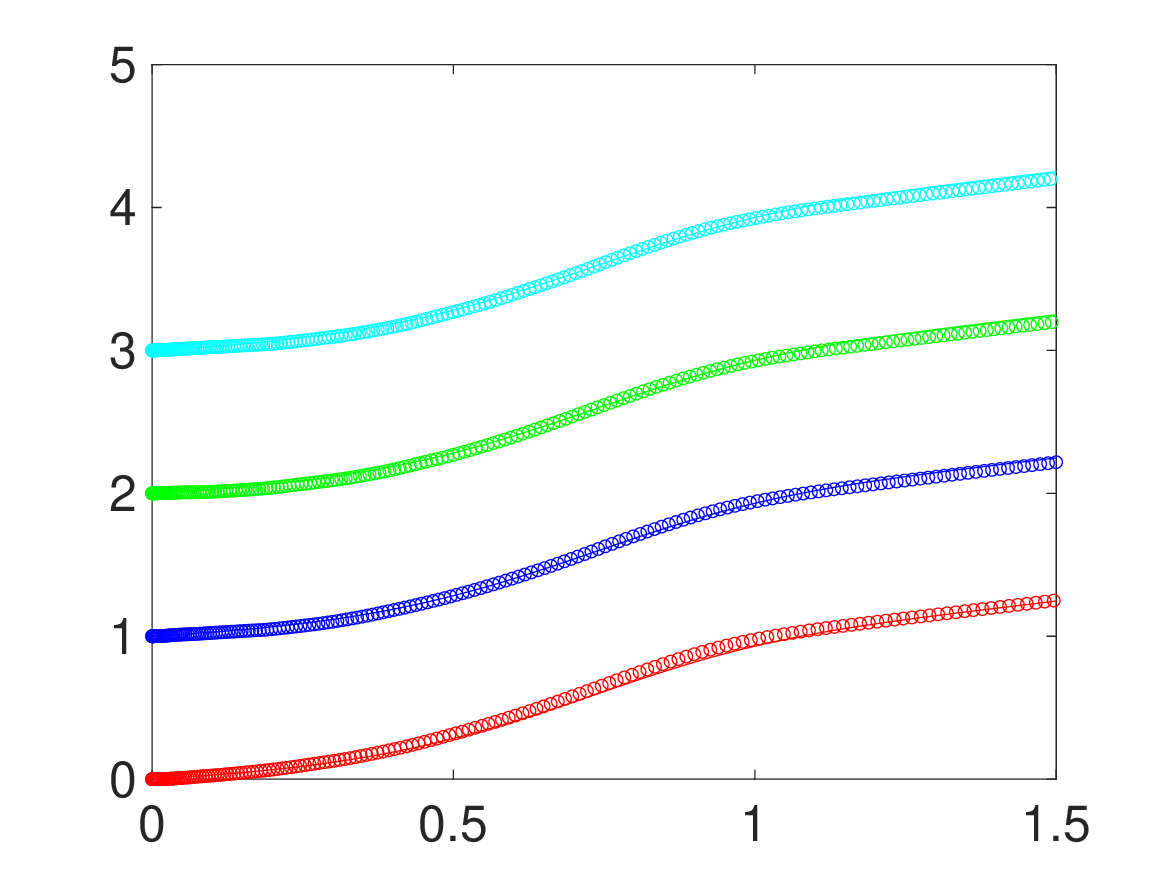}
\put(-220,140){$(b)$}
\put(-220,90){$V/V_e$}
\put(-110,-5){$\eta$} 
\end{center}
\caption{$V/V_e$ model performance is shown for $m \approx -0.13$ ($a$) and $-0.16$ ($b$) respectively listed in table \ref{tab:data1}. Note that the profiles are shifted upwards by 1 from $m131-m134$ ($a$) and $m161-m164$ ($b$) respectively for clarity.}
\label{fig:Vmodel2}
\end{figure}

Next, the performance of the $V$ model is assessed in figure \ref{fig:Vmodel1} using the same data used in figure \ref{fig:V}. The model shows good agreement with the reference data for all the profiles. The model is also tested for constant $m$ cases listed in table \ref{tab:data1} as shown in figure \ref{fig:Vmodel2} showing good agreement for all the cases.

\subsection {Discussion}
\begin{figure}
\begin{center}
\includegraphics[height=55mm]{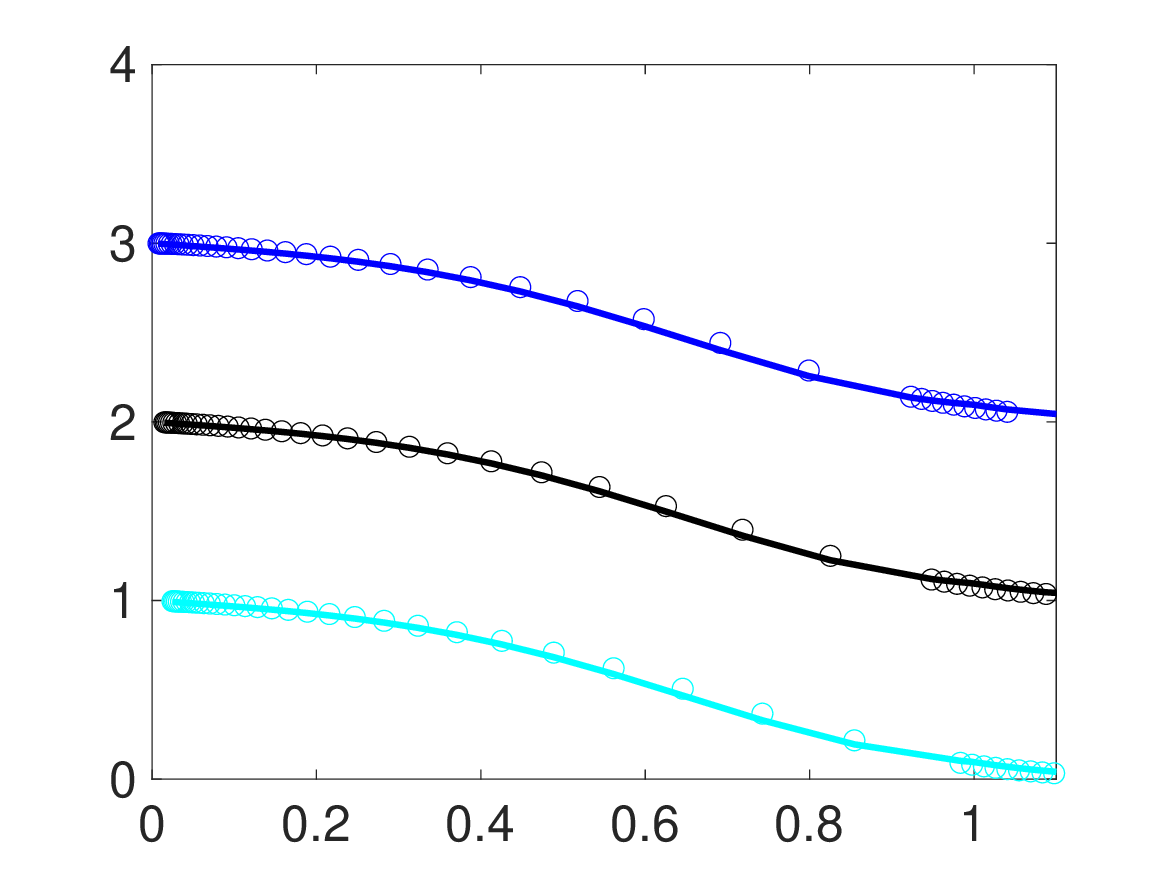}
\put(-220,90){$T^+$}
\put(-110,-5){$\eta$} 
\end{center}
\caption{$T^+$ profiles predicted by Eq. \eqref{eq:Tfinal} (symbols) are compared to that obtained from Eq. \eqref{eq:Tmodelin} showing good agreement for R0, R- and R+ cases listed in Table \ref{tab:data}. Note that the profiles are shifted upwards by 1 for clarity.}
\label{fig:modelcomp}
\end{figure}

Before proceeding further, it is important to understand key differences in the underlying assumptions of Eq. \eqref{eq:Tfinal} compared to that of Eq. \eqref{eq:Tmodelin}. Note that while deriving the latter, both the pressure gradient and the wall-normal velocity terms were ignored and the resulting equation was integrated from a generic $y$ to $\delta$. However, this is not the case for Eq. \eqref{eq:newT}, where both the aforementioned terms are retained and the resulting equation is integrated from $0$ to a generic $y$. This is the reason why setting $\beta = 0$ in Eq. \eqref{eq:Tfinal} yields a model for $T$ in ZPG TBL as 
\begin{eqnarray}
T^+ = 1 -  \frac{UV}{U_eV_e},
\end{eqnarray}
which is different than Eq. \eqref{eq:Tmodelin}. However, both these expressions can be expected to give very similar results for any ZPG TBL. For example, Figure \ref{fig:modelcomp} shows comparison between the prediction of Eqs. \eqref{eq:Tmodelin} and \eqref{eq:Tfinal} for rough wall ZPG TBL data used in figure \ref{fig:KF}. The smooth wall ZPG TBL cases (not shown here) show similar agreement between the two models. 

\subsection{Method performance for pressure gradient TBL}

Two popular methods in literature to determine $u_{\tau}$ from profile data are corrected Clauser chart method and fitting the data to the inner layer profile of \cite{nickels2004}. Both these approaches require accurate near-wall data from the inner layer. For example, \cite{knopp2021} performed experiments on pressure gradient TBL, where they measured $u_{\tau}$ directly and compared it to that obtained using indirectly using both the aforementioned methods. Their profile measurements had fine resolution near wall as they were able to resolve the buffer layer adequately. They reported an error of approximately $6 \%$ using the corrected Clauser chart method  and $4 \%$ using the latter method. The correction used for the Clauser chart method were empirical. The readers are referred to   \cite{knopp2021} for a detailed discussion on uncertainties of the methods, measurements etc. 

\begin{figure}
\begin{center}
\includegraphics[height=55mm]{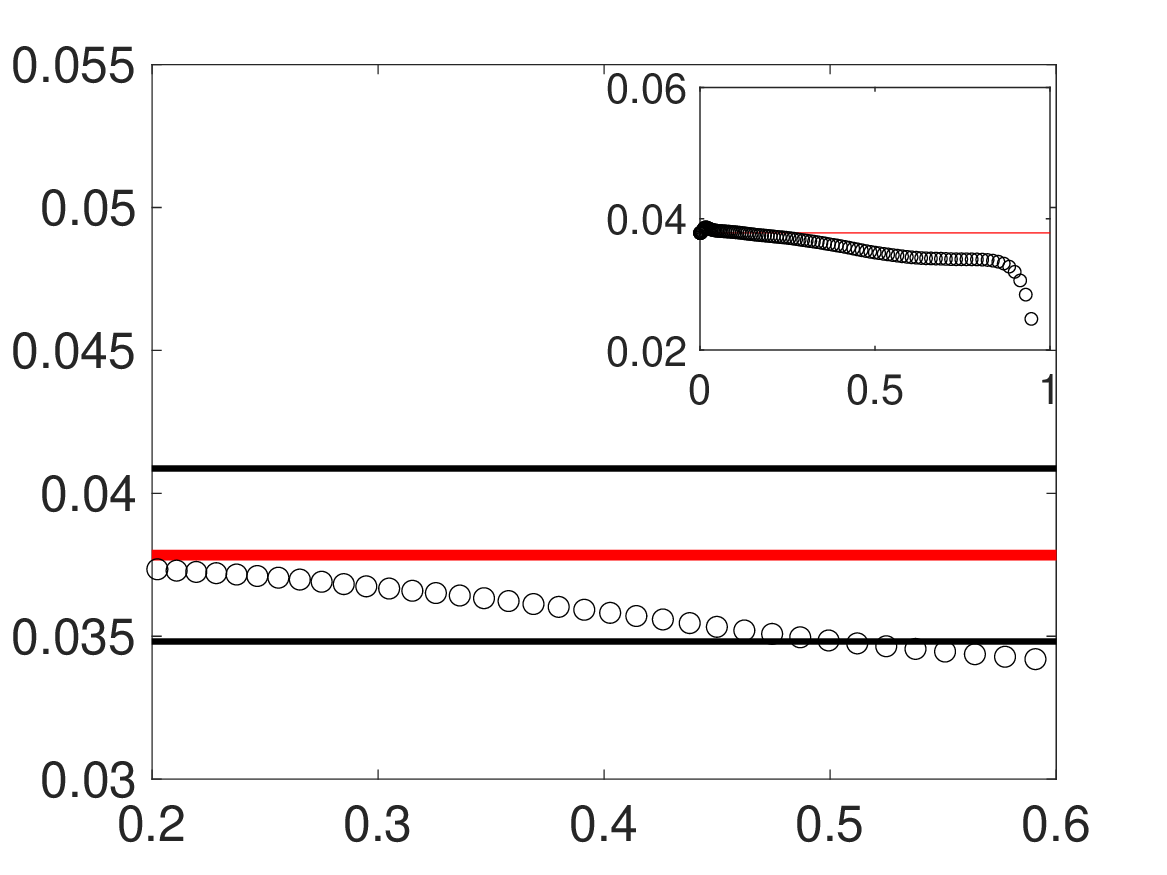}
\put(-220,80){$u_{\tau}/U_e$}
\put(-220,140){$(a)$} \\
\includegraphics[height=55mm]{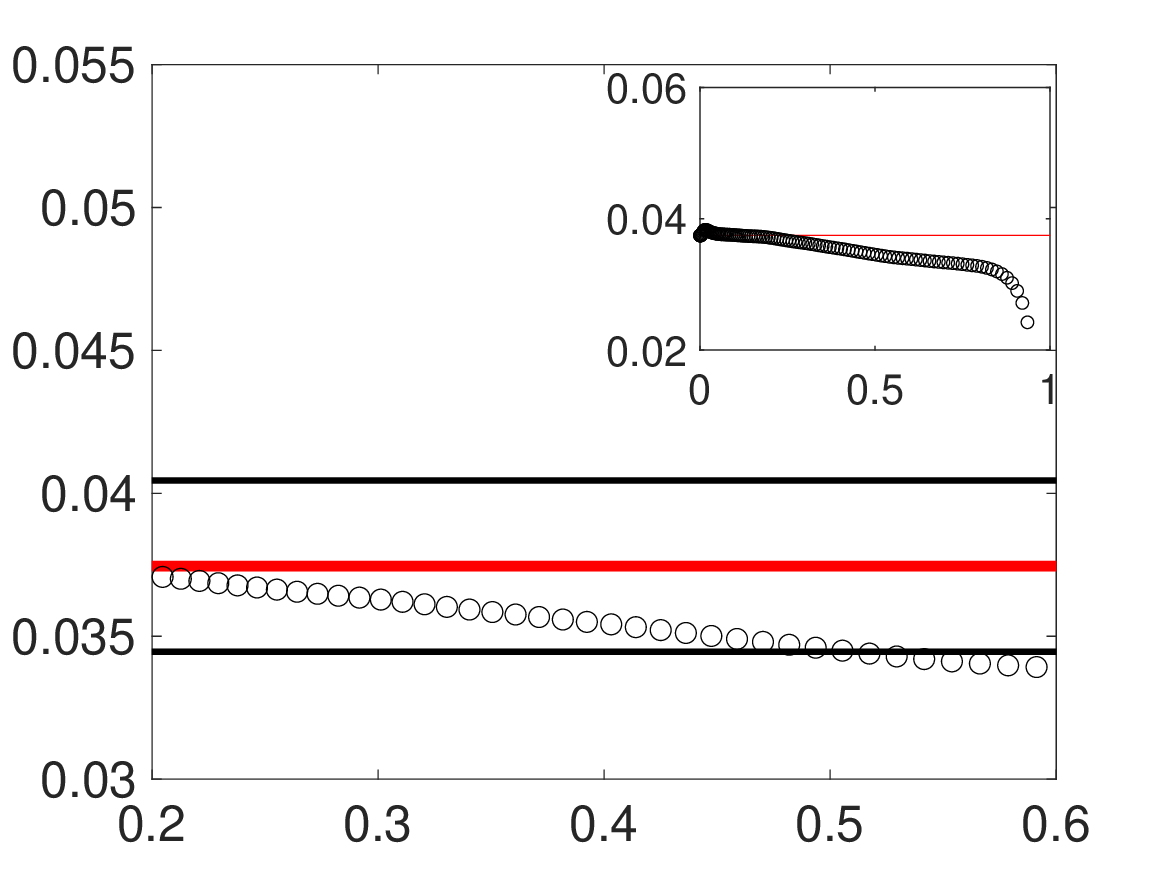}
\put(-220,80){$u_{\tau}/U_e$}
\put(-220,140){$(b)$} \\
\includegraphics[height=55mm]{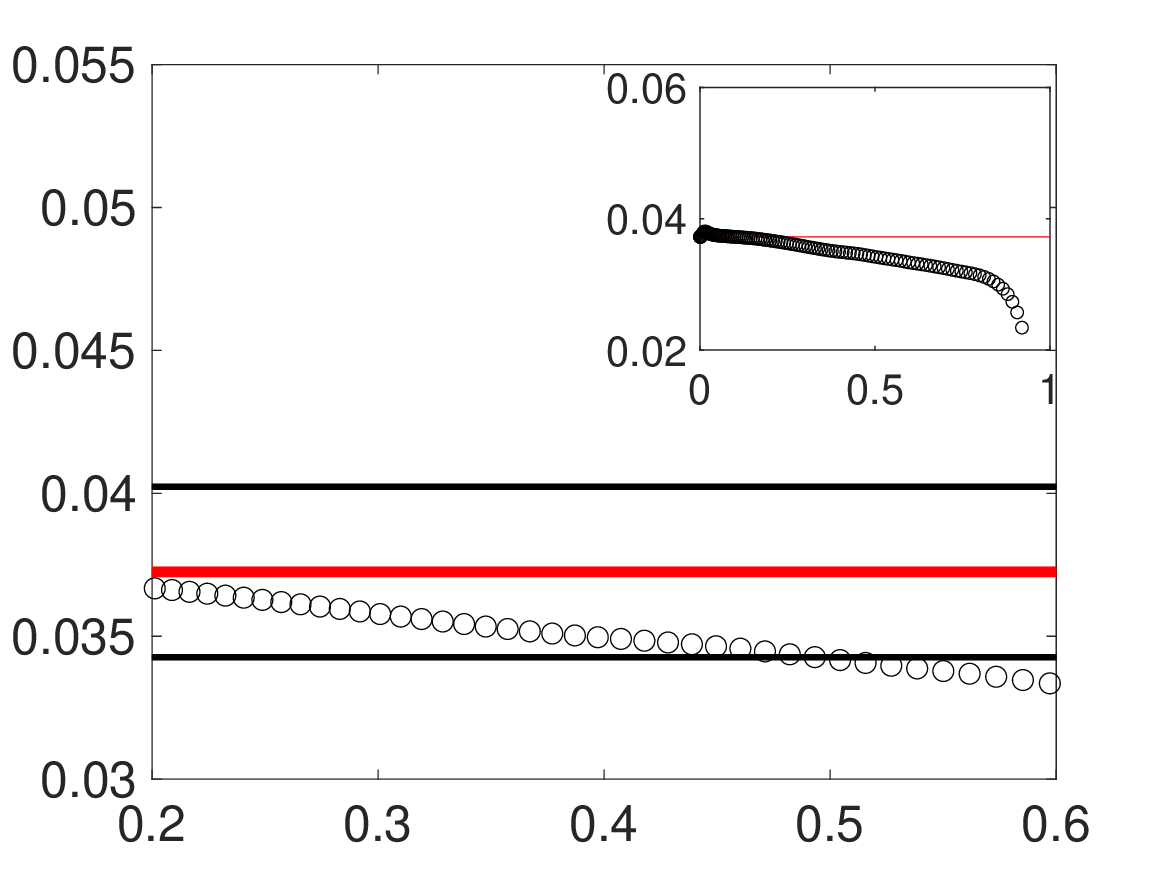}
\put(-220,80){$u_{\tau}/U_e$}
\put(-220,140){$(c)$} \\
\includegraphics[height=55mm]{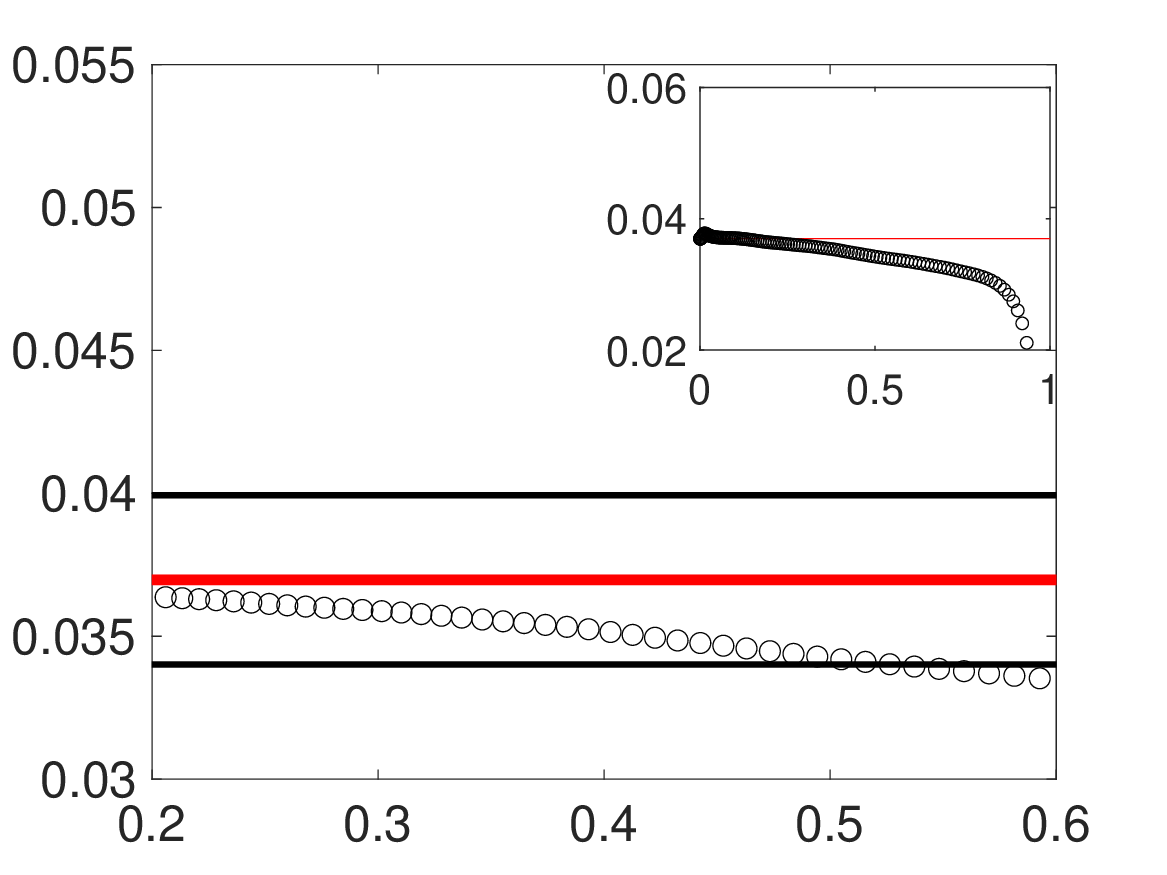}
\put(-110,-5){$\eta$}
\put(-220,80){$u_{\tau}/U_e$}
\put(-220,140){$(d)$}
\end{center}
\caption{Predicted $u_{\tau}$ (symbol) is compared to true $u_{\tau}$ (line) for cases $\beta 11$ $(a)$, $\beta 12$ $(b)$, $\beta 13$ $(c)$, $\beta 14$ $(d)$. Note that the black lines show $\pm 8$ \% of the red line.}
\label{fig:utaub1}
\end{figure}

\begin{figure}
\begin{center}
\includegraphics[height=55mm]{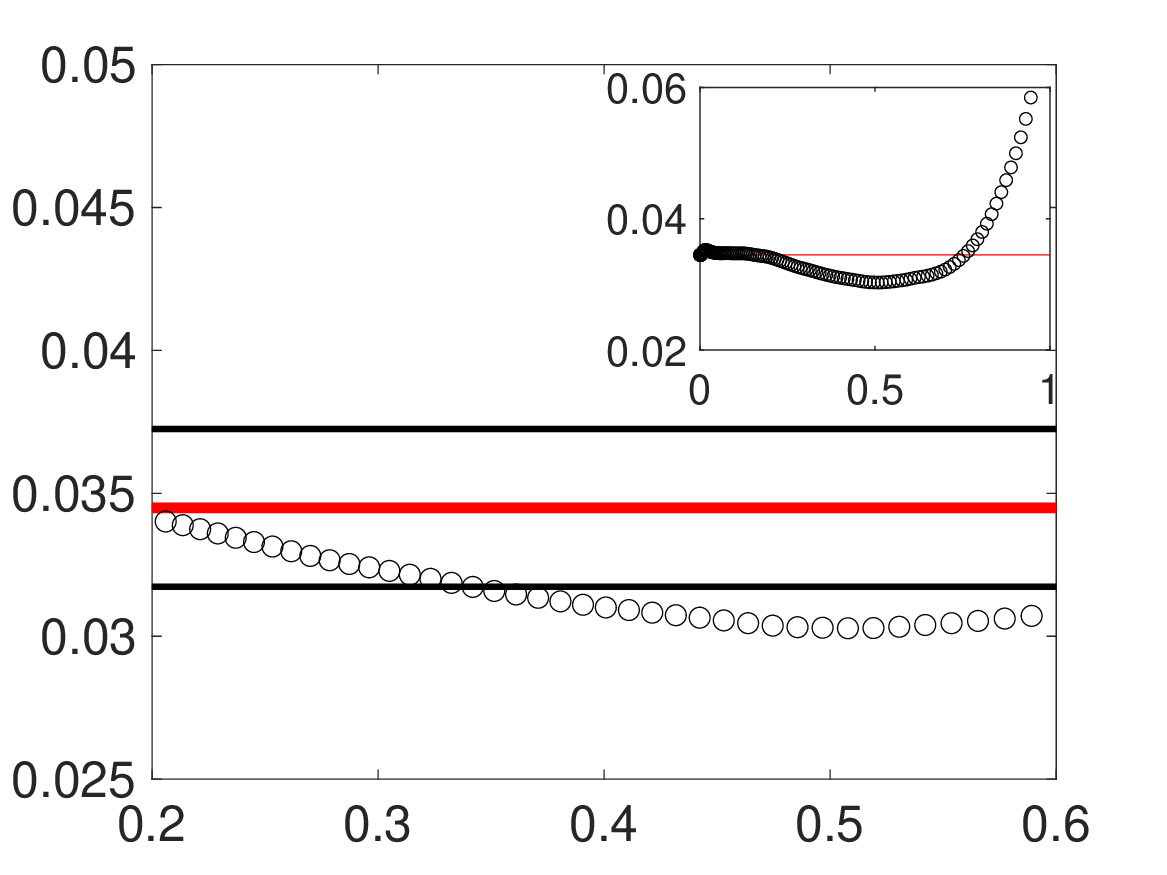}
\put(-220,80){$u_{\tau}/U_e$}
\put(-220,140){$(a)$} \\
\includegraphics[height=55mm]{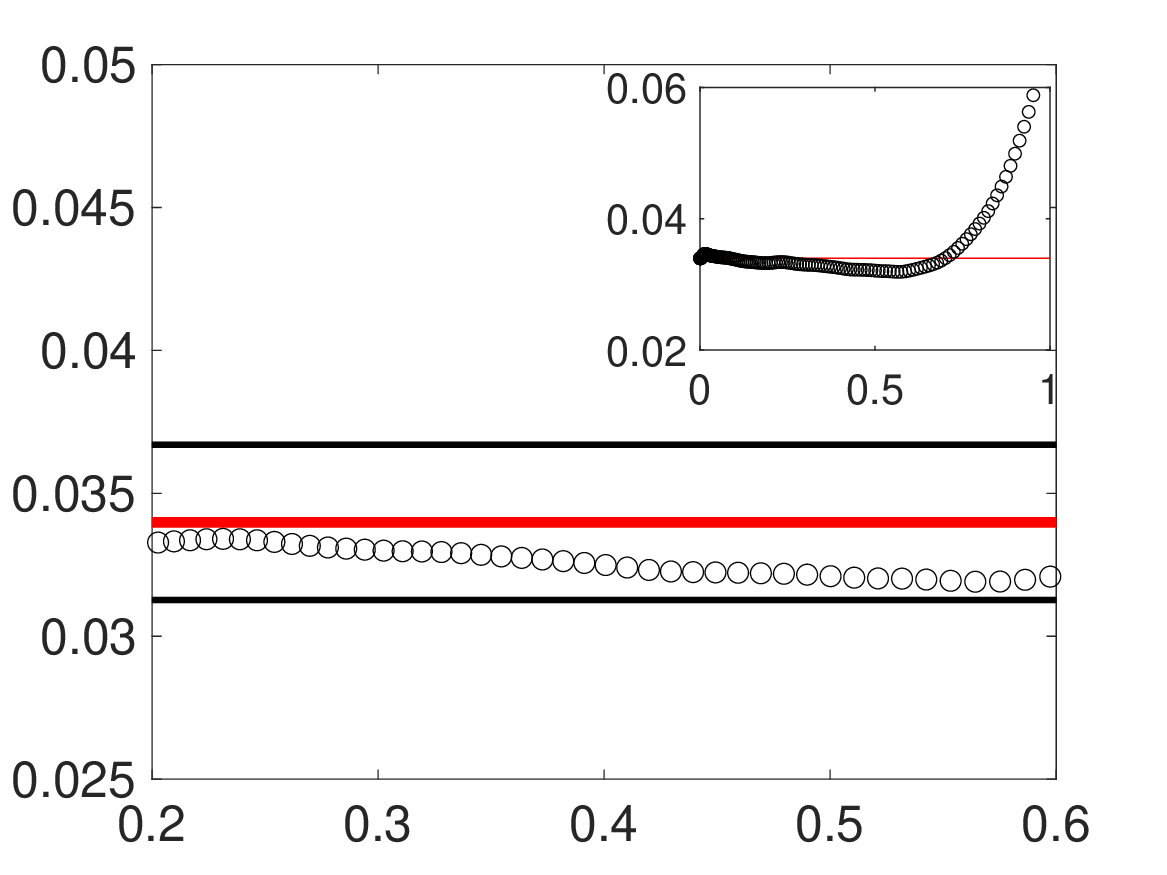}
\put(-220,80){$u_{\tau}/U_e$}
\put(-220,140){$(b)$} \\
\includegraphics[height=55mm]{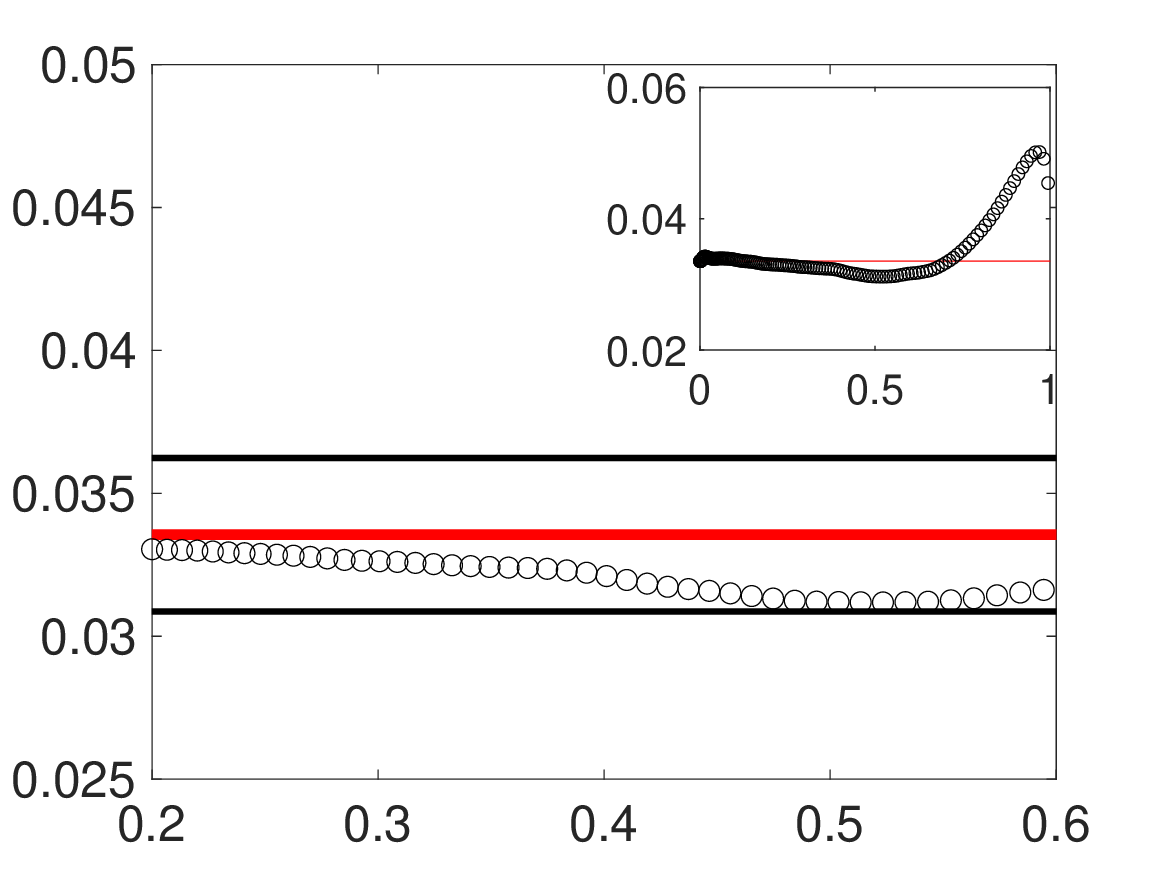}
\put(-220,80){$u_{\tau}/U_e$}
\put(-220,140){$(c)$} \\
\includegraphics[height=55mm]{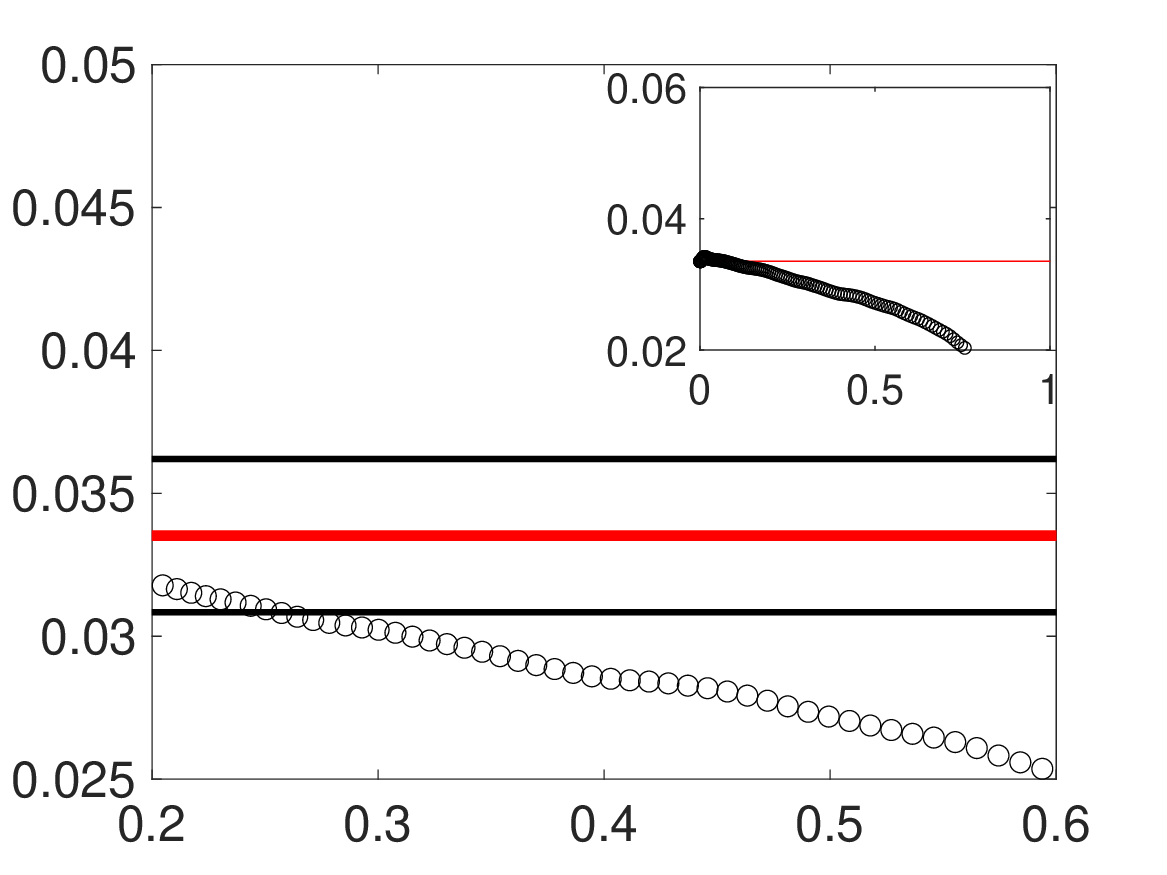}
\put(-110,-5){$\eta$}
\put(-220,80){$u_{\tau}/U_e$}
\put(-220,140){$(d)$}
\end{center}
\caption{Predicted $u_{\tau}$ (symbol) is compared to true $u_{\tau}$ (line) for cases $\beta 21$ $(a)$, $\beta 22$ $(b)$, $\beta 23$ $(c)$, $\beta 24$ $(d)$. Note that the black lines show $\pm 8$ \% of the red line.}
\label{fig:utaub2}
\end{figure}

Figures \ref{fig:utaub1} and  \ref{fig:utaub2} show the predicted $u_{\tau}$ using Eq. \eqref{eq:pgmethod} for $\beta11-\beta14$ and $\beta21-\beta24$ cases respectively, compared to the reference values. The black lines show the error bounds of $\pm 8$. Despite simplicity of the model, the prediction is reasonable. It is important to stress again on the fact that the proposed method does not require any near-wall data from the region $\eta < 0.2$. However, if accurate near-wall data are available, the prediction accuracy is improved. For example, if data in the range $0.1 < \eta < 0.2$ were to be used, it appears that the accuracy will be comparable to the existing methods without requiring ad-hoc corrections and parameter tuning. 

\begin{figure}
\begin{center}
\includegraphics[height=55mm]{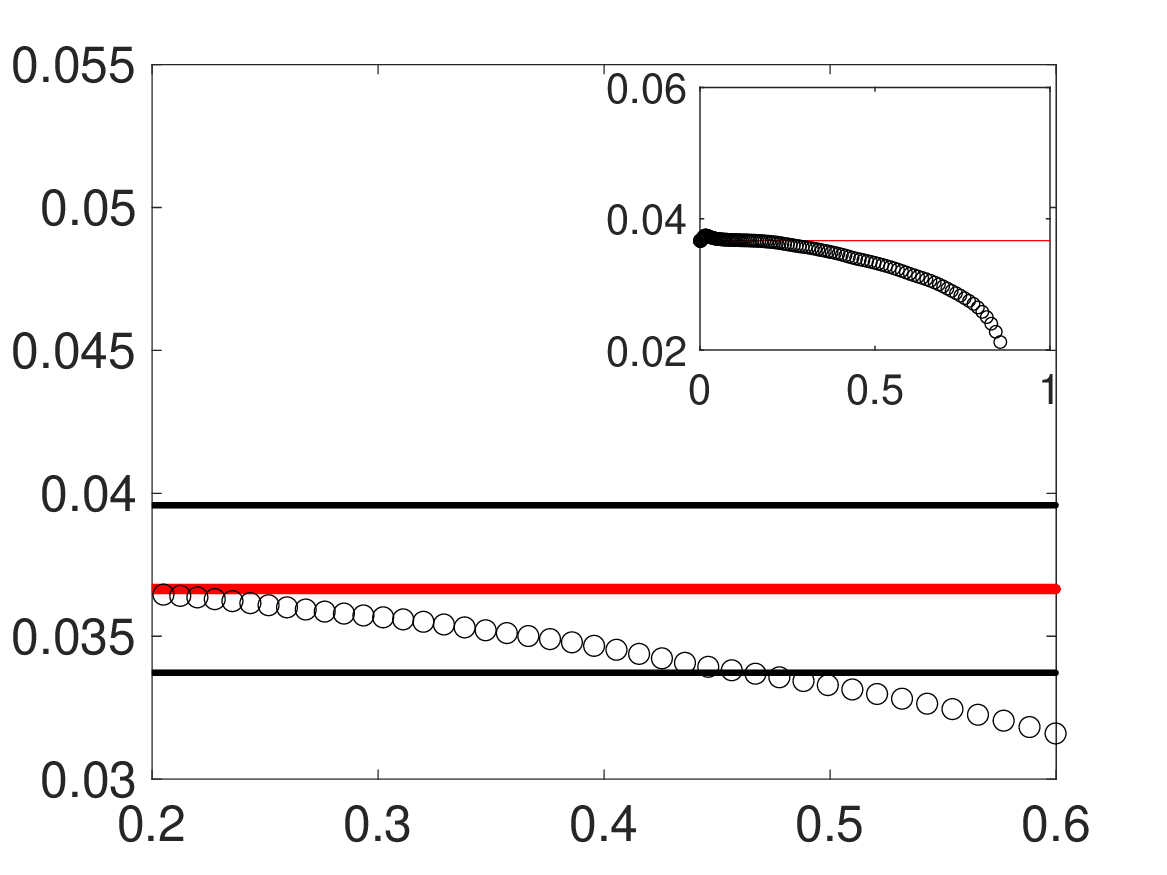}
\put(-220,80){$u_{\tau}/U_e$}
\put(-220,140){$(a)$} \\
\includegraphics[height=55mm]{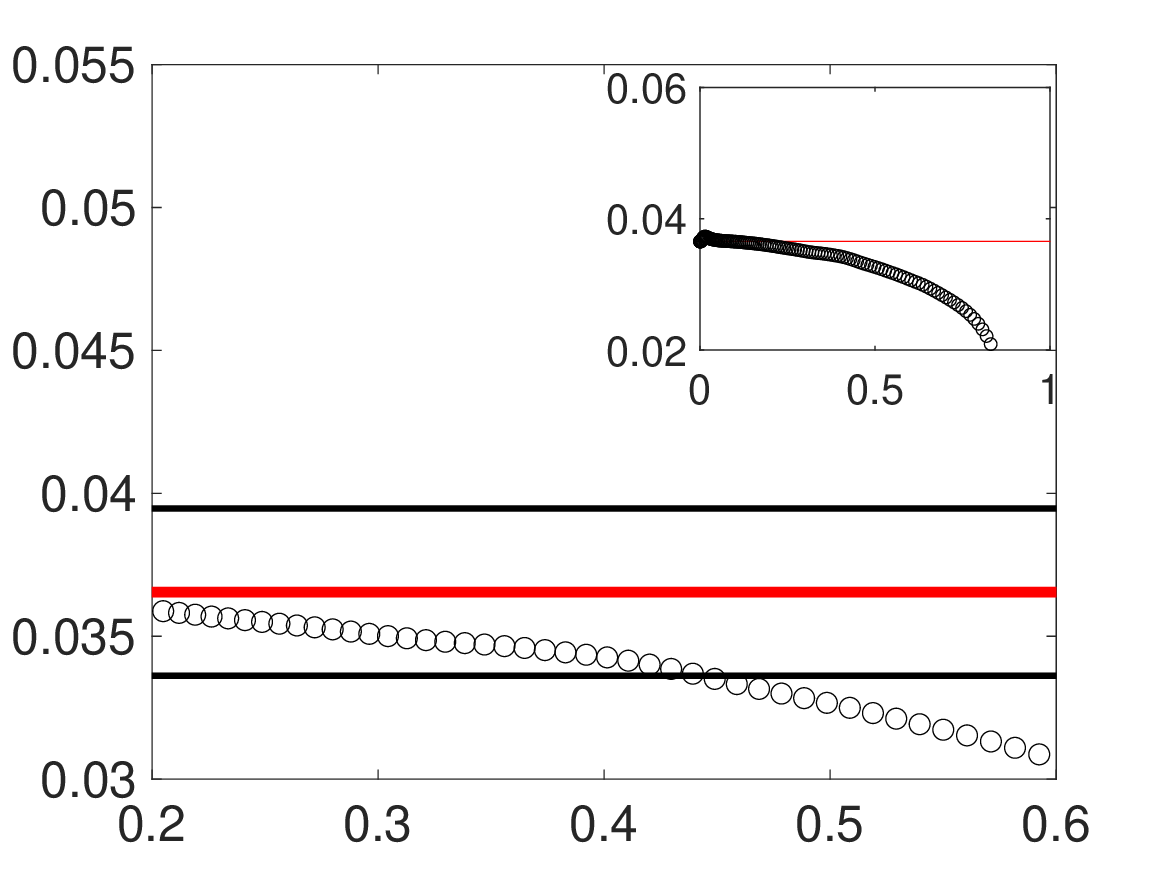}
\put(-220,80){$u_{\tau}/U_e$}
\put(-220,140){$(b)$} \\
\includegraphics[height=55mm]{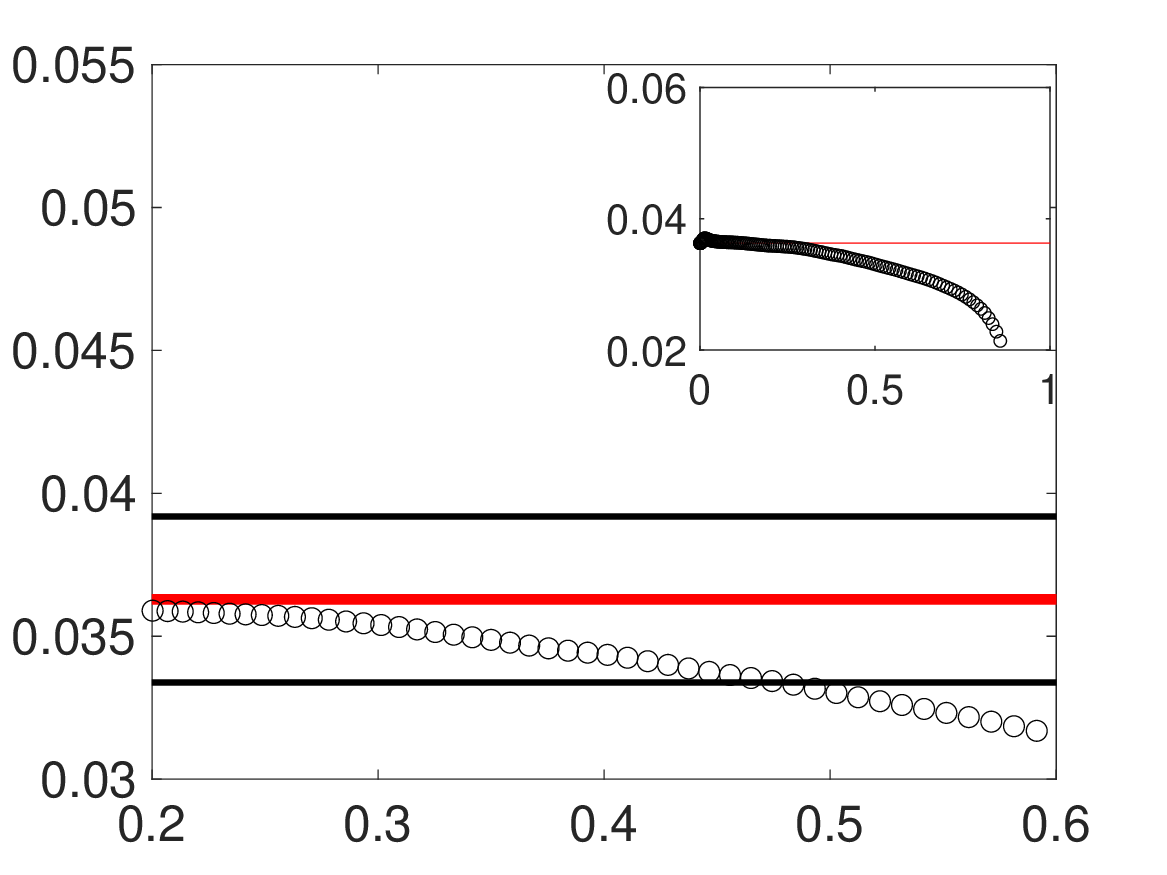}
\put(-220,80){$u_{\tau}/U_e$}
\put(-220,140){$(c)$} \\
\includegraphics[height=55mm]{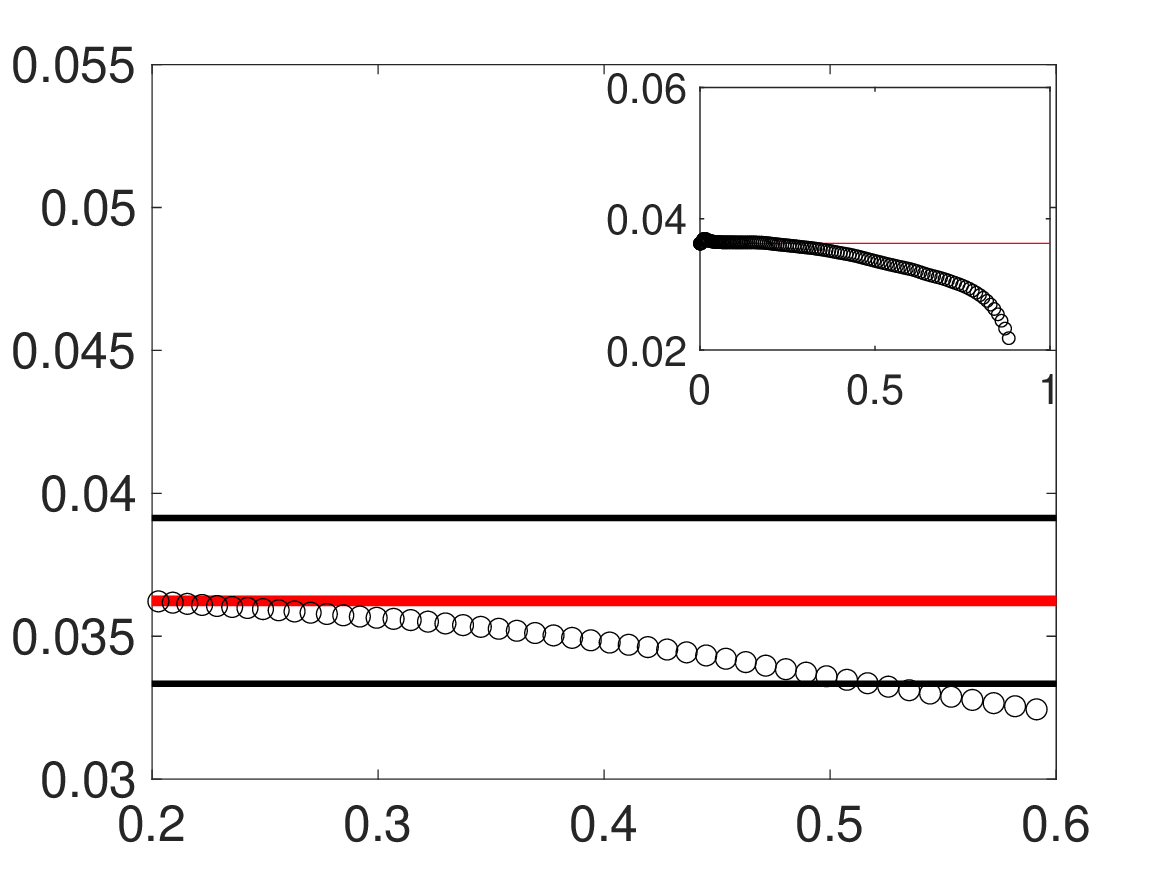}
\put(-110,-5){$\eta$}
\put(-220,80){$u_{\tau}/U_e$}
\put(-220,140){$(d)$}
\end{center}
\caption{Predicted $u_{\tau}$ (symbol) is compared to true $u_{\tau}$ (line) for cases $m131$ $(a)$, $m132$ $(b)$, $m133$ $(c)$, $m134$ $(d)$. Note that the black lines show $\pm 8$ \% of the red line.}
\label{fig:utaum13}
\end{figure}

\begin{figure}
\begin{center}
\includegraphics[height=55mm]{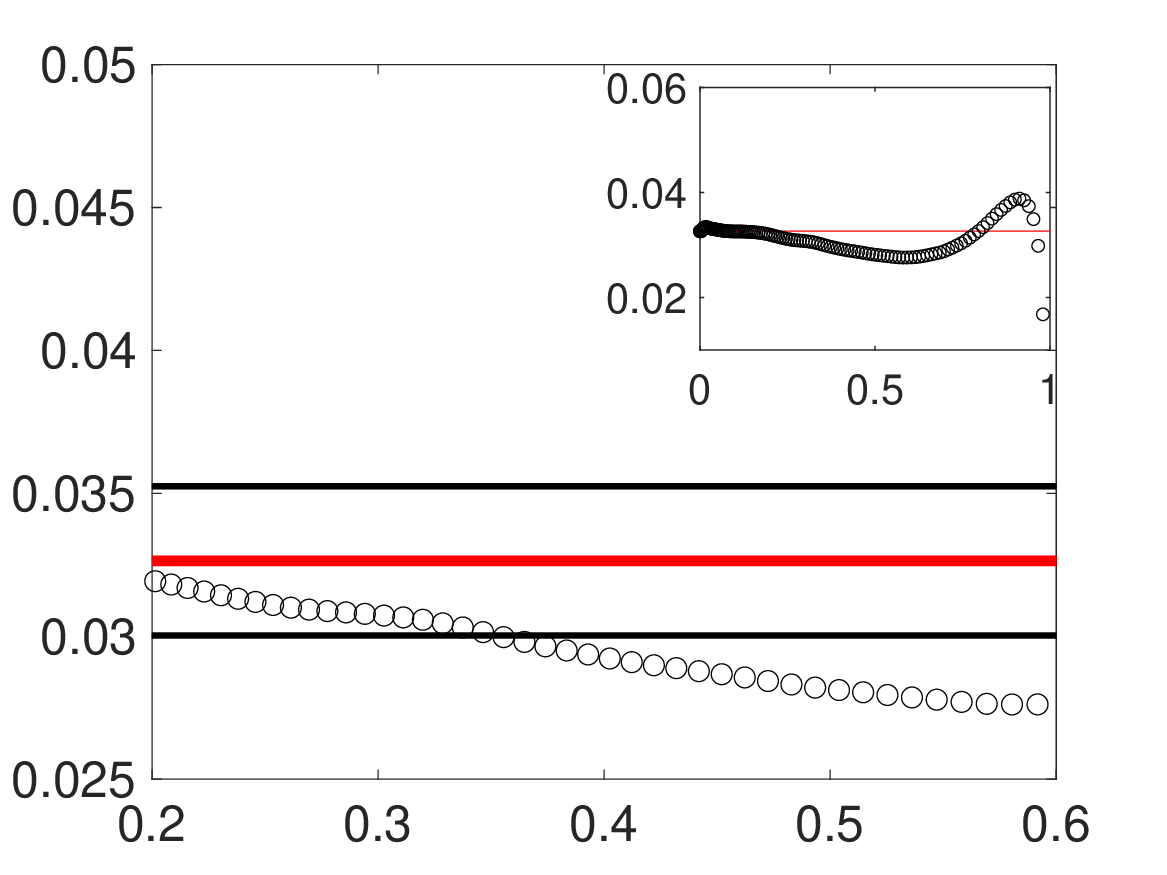}
\put(-220,80){$u_{\tau}/U_e$}
\put(-220,140){$(a)$} \\
\includegraphics[height=55mm]{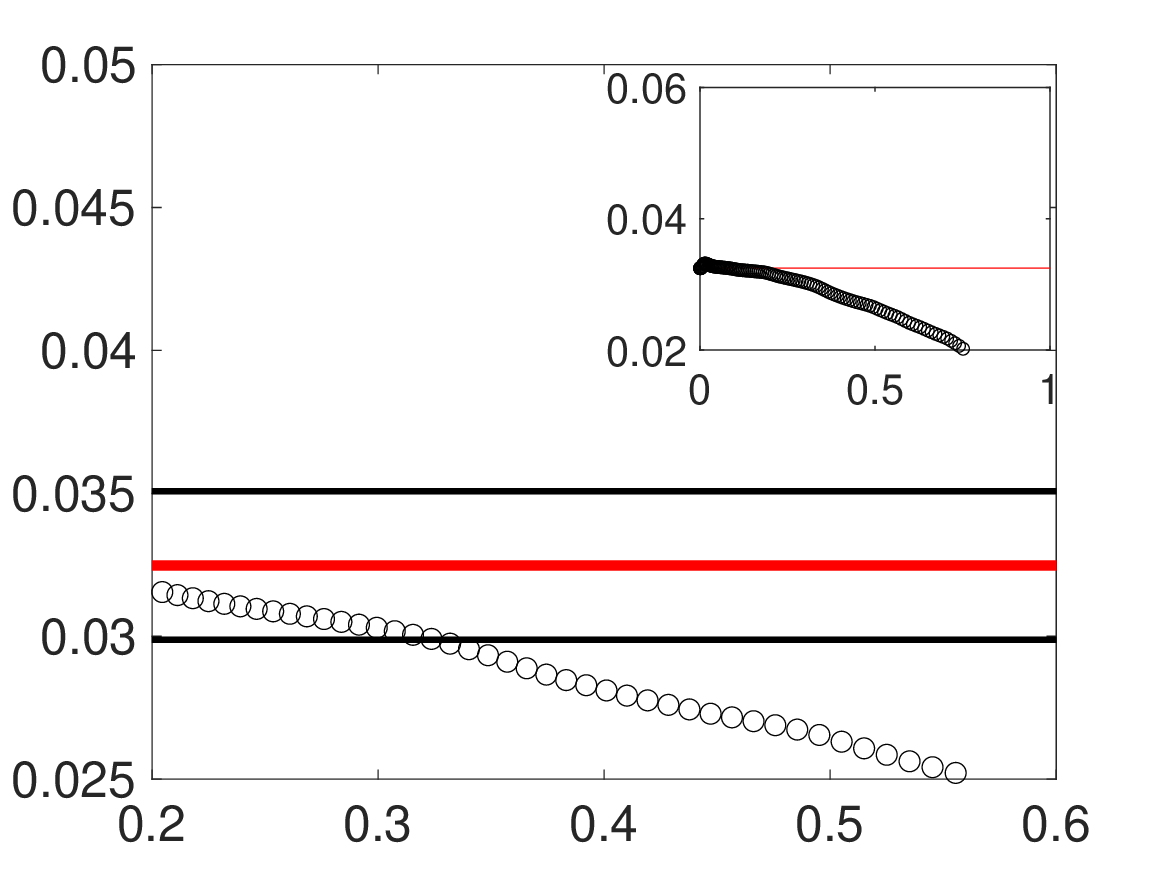}
\put(-220,80){$u_{\tau}/U_e$}
\put(-220,140){$(b)$} \\
\includegraphics[height=55mm]{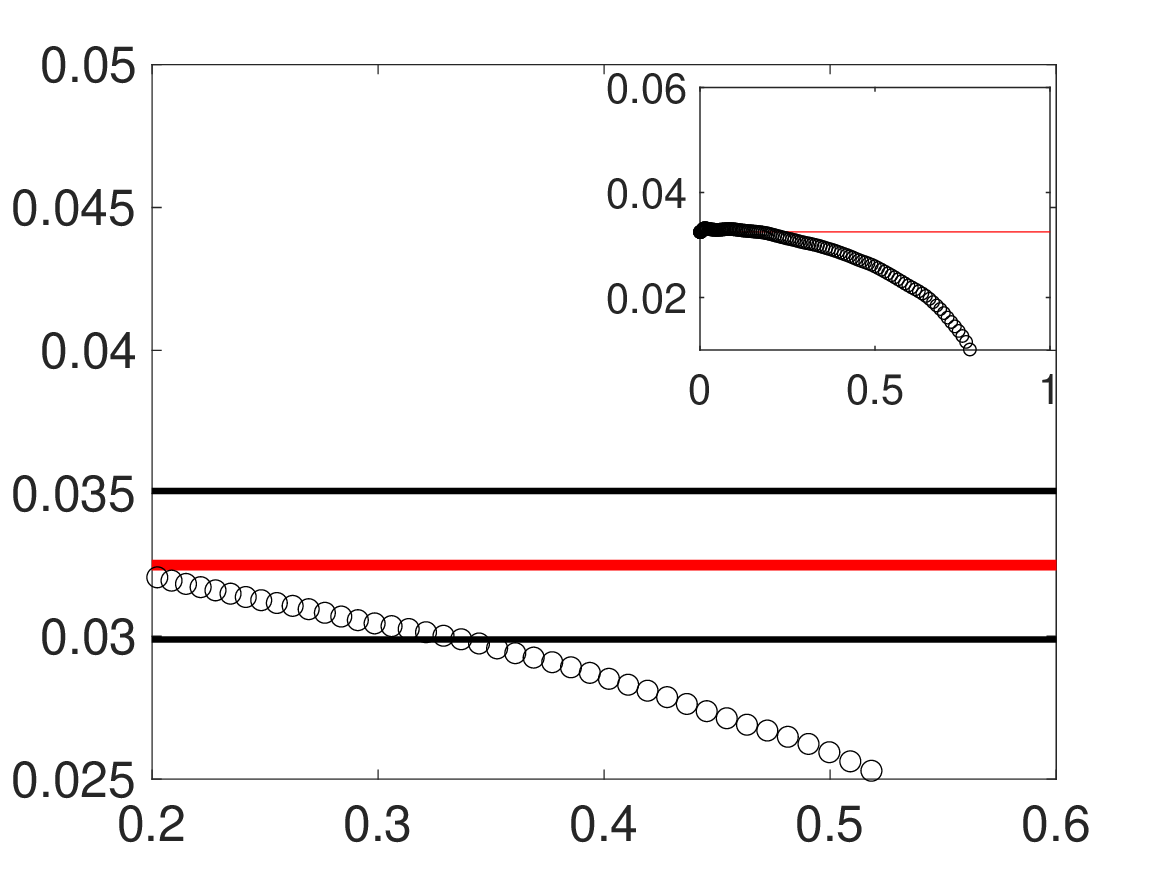}
\put(-220,80){$u_{\tau}/U_e$}
\put(-220,140){$(c)$} \\
\includegraphics[height=55mm]{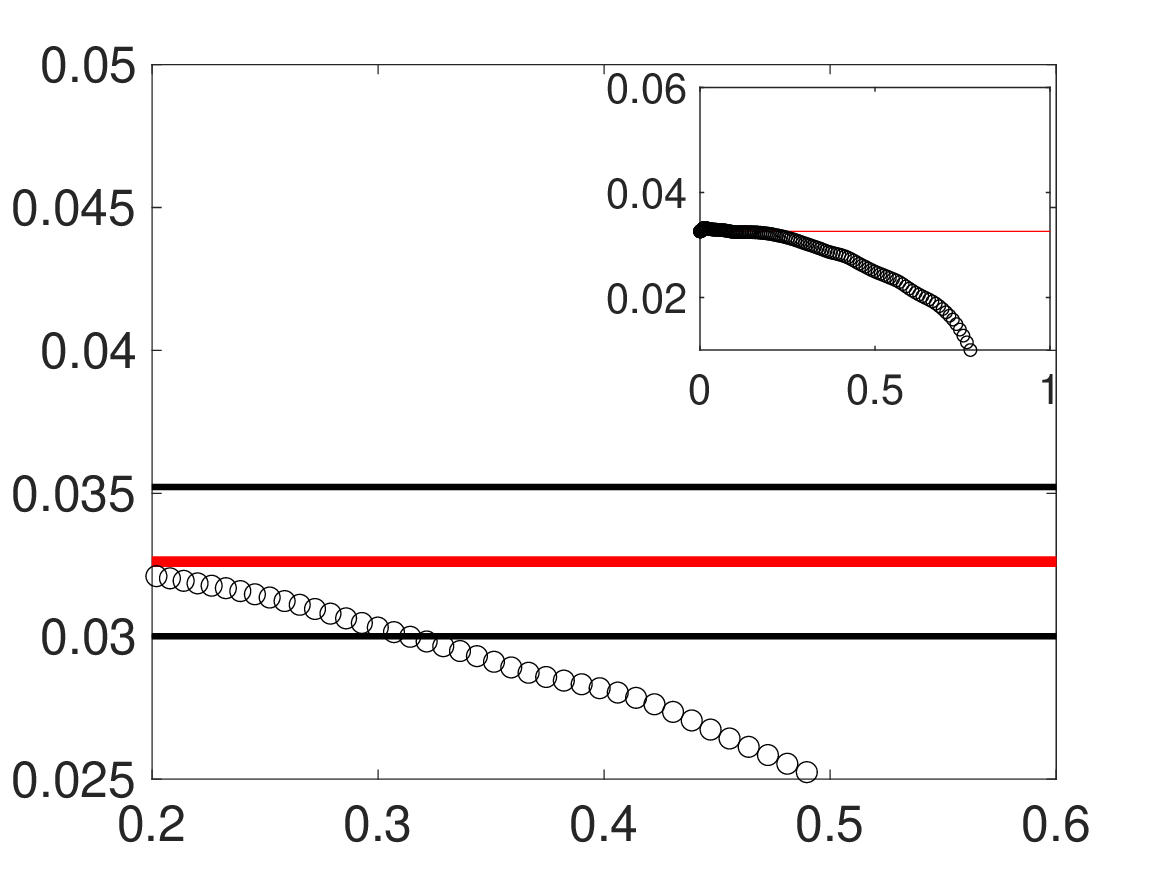}
\put(-110,-5){$\eta$}
\put(-220,80){$u_{\tau}/U_e$}
\put(-220,140){$(d)$}
\end{center}
\caption{Predicted $u_{\tau}$ (symbol) is compared to true $u_{\tau}$ (line) for cases $m161$ $(a)$, $m162$ $(b)$, $m163$ $(c)$, $m164$ $(d)$. Note that the black lines show $\pm 8$ \% of the red line.}
\label{fig:utaum16}
\end{figure}

In order to test robustness, the proposed method is also tested for more challenging constant-$m$ TBL cases in Figures \ref{fig:utaum13} and \ref{fig:utaum16}. The method gives good accuracy for $m131-m134$. The TBL with $m=-0.16$ is the most challenging pressure gradient case considered in the present work, due to rapid variation of $\beta$ (Figure \ref{fig:pgtbl2}). However, the overall accuracy is still reasonable for $\eta < 0.35$ for $m161-m164$ profiles.

\subsection{Sensitivity to TBL parameters} 
Boundary layer parameters used as inputs to the proposed method for pressure gradient TBL (Eq. \eqref{eq:pgmethod}) are $\eta$, $Re_{\delta}$ and $K$. It appears that the only source of error can be from the approximation $\delta \approx \delta_{99}$. However, it can be expected that the model performance would not degrade as long as data points are available below $\eta < 0.35$. Note that $\delta$ used in the present work is that reported by \cite{bobke2017}, who used the method developed by \cite{vinuesa2016} to obtain $\delta$ from their LES results. $H$ does not appear in Eq. \eqref{eq:pgmethod}, and hence is not required as input to the method.  


\section{Conclusions} \label{sec:conc}
A novel method to determine $u_{\tau}$ from measured profile data is proposed based on the mean stress model of \cite{kumar2021}. The method is based on integral analysis of the governing equations of the mean flow and like all other such methods, requires shear stress profile to determine $u_{\tau}$. However, unlike all the other methods, the proposed method does not require any near-wall ($\eta < 0.2$) data, thereby avoiding the problems associated with presumption of a universal mean profile and the associated parameters. Unlike most existing methods, the proposed method can handle wall roughness without any modification, and is shown to predict $u_{\tau}$ accurately for a range of $Re$ for both smooth and rough wall TBL. The method requires $H$ and $\delta$ as inputs, and it is shown to be relatively insensitive to the uncertainty in these parameters. 

Since, the method is based on the mean stress model of \cite{kumar2021} which was derived for ZPG TBL, it requires extension to include pressure gradient effects. Hence, the model derivation of \cite{kumar2021} is revisited to obtain novel model for mean stress and wall-normal velocity in TBL including the pressure gradient effects,  which is subsequently used to propose a method to reliably determine $u_{\tau}$ from profile data. Therefore, the proposed method provides an alternative to current popular methods to indirectly determine $u_{\tau}$ in TBL within an acceptable uncertainty, and is general enough to handle wall roughness and pressure gradient effects. 

A step-by-step recipe to obtain $u_{\tau}$ from measurement data is as follows:
\begin{enumerate}
  \item Using measured $U$ profile, obtain $\delta$ and hence $\eta$, $K$, $Re_{\delta}$ and $V/V_e$.
  \item Plot the rhs of Eq. \eqref{eq:pgmethod} in the range $0.2 < \eta < 0.5$ for ZPG or $0.2 < \eta < 0.4$ for pressure gradient TBL.
  \item Draw a horizontal line which best-fits the plotted data. Predicted $u_{\tau}$ is the value where this line intersects the $y-$axis. 
\end{enumerate}

The proposed method requires the knowledge of $K$, $U_e$ and ${\delta}$, which can be challenging to obtain in pressure gradient TBL. Also, an X-wire probe or PIV might be required to measure $T$. The method is developed for a spatially growing TBL and hence, it does not work for internal flows such as channel and pipe flows. The method incorporates wall roughness and pressure gradient effects, but requires modifications to account for additional complications like transverse curvature, wall injection or suction etc. Lastly, the accuracy of the method relies on the accuracy of the proposed model for $I$ (Eq. \eqref{eq:Imodel}), which appears to be acceptable for the TBL cases shown in the present work. A more accurate model for $I$ has the potential to improve the accuracy of the proposed method further.

\begin{acknowledgements}
This  work  is  supported  by  the  United  States  Office  of  Naval  Research  (ONR)  under ONR  Grant  N00014-20-1-2717 with Dr. Peter Chang as technical monitor.  The  authors  thank  Prof. J.  Klewicki  for  providing the experimental data published in \citet{morrill2015}.
\end{acknowledgements}

\section*{Conflict of interest}
The authors declare that they have no conflict of interest.

\bibliographystyle{spbasic}
\bibliography{draft}

\end{document}